\documentclass[sigconf]{acmart}
\usepackage{subfiles}
\usepackage{paralist}
\usepackage{amsmath,amsfonts,amsthm}
\usepackage{mathrsfs}
\usepackage{amsbsy}
\usepackage[mathscr]{eucal} 
\usepackage{algorithmic}

\usepackage[ruled,vlined,linesnumbered,resetcount]{algorithm2e}
\usepackage{booktabs}
\usepackage{graphicx}
\usepackage[T1]{fontenc}
\usepackage{textcomp}
\usepackage{xcolor}
\usepackage{multirow}
\usepackage[center]{subfigure}
\usepackage[font=normalsize]{caption}
\usepackage{bbold}
\usepackage{mathtools}
\usepackage{enumitem}
\SetKwComment{Comment}{$\triangleright$\ }{}
\SetKwInput{KwData}{Input}
\SetKwInput{KwResult}{Output}
\DeclarePairedDelimiter{\abs}{\lvert}{\rvert}
\newcommand{\rom}[1]{\uppercase\expandafter{\romannumeral #1\relax}}
\def\BibTeX{{\rm B\kern-.05em{\sc i\kern-.025em b}\kern-.08em
		T\kern-.1667em\lower.7ex\hbox{E}\kern-.125emX}}
\makeatletter
\def\endthebibliography{%
	\def\@noitemerr{\@latex@warning{Empty `thebibliography' environment}}%
	\endlist
}
\makeatother

\newcommand{\method}{\textsc{SSumM}\xspace}

\newcommand{\kGs}{\textsc{k-Gs}\xspace}
\newcommand{\Gs}{\textsc{Gs}\xspace}

\newcommand{\SAAGs}{\textsc{SAA-Gs}\xspace}

\newcommand{\linearSAAGs}{\textsc{SAA-Gs} (linear sample)\xspace}

\newcommand{\SL}{\textsc{S2L}\xspace}

\newcommand{\summary}{$\overline{G}$\xspace}

\newcommand{\nsummary}{\overline{G}\xspace}

\newcommand{\graph}{$G$\xspace}

\newcommand{\adj}{$A$\xspace}

\newcommand{\REadj}{$\hat{A}$\xspace}

\newcommand{\size}{$k$\xspace}

\newcommand{\fgraph}{$G = (V, E)$\xspace}

\newcommand{\fsummary}{$\overline{G} = (S, P, \omega)$\xspace}

\newcommand{\rgraph}{$\hat{G} = (V, \hat{E}, \hat{\omega})$\xspace}

\newcommand{\F}{$\omega : P \rightarrow \mathbb{Z}^{+}$\xspace}

\newcommand{\snode}{$S$\xspace}

\newcommand{\sedge}{$P$\xspace}

\newcommand{\reconstruction}{$\ell_p$ \textit{reconstruction error} ($RE_p$)\xspace}

\newcommand{\neighbor}{$\ST$\xspace}

\newcommand{\lpp}{$\ell_{p}$\xspace}

\newcommand{\lone}{$\ell_{1}$\xspace}

\newcommand{\ltwo}{$\ell_{2}$\xspace}

\newcommand{\re}{reconstruction error\xspace}

\newtheoremstyle{problemstyle}  
{3pt}                                               
{3pt}                                               
{\normalfont}                               
{}                                                  
{\bfseries\itshape}                 
{\normalfont\bfseries:}         
{.5em}                                          
{}                                                  
\theoremstyle{problemstyle}

\newtheorem{problem}{Problem}

\newcommand{\smallsection}[1]{{\vspace{0.05in} \noindent {\bf{\underline{\smash{#1}}}}}}

\newcommand\blue[1]{\textcolor{blue}{#1}}

\newcommand{\Vu}{V_{u}}
\newcommand{\Vv}{V_{v}}
\newcommand{\EAB}{E_{AB}}

\newcommand{\Euv}{E_{\Vu\Vv}}

\newcommand{\VuVv}{\{V_{u},V_{v}\}}
\newcommand{\GH}{\hat{G}}
\newcommand{\GB}{\overline{G}}
\newcommand{\EH}{\hat{E}}
\newcommand{\wH}{\hat{\omega}}
\newcommand{\AH}{\hat{A}}
\newcommand{\wmax}{\omega_{max}}
\newcommand{\PIAB}{\Pi_{AB}}
\newcommand{\PIuv}{\Pi_{\Vu\Vv}}
\newcommand{\AB}{\{A,B\}}
\newcommand{\PS}{\Pi_{S}}
\newcommand{\ST}{\mathcal{S}_{t}}
\newcommand{\PSTAR}[1]{P^{\star}(#1)}
\newcommand{\CBAR}{\bar{C}}
\newcommand{\GBSTAR}[1]{\bar{G}^{\star}(#1)}
\newcommand{\CSTAR}[1]{Cost^{\star}(#1)}
\newcommand{\CSTARSUB}[2]{Cost^{\star}_{#1}(#2)}
\newcommand{\CSUB}[2]{Cost_{#1}(#2)}
\newcommand{\SPRIME}{S^{\prime}}
\newcommand{\APRIME}{A^{\prime}}

\newcommand{\anneal}{\theta(t)}
\newcommand{\sneighbor}[1]{\bar{N}_{#1}}
\newcommand{\reductionAB}{Reduction(A,B)}

\newcommand{\EAA}{E_{AA}}
\newcommand{\EBB}{E_{BB}}
\newcommand{\EAPRIME}{E_{\APRIME\APRIME}}

\newcommand{\PAPRIME}{\sigma_{\APRIME\APRIME}}
\newcommand{\PIAC}{\Pi_{AC}}
\newcommand{\PIAA}{\Pi_{AA}}
\newcommand{\PIBB}{\Pi_{AB}}
\newcommand{\PIAPRIME}{\Pi_{\APRIME\APRIME}}
\newcommand{\PIAPRIMEC}{\Pi_{\APRIME C}}
\newcommand{\EAC}{E_{AC}}
\newcommand{\EAPRIMEC}{E_{\APRIME C}}
\newcommand{\PAC}{\sigma_{AC}}
\newcommand{\PAPRIMEC}{\sigma_{\APRIME C}}

\newcommand{\Pedge}{\{A,C\} \in \PSTAR{S}}
\newcommand{\NPedge}{\{A,C\} \notin \PSTAR{S}}
\newcommand{\APedge}{\{\APRIME,C\} \in \PSTAR{\SPRIME}}
\newcommand{\ANPedge}{\{\APRIME,C\} \notin \PSTAR{\SPRIME}}

\AtBeginDocument{%
  \providecommand\BibTeX{{%
    \normalfont B\kern-0.5em{\scshape i\kern-0.25em b}\kern-0.8em\TeX}}}

\setcopyright{acmcopyright}
\copyrightyear{2020}
\acmYear{2020}
\acmDOI{10.1145/3394486.3403057}

\acmConference[KDD '20]{Proceedings of the 26th ACM SIGKDD Conference on Knowledge Discovery and Data Mining}{August 23--27, 2020}{Virtual Event, USA}
\acmPrice{15.00}
\acmISBN{978-1-4503-7998-4/20/08}

\begin{document}
\title{SSumM: Sparse Summarization of Massive Graphs}

\author{Kyuhan Lee}
\authornote{Equal Contribution. \footnotemark[2]Corresponding author.}
\affiliation{%
	\institution{KAIST AI}
}
\email{kyuhan.lee@kaist.ac.kr}

\author{Hyeonsoo Jo}
\authornotemark[1]
\affiliation{%
	\institution{KAIST AI}
}
\email{hsjo@kaist.ac.kr}

\author{Jihoon Ko}
\affiliation{%
	\institution{KAIST AI}
}
\email{jihoonko@kaist.ac.kr}

\author{Sungsu Lim}
\affiliation{%
	\institution{CNU CSE}
}
\email{sungsu@cnu.ac.kr}

\author{Kijung Shin}
\authornotemark[2]
\affiliation{%
	\institution{KAIST AI \& EE}
}
\email{kijungs@kaist.ac.kr}


\begin{abstract}
\vspace{-1mm}
Given a graph $G$ and the desired size $k$ in bits,
how can we summarize $G$ within $k$ bits, while minimizing the information loss?

Large-scale graphs have become omnipresent, posing considerable computational challenges.
Analyzing such large graphs can be fast and easy if they are compressed sufficiently to fit in main memory or even cache. 
Graph summarization, which yields a coarse-grained summary graph with merged nodes, stands out with several advantages among graph compression techniques. 
Thus, a number of algorithms have been developed for obtaining a concise summary graph with little information loss or equivalently small reconstruction error.
However, the existing methods focus solely on reducing the number of nodes, and they often yield dense summary graphs, failing to achieve better compression rates.
Moreover, due to their limited scalability, they can be applied only to moderate-size graphs.

In this work, we propose \method, a scalable and effective graph-summarization algorithm that yields a sparse summary graph. 
\method not only merges nodes together but also sparsifies the summary graph, and the two strategies are carefully balanced based on the minimum description length principle.
Compared with state-of-the-art competitors, \method is 
\textbf{\textit{(a) Concise:}} yields up to $11.2\times$ smaller summary graphs with similar reconstruction error, 
\textbf{\textit{(b) Accurate:}} achieves up to $4.2\times$ smaller reconstruction error with similarly concise outputs,
and \textbf{\textit{(c) Scalable:}} summarizes $26\times$ larger graphs while exhibiting linear scalability.
We validate these advantages through extensive experiments on $10$ real-world graphs. 
\vspace{-4mm}
\end{abstract}

\maketitle


\begin{figure}[t]
	\centering
	\includegraphics[width=\linewidth]{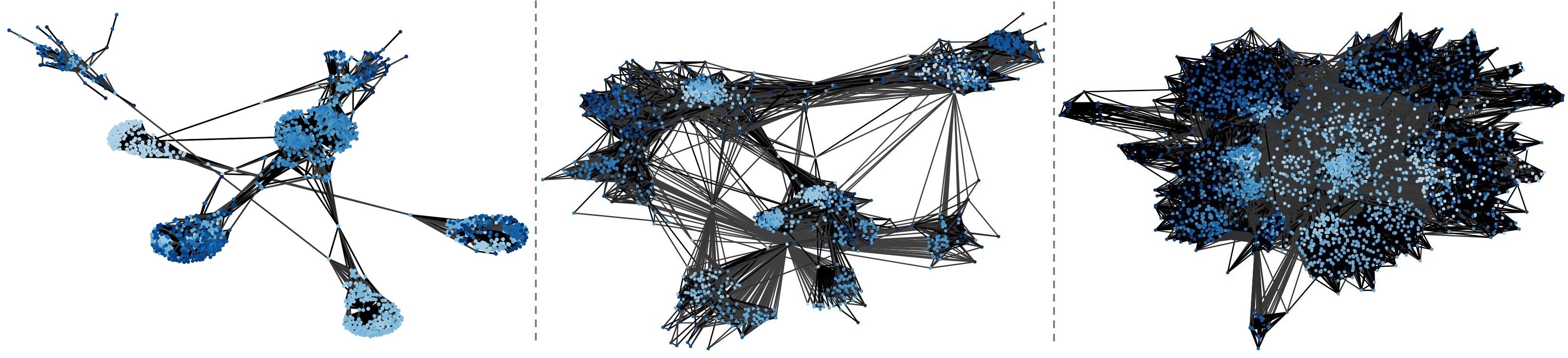}
	\includegraphics[width=\linewidth]{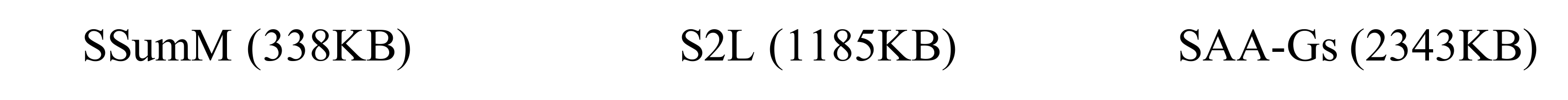} \\
	\vspace{-0.5mm}
	\caption{\label{fig:vis} \underline{\smash{\method gives sparse and concise summary graphs.}} The reconstruction errors of the above summary graphs, obtained by different algorithms, are similar ($\pm5\%$).  
	}
\end{figure}

\vspace{-1mm}
\section{Introduction}
\label{intro}
\vspace{-1mm}

Graphs are a fundamental abstraction that is widely used to represent a variety of relational datasets.
As the underlying data are accumulated rapidly, massive graphs have appeared, such as (a) $3.5$ billion web pages with $128$ billion hyperlinks \cite{JWS-0003}, (b) professional networks with more than $20$ billion connections \cite{shin2019sweg}, and (c) social networks with hundreds of billions of connections \cite{ching2015one}.

Despite the abundance of massive graphs, 
many existing graph-analysis tools 
are inapplicable to such graphs since their computational costs grow rapidly with the size of graphs.
Moreover, massive graphs often do not fit in main memory, causing I/O delays over the network or to the disk.

These problems can be addressed by \textit{graph summarization}, which aims to preserve the essential structure of a graph while shrinking its size by removing minor details.
Given a graph \fgraph and the desired size $k$, the objective of the graph summarization problem is to find a {\it summary graph} \fsummary of size $k$ from which $G$ can be accurately reconstructed. 
The set \snode is a set of {\it supernodes}, which are distinct and exhaustive subsets of nodes in $G$, and the set \sedge is a set of {\it superedges} (i.e., edges between supernodes).
The weight function $\omega$ assigns an integer to each superedge. 
Given the summary graph $\GB$, we reconstruct a graph $\GH$ by connecting all pairs of nodes belonging to the source and destination supernodes of each superedge and assigning a weight, computed from the weight of the superedge, to each created edge.
Note that $\GH$ is not necessarily the same with $G$, and we call their similarity the {\it accuracy} of $\GB$.
\par


Graph summarization stands out among a variety of graph-compression techniques (relabeling nodes \cite{boldi2004webgraph, apostolico2009graph, chierichetti2009compressing}, encoding frequent substructures with few bits \cite{buehrer2008scalable,khan2015set}, etc.) due to the following benefits:  
\textbf{(a)\,Elastic}: we can reduce the size of outputs (i.e., a summary graph) as much as we want at the expense of increasing reconstruction errors. \textbf{(b)\,Analyzable}: since the output of graph summarization is also a graph, existing graph analysis and visualization tools can easily be applied. For example, \cite{riondato2017graph, lefevre2010grass, beg2018scalable} compute adjacency queries, PageRank \cite{page1999pagerank}, and triangle density \cite{tsourakakis2015k} directly from summary graphs, without restoring the original graph. \textbf{(c)\,Combinable for Additional Compression}: due to the same reason,  the output summary graph can be further compressed using any graph-compression techniques.\par




While a number of graph-summarization algorithms \cite{lefevre2010grass, riondato2017graph, beg2018scalable} have been developed for finding accurate summary graphs (i.e., those with low reconstruction errors) and eventually realizing the above benefits, they share common limitations.
First, their scalability is severely limited, and they cannot be applied to billion-scale graphs for which graph summarization can be extremely useful. Specifically, the largest graph to which they were applied has only about $3$ million nodes and $34$ million edges, which take only about $23.8$MB \cite{beg2018scalable}.
More importantly, existing algorithms are not effective in reducing the size in bits of graphs since they solely focus on reducing the number of nodes (see Fig.~\ref{fig:vis}).
Surprisingly, the size in bits of summary graphs often exceeds that of the original graphs, as reported in \cite{riondato2017graph} and shown in our experiments.  

\par

To address these limitations, we propose \method(\textbf{S}parse \textbf{Sum}mar-ization of \textbf{M}assive Graphs), a scalable graph-summarization algorithm that yields concise but accurate summary graphs. 
\method focuses on minimizing reconstruction errors while limiting the size in bits of the summary graph, instead of the number of nodes.
Moreover, to co-optimize the compactness and accuracy, \method carefully combines nodes and at the same time sparsifies edges. 
Lastly, for scalability, \method rapidly searches promising candidate pairs of nodes to be merged. 
As a result, \method significantly outperforms its state-of-the-art competitors in terms of scalability and the compactness and accuracy of outputs. \par

\begin{figure}[t]
	\vspace{-5mm}
	\centering
	\subfigure[Compactness and Accuracy]{
	\includegraphics[width= 0.42\linewidth]{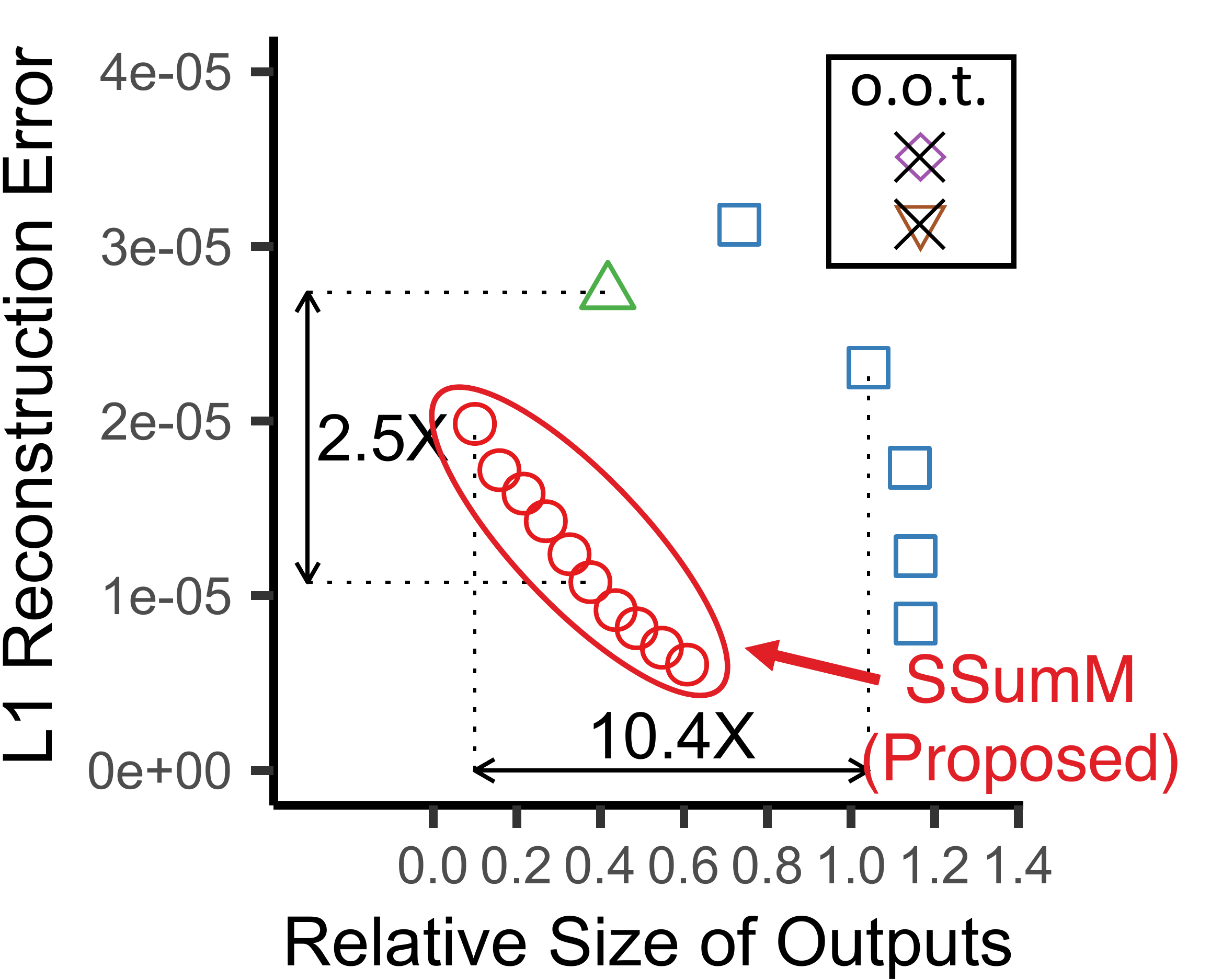} 
	}
	\subfigure[Scalability]{
	\includegraphics[width= 0.4\linewidth]{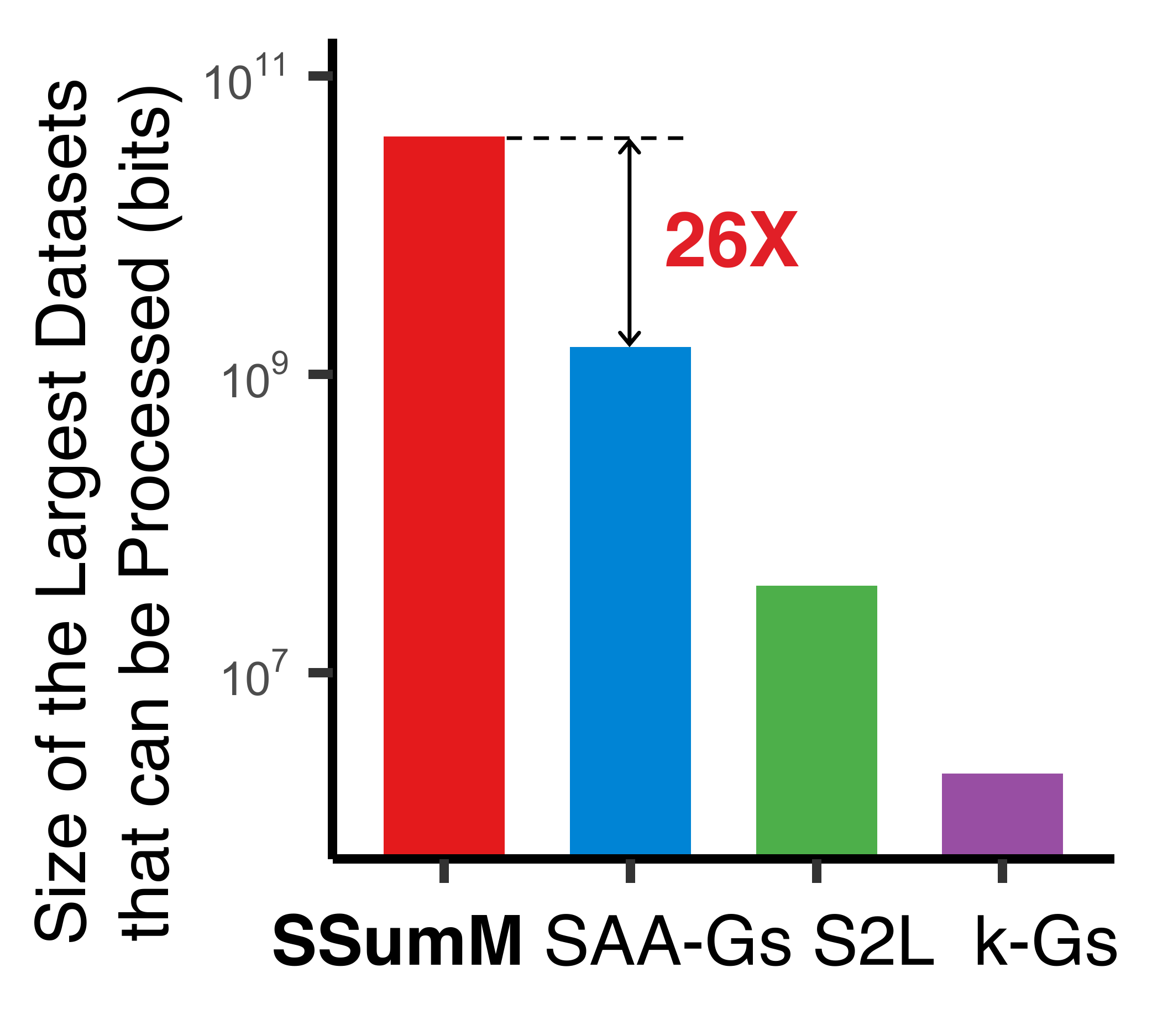} \label{fig:crown:scal} 
	} \\
	\vspace{-2mm}
	\caption{\label{fig:crown}\underline{\smash{Advantages of \method.}} Compared to its state-of-the-art competitors, \method yields more compact and accurate summary graphs, and it successfully processes a $\mathbf{26 \times}$ larger dataset with {\bf $\mathbf{0.8}$ billion edges}. See Sect.~\ref{sec:exp} for details.
	}
\end{figure}

In summary, our contributions in this work are as follows: 
\vspace{-1mm}
\begin{itemize}[leftmargin=*]
\item \textbf{Practical Problem Formulation}: We introduce a new practical variant (Problem \ref{problem}) of the graph summarization problem, where the size in bits of outputs (instead of the number of supernodes) is constrained so that the outputs easily fit target storage. 
\item \textbf{Scalable and Effective Algorithm Design}: 
We propose \method for the above problem. Compared to its state-of-the-art competitors, \method handles up to $\mathbf{26\times}$ larger graphs with linear scalability, and it yields up to $\mathbf{11.2\times}$ smaller summary graphs with similar reconstruction errors (Fig.~\ref{fig:crown} and Thm.~\ref{thm:time_complexity}).
\item \textbf{Extensive Experiments}: Throughout extensive experiments on 10 real-world graphs, we validate the advantages of \method over its state-of-the-art competitors.
\end{itemize}
\vspace{-1mm}
\noindent \textbf{Reproducibility}: The source code and datasets used in the paper can be found at \url{http://dmlab.kaist.ac.kr/ssumm/}.

In Sect.~\ref{sec:pfn}, we introduce some notations and concepts, and we formally define the problem of graph summarization within the given size in bits. 
In Sect.~\ref{sec3}, we present \method, our proposed algorithm for the problem, and we analyze its time and space complexity. 
In~Sect. \ref{sec:exp}, we evaluate \method through extensive experiments.
After discussing related work in Sect.~\ref{sec:rel}, 
we draw conclusions in Sect.~\ref{sec:con}. 



\begin{table}[t]
	\vspace{-5mm}
	\small
	\begin{center}
		\caption{Symbols and Definitions.}
		\label{tab:symndef}
		\scalebox{0.9}{
			\begin{tabular}{l|l}
				\toprule 
				\textbf{Symbol}  & \textbf{Definition}\\
				\midrule
				\multicolumn{2}{l}{\bf Symbols for the problem definition (Sect.~\ref{sec:pfn})} \\
				\midrule 
				\fgraph & input graph with subnodes $V$ and subedges $E$\\
				\adj & adjacency matrix of \graph\\
				\midrule
				\fsummary &summary graph with supernodes \snode, superedges \sedge, \\
				&and a weight function $\omega$\\
				$\Vu$ & supernode with the subnode $u$ \\
				$\PS$ & set of all unordered pairs of supernodes \\
				$\EAB$ & set of subedges between the supernodes $A$ and $B$ \\
				$\PIAB$ & set of all possible subedges between the supernodes $A$ and $B$ \\ 
				\midrule
				\rgraph &  reconstructed graph with subnodes $V$, subedges $\EH$, \\ & and a  weight function $\wH$ \\ 
				$\AH$  &  weighted adjacency matrix of $\GH$\\		
				\midrule
				\size  &  desired size in bits of the output summary graph\\ 
				\midrule
				\multicolumn{2}{l}{\bf Symbols for the proposed algorithm (Sect. \ref{sec:method})} \\
				\midrule  
				$T$  &  given number of iterations\\ 
				$\anneal$  &  threshold at the $t$-th iteration\\
				\neighbor &set of candidate sets at the $t$-th iteration\\   
				\bottomrule
			\end{tabular}
		}
	\end{center}
\end{table}

\begin{figure*}[t]
	\vspace{-6mm}
	\centering
	\subfigure[Input graph \graph]{
		\label{inputgraph}
		\includegraphics[width= 0.18\linewidth]{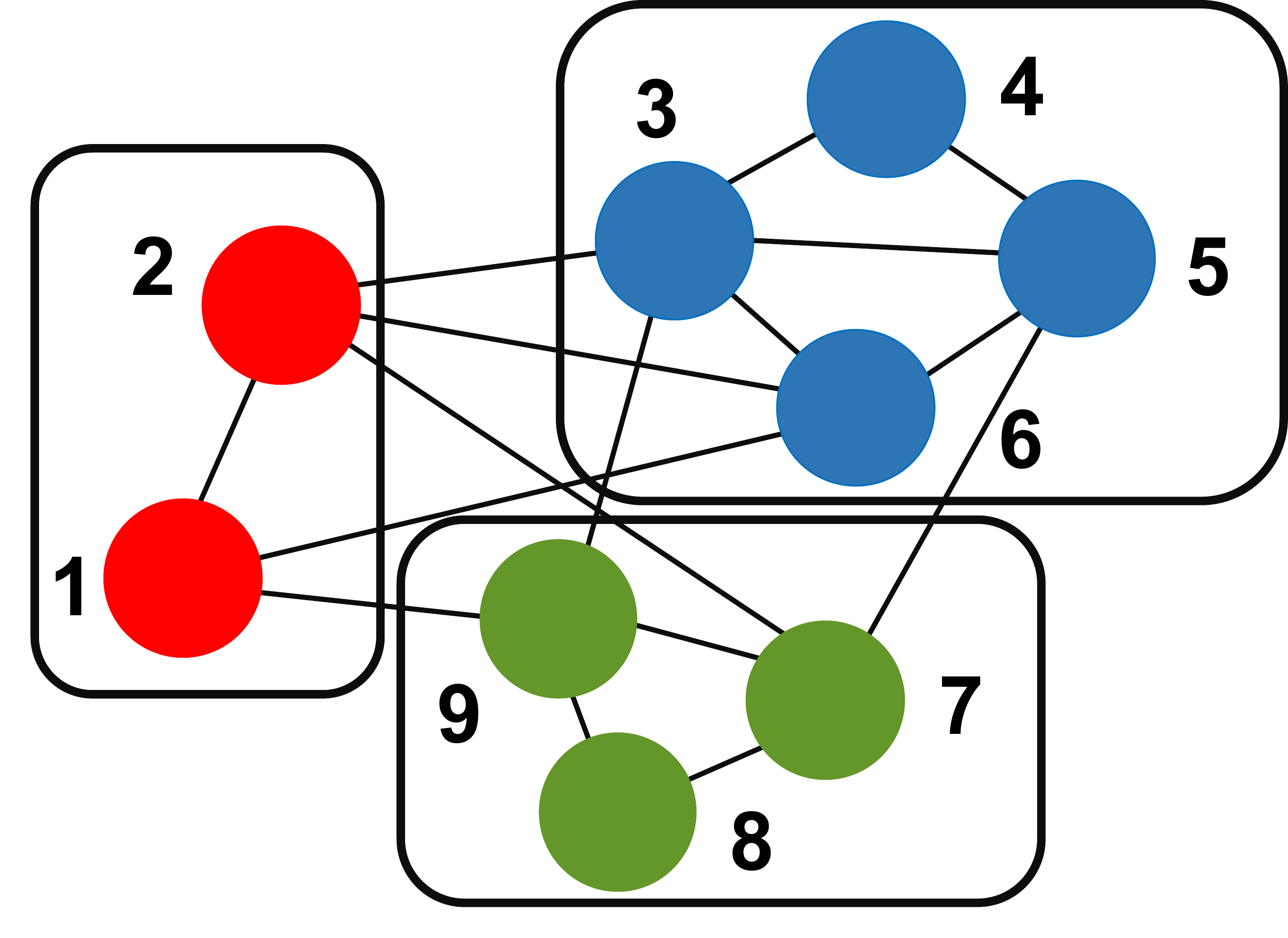}
	} 
	\subfigure[Adjacency matrix \adj]{
		\label{adj}
		\includegraphics[width= 0.262\linewidth]{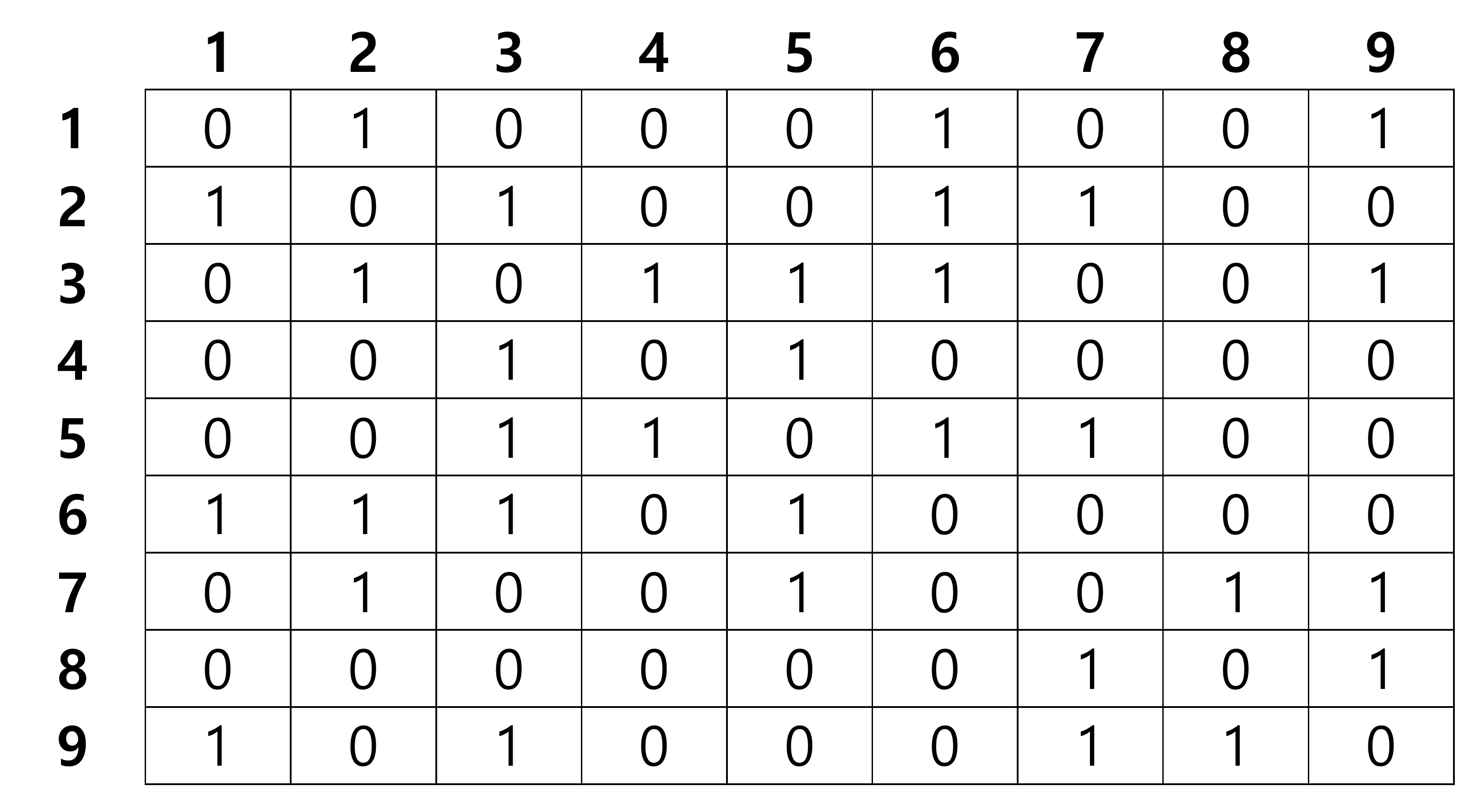}
	} 
	\subfigure[Summary graph \summary]{
		\label{summarygraph}
		\includegraphics[width= 0.24\linewidth]{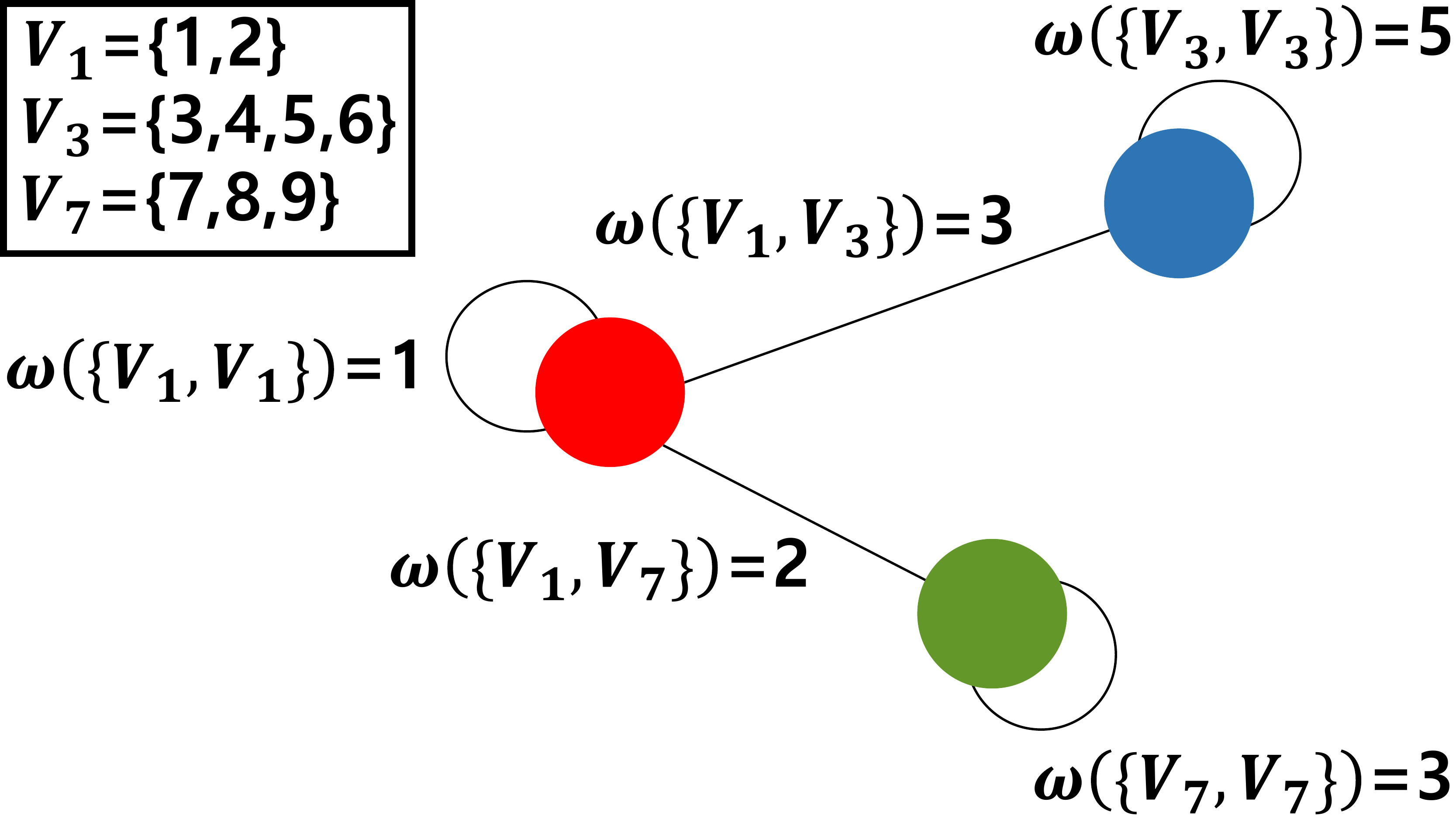}
	} 
	\subfigure[Reconstructed adjacency matrix \REadj]{
		\label{readj}
		\includegraphics[width= 0.265\linewidth]{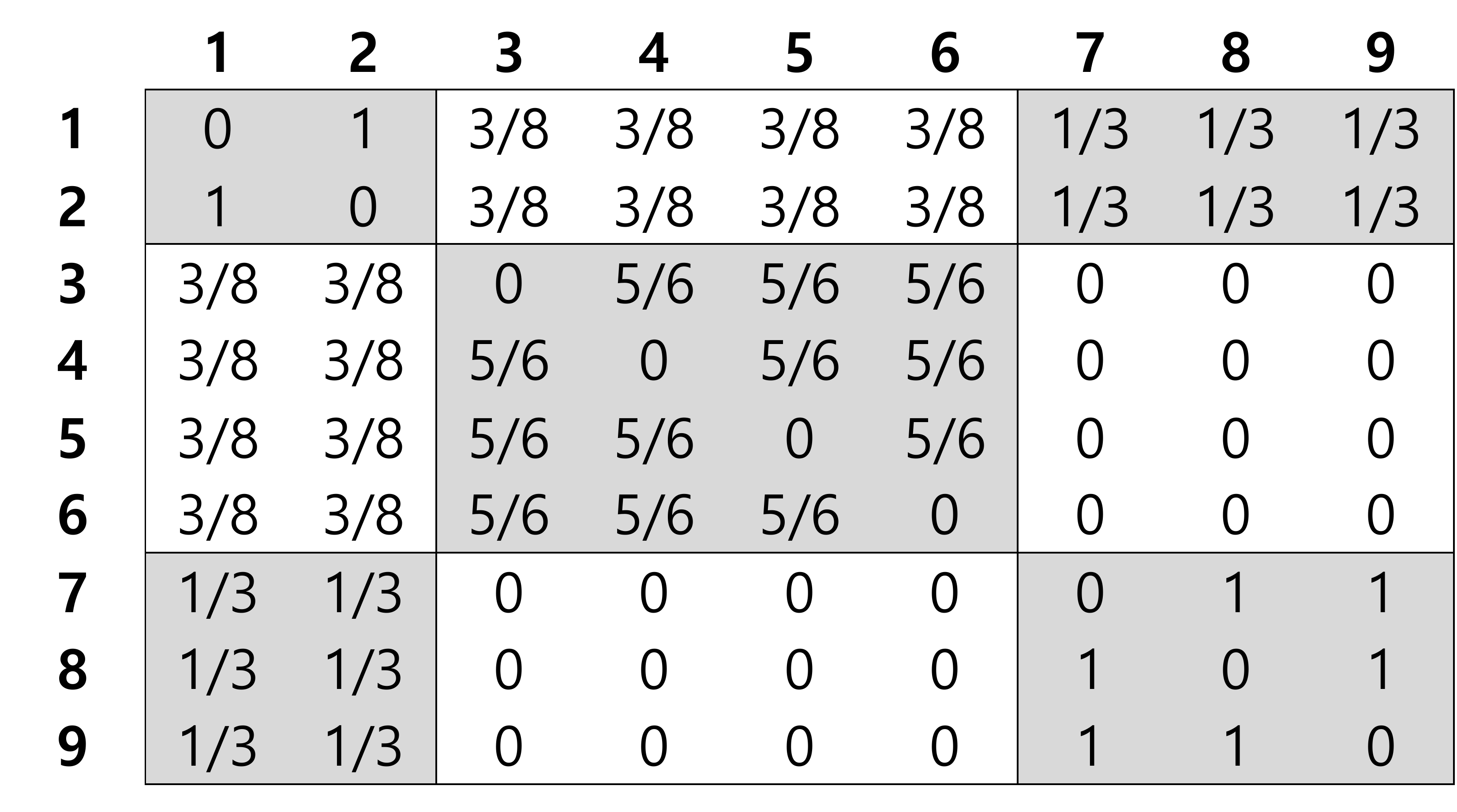}
	} \\ \vspace{-2mm}
	\caption{\label{fig:concept} 
		\underline{\smash{Illustration of graph summarization.}} An example graph $G$ in (a) has the adjacency matrix $A$ in (b).
		From a summary graph $\GB$ in (c), we restore a graph $\GH$, whose weighted adjacency matrix is $\AH$ in (d).
		Each subnode in $G$ belongs to one supernode in $\GH$, and the weight of each superedge corresponds to the number of subedges between the two supernodes. For example, since there are $3$ subedges (i.e., $\{1,6\}$, $\{2,3\}$, $\{2,6\}$) between supernodes $V_{1}$ and $V_{3}$, the weight $\omega(\{V_1,V_3\})$ of the superedge $\{V_{1},V_{3}\}$ is $3$.
		Note that two supernodes do not have to be connected by a superedge even when there are subedges between them (see $V_{3}$ and $V_{7}$).
		Eq.~\eqref{eq:weight} is used for the weights of subedges in $\GH$. For example, the weight $\wH(\{1,3\})$ of the subedge $\{1,3\}$ in $\GH$ is $\frac{\omega(\{V_1, V_3\})}{|\Pi_{V_1V_3}|}=\frac{3}{8}$. 
	}
\end{figure*}

\vspace{-1mm}
\section{Preliminaries \& Problem Definition}
\label{sec:pfn}
\vspace{-1mm}
We introduce some notations and concepts that are used throughout this paper. Then, we define the problem of summarizing a graph within the given size in bits. Table~\ref{tab:symndef} lists some frequently-used notations, and Fig.~\ref{fig:concept} illustrates some important concepts. 

\vspace{-1mm}
\subsection{Notations and Concepts}
\label{sec:pfn:concept}
\vspace{-1mm}

\smallsection{Input graph:}
Consider an undirected graph \fgraph with nodes $V$ and edges $E$. 
Each edge $\{u,v\}\in E$ joins two distinct nodes $u\neq v\in V$. 
We assume that $G$ is undirected without self-loops for simplicity, while the considered problem and our proposed algorithm can easily be extended to directed graphs with self-loops. 
We call nodes and edges in $G$ {\it subnodes} and {\it subedges}, respectively, to distinguish them from those in summary graphs, described below.
\par

\smallsection{Summary graph:}
A {\it summary graph} \fsummary of a graph \fgraph consists of a set \snode of {\it supernodes}, a set \sedge of {\it superedges}, and a weight function $\omega$.
The supernodes \snode are distinct and exhaustive subsets of $V$, i.e., $\bigcup_{A \in S} A = V$ and $A \cap B = \emptyset$ for all $A \neq B \in S$.
Thus, every subnode in $V$ is contained in exactly one supernode in $S$, and we denote the supernode that each subnode $v \in V$ belongs to as $\Vv\in S$. 
Each superedge $\{A,B\}\in P$ connects two supernodes $A,B\in S$, and if $A=B$, then $\{A,B\}=\{A,A\}$ indicates the self-loop at the supernode $A\in S$.  
We use $\PS:={S \choose 2}\cup\{\{A,A\}:A\in S\}$ to denote the all unordered pairs of supernodes, and then $P\subseteq \PS$.
The weight function \F assigns to each superedge $\{A,B\}\in P$ the number of subedges between two subnodes belonging to $A\in S$ and $B\in S$, respectively. 
Let the set of such subedges as $\EAB:=\{\{u,v\}\in E:u\in A, v\in B\}$. Then, $\omega(\{A,B\}):=|\EAB|$ for each superedge $\{A,B\}\in P$. 
See Fig.~\ref{fig:concept} for an example summary graph.


\smallsection{Reconstructed graph:}
Given a summary graph \fsummary, we obtain a {\it reconstructed graph} \rgraph conventionally as in \cite{riondato2017graph, lefevre2010grass, beg2018scalable}.
The set $V$ of subnodes is recovered by the union of all supernodes in $S$. 
The set $\EH$ of subedges is defined as the set of all pairs of distinct subnodes belonging to two supernodes connected by a superedge in $P$.
That is, $\EH:= \{\{u,v\}\in V\times V: u \neq v,~\VuVv \in P \}$. 
The weight function $\wH:\EH \rightarrow \mathbb{R}^{+}$ is defined as follows:
\begin{equation}
\wH(\{u,v\}):= 
\frac{\omega(\{\Vu,\Vv\})}{|\PIuv|}=\frac{|\Euv|}{|\PIuv|}, \label{eq:weight}
\end{equation}
where $\PIuv:= \{\{u,v\}:u\neq v, u\in\Vu, v\in\Vv\}$ is the set of all possible subedges between two supernodes.
That is, in Eq.~\eqref{eq:weight}, the denominator is the maximum number of subedges between two supernodes, and each nominator is the actual number of subedges between two supernodes.
Note that the graph $\GH$ reconstructed from $\GB$ is not necessarily the same with the original graph $G$, and we discuss how to measure their difference in the following section. 

\smallsection{Adjacency matrix:}
We use $A$ to denote the adjacency matrix of the original graph $G$, and we use $\AH$ to denote the weighted adjacency matrix of a reconstructed graph $\GH$.
See Fig.~\ref{fig:concept} for examples.
\vspace{-1mm}
\subsection{Problem Definition}\label{sec:pfn:probdef}
\vspace{-1mm}

Now that we have introduced necessary concepts, 
we formally define, in Problem~\ref{problem}, the problem of graph summarization within the given size in bits. 
Then, we discuss how we measure the reconstruction error and size of summary graphs.
Lastly, we compare the defined problem with the original graph summarization problem.

\begin{problem}[\label{problem}Graph Summarization within a Budget in Bits]~
\begin{itemize}[leftmargin=*]
	\item \textbf{Given:} a graph \fgraph and the desired size \size in bits
	\item \textbf{Find:} a summary graph \fsummary
	\item \textbf{to Minimize:} the reconstruction error
	\item \textbf{Subject to:} \textit{$Size(\overline{G}) \leq $ \size}.
\end{itemize}
\end{problem}

\smallsection{Reconstruction error:}
The {\it reconstruction error} corresponds to the difference between the original graph $G$ and the graph $\GH$ reconstructed from the summary graph $\GB$.
While there can be many different ways of measuring the reconstruction error, as in the previous studies of graph summarization \cite{riondato2017graph, lefevre2010grass, beg2018scalable}, we use the \reconstruction, defined as
\begin{align}
RE_p(G|\nsummary):=\left(\sum_{i=1}^{\abs{V}}{\sum_{j=1}^{\abs{V}}{\abs{A(i,j)-\hat{A}(i,j)}^p}}\right)^{1/p},
\label{eq: lpNorm}
\end{align}
where $A(i,j)$ is the $(i,j)$-th entry of the matrix $A$.
Recall that $A$ and $\AH$ are the (weighted) adjacency matrices of $G$ and $\GH$, respectively.
 
\smallsection{Size of summary graphs:} 
As in the previous studies  \cite{riondato2017graph, lefevre2010grass, beg2018scalable}, we define the size in bits of (summary) graphs based on the assumption that they are stored in the edge list format.
Specifically, the size in bits of the input graph \fgraph is defined as
\begin{align}
Size(G) := 2|E|\log_{2}|V|,
\label{eq: inputsize}
\end{align}
since each of $|E|$ subedges consists of two subnodes, each of which is encoded using $\log_{2}|V|$ bits.
Note that, in order to distinguish $|V|$ items, at least $\log_{2}|V|$ bits per item are required.
Similarly, the size in bits of the summary graph \fsummary is defined as
\begin{align}
	Size(\overline{G}) := |P|(2\log_2 |S|+\log_2 \wmax) + |V|\log_2 |S|,
	\label{eq: summarysize}
\end{align}
where $\wmax:=\max_{\{A,B\}\in P}\omega(\{A,B\})$ is the maximum superedge weight. 
The first term in Eq.~\eqref{eq: summarysize} 
is for $|P|$ superedges, each of which consists of two supernodes and an edge weight, which are encoded using $2\log_{2}|S|$ bits and $\log_{2} \wmax$ bits, respectively.
Again, in order to distinguish $|S|$ (or $\wmax$) items, at least $\log_{2}|S|$ (or $\log_{2} \wmax$) bits are required for encoding each item.\footnote{Since $\omega$ cannot be zero in our algorithm, we need to distinguish $\wmax$ potential distinct values, i.e., $\{1,2,...,\wmax\}$.}
The second term in Eq.~\eqref{eq: summarysize} 
is for the membership information. Each of $|V|$ nodes belongs to a single supernode, which is encoded using $\log_{2}|S|$ bits. 
%

\smallsection{Comparison with the original problem:}
Different from Problem~\ref{problem}, where we constrain the size in bits of a summary graph,
the number of supernodes is constrained in the original graph summarization problem \cite{lefevre2010grass}.
By constraining the size in bits, we can easily make summary graphs tightly fit in target storage (main memory, cache, etc.).
On the other hand, it is not trivial to control the number of nodes so that a summary graphs tightly fits in target storage. 
This is because how the size of summary graphs changes depending on the number of supernodes varies across datasets.



\SetKwRepeat{Do}{do}{while}

\vspace{-1.5mm}
\section{Proposed Method} 
\label{sec3}
\label{sec:method}
\vspace{-1mm}


We propose \method (\textbf{S}parse \textbf{Sum}marization of \textbf{M}assive Graphs), a scalable and effective algorithm for Problem~\ref{problem}.
\method is a randomized greedy search algorithm equipped with three novel ideas.

One main idea of \method is to carefully balance the changes in the reconstruction error and size of the summary graph at each step of the greedy search.
To this end, \method adapts the minimum description length principle (the MDL principle) \cite{rissanen1978modeling} to measure both the reconstruction error and size commonly in the number of bits.
Then, \method performs a randomized greedy search, aiming to minimize the total number of bits.

Another main idea of \method is to tightly combine two strategies for summarization: merging supernodes into a single supernode, and sparsifying the summary graph.
Specifically, instead of creating all possible superedges as long as their weight is not zero, 
\method selectively creates superedges so that its cost function is minimized.
\method also takes this selective superedge creation into consideration when deciding supernodes to be merged.

Lastly, \method achieves linear scalability by rapidly but effectively finding promising candidate pairs of supernodes to be merged.


In this section, we present the cost function (Sect.~\ref{sec:method:objective}) and the search method (Sect.~\ref{sec:method:search}) of \method. After that, we analyze its time and space complexity (Sect.~\ref{sec:method:analysis}).


\vspace{-1mm}
\subsection{Cost Function in \method}
\label{sec:method:objective}
\vspace{-1mm}






In this subsection, we introduce the cost function, which is used at each step of the greedy search in \method.
The cost function is for measuring the quality of candidate summary graphs by balancing the size and the reconstruction error, which is important since \method aims to reduce the size of the output summary graph while increasing the reconstruction error as little as possible.

For balancing the size and reconstruction error, they need to be directly comparable. To this end, we measure both in terms of the number of bits by adapting the minimum description length principle.
The principle states that given data, which is the input graph $G$ in our case, the best model, which is a summary graph $\GB$, for the data is the one that minimizes $Cost(\GB,G)$, the description length in bits of $G$ defined as
\begin{align}
Cost(\GB,G) := Cost(\GB) + Cost(G|\GB),
\label{eq:savingcost}
\end{align}
where the description length is divided into the model cost $Cost(\GB)$ and the data cost $Cost(G|\GB)$.
The model cost measures the number of bits required to describe $\GB$.
The data cost measures the number of bits required to describe $G$ given $\GB$ or equivalently to describe the difference between $G$ and $\GH$, which is reconstructed from $\GB$. Thus, the data cost is naturally interpreted as the reconstruction error of $\GB$ in bits.
Note that, if $G=\GH$ without any reconstruction error, then $Cost(G|\GB)$ becomes zero.

Eq.~\eqref{eq:savingcost} is the cost function that \method uses to balance the size and reconstruction error and thus to measure the quality of candidate summary graphs.
Below, we describe each term of Eq.~\eqref{eq:savingcost} in detail, and then we divide it into the cost for each supernode.  




\smallsection{Model cost:} 
For the model cost, we use Eq.~\eqref{eq:model}. It is an upper bound of Eq.~\eqref{eq: summarysize} which measures the size of a summary graph in bits. In Eq.~\eqref{eq:model}, $\log_{2}|V|$ ($\geq \log_{2}|S|$) and $\log_{2}|E|$ ($\geq \log_{2}\wmax$) bits are used to distinguish supernodes and superedges, respectively.
That is,
\begin{align}
Cost(\GB) & :=|P|(2\log_2 |V|+\log_2 |E|) + |V|\log_2 |V|. \label{eq:model}
\end{align}
We divide the total model cost into the model cost for each supernode pair as follows:
\begin{equation}
Cost(\GB) = |V|\log_2 |V| + \sum\nolimits_{\{A,B\}\in\PS} Cost(\{A,B\}|\GB), \label{eq:model:divide}
\end{equation}
where $Cost(\{A,B\}|\GB) := \mathbb{1}(\AB\in P)\times (2\log_2 |V|+\log_2 |E|)$ is the model cost for each supernode pair $\{A,B\}\in\PS$.


\smallsection{Data cost:} 
The data cost $Cost(G|\GB)$ is the number of bits required to exactly describe $G$, or equivalently all subedges in $G$, given $\GB$.
As explained above, $Cost(G|\GB)$ is naturally interpreted as the reconstruction error of $\GB$ in bits.
We divide the total data cost into the data cost for each supernode pair as follows:
\begin{equation}
Cost(G|\GB) = \sum\nolimits_{\{A,B\}\in\PS} Cost(\EAB|\GB), \label{eq:data:divide}
\end{equation}
where $Cost(\EAB|\GB)$ is the number of bits required to describe the subedges between the supernodes $A$ and $B$ (i.e., $\EAB$).

For each $Cost(\EAB|\GB)$, we assume a {\it dual-encoding method} to take into consideration both cases where the superedge $\AB$ exists or not.
Specifically, one between two encoding methods is used depending on whether the superedge $\AB$ exists in $\GB$ or not.
In a case where $\AB$ exists in $\GB$, the first encoding method is used, and it optimally assigns bits to denote whether each possible subedge in $\PIAB$ exists or not. Then, the number of bits required is tightly lower bounded by the Shannon entropy \cite{shannon1998mathematical}. Thus, we define $Cost(\EAB|\GB)$ as
\begin{equation}
	Cost_{(1)}(\EAB|\GB) := -|\PIAB|(\sigma \log_{2}\sigma +(1-\sigma) \log_{2}(1-\sigma)),
	\label{eq:entropy}
\end{equation}
where $\sigma:=\frac{|\EAB|}{|\PIAB|}$ is the proportion of existing subedges in $\PIAB$.
Note that in order to compute $\sigma$, the superedge $\AB$ and its weight $\omega(\AB)$ need to be retained in $\GB$.

In a case where $\AB$ does not exist in $\GB$, the second encoding method is used, and it simply lists all existing subedges in $\EAB$. Then, the number of required bits is
\begin{equation}
Cost_{(2)}(\EAB|\GB) := 2|\EAB|\log_{2}|V|,
\label{eq:corrections}
\end{equation}
where $2\log_{2}|V|$ is the number bits required to encode an subedge. 
Note that, for this encoding method, the superedge $\AB$ and its weight $\omega(\AB)$ do not need to be retained in $\GB$.

Then, the final number of bits required to describe $\EAB$ is
\begin{equation}
Cost(\EAB|\GB):=\left\{
\begin{array}{ll}
Cost_{(1)}(\EAB|\GB) & \text{if} ~\AB \in P \\ 
Cost_{(2)}(\EAB|\GB) & \text{otherwise}.\\
\end{array}
\right. \label{eq:min_cost}
\end{equation}


\smallsection{Cost decomposition:} 
By combining Eq.~\eqref{eq:savingcost}$-$ Eq.~\eqref{eq:min_cost}, the total description cost $Cost(\GB,G)$ can be divided into that for each supernode pair as follows:
$$Cost(\GB,G)=|V|\log_2 |V| + \sum\nolimits_{\{A,B\}\in\PS} Cost_{AB}(\GB,G),$$
where $Cost_{AB}(\GB,G)$, the total description cost for each supernode pair $\AB\in\PS$, is defined as
\begin{equation}
Cost_{AB}(\GB,G):=Cost(\AB|\GB)+Cost(\EAB|\GB).\label{eq:cost:edge}
\end{equation}
Based on this cost, we also define the total description cost of each supernode $A$ by summing the costs for the pairs containing $A$, i.e.,
\begin{equation}
Cost_{A}(\GB,G):=\sum\nolimits_{B\in S}Cost_{AB}(\GB,G). \label{eq:cost:node}
\end{equation}
Eq.~\eqref{eq:cost:node} is used by \method when deciding supernodes to be merged, as described in detail in the following subsection.

\smallsection{Optimal encoding given a set of supernodes:} 
Once a set $S$ of supernodes is fixed, then the set $P$ of superedges that minimizes Eq.~\eqref{eq:savingcost} is easily obtained by minimizing Eq.~\eqref{eq:cost:edge} for each pair $\{A,B\}\in \PS$ of supernodes.
That is, the superedge between each pair $\{A,B\}$ is created if and only if it reduces Eq.~\eqref{eq:cost:edge}.	
We let $\PSTAR{S}$ be the set of superedges that minimizes Eq.~\eqref{eq:savingcost} given $S$, and we let $\GBSTAR{S}=(S,\PSTAR{S},\omega)$ be the summary graph consisting of $S$ and $\PSTAR{S}$. 
Then, minimizing Eq.~\eqref{eq:savingcost} is equivalent to finding $S$ that minimizes 
\begin{equation}
\CSTAR{S} := Cost(\GBSTAR{S}) + Cost(G|\GBSTAR{S}). \label{eq:savingcost_simple}
\end{equation}
Similarly, as in Eq.~\eqref{eq:cost:edge} and Eq.~\eqref{eq:cost:node}, we let the description costs of each supernode pair $\{A,B\}\in\PS$ and supernode $A\in S$ in $\GBSTAR{S}$ be
\begin{align}
& \CSTARSUB{AB}{S}  := Cost_{AB}(\GBSTAR{S},G), \label{eq:cost:edge_simple}\\
& \CSTARSUB{A}{S} :=\sum\nolimits_{B\in S}\CSTARSUB{AB}{S}. \label{eq:cost:node_simple}
\end{align}

\begin{algorithm}[t]
	\SetAlgoLined
	\LinesNumbered
	\KwData{
		(a) input graph: \fgraph  \\
		\quad \quad \quad \ (b) the number of iterations: $T$ \\
		\quad \quad \quad \ (c) desired size of $\GB$: \size}
	\KwResult{summary graph: \fsummary}
	$S\leftarrow \{\{u\}:u\in V \}$  \label{alg:line:init1} \Comment*[r]{\blue{initialize $\GB$ to $G$}}  
	$P\leftarrow \{\{\Vu,\Vv\}\in \PS:\{u,v\}\in E\}$ ; \label{alg:line:init2} \\
	$t$ $\leftarrow$ 1 \Comment*[r]{\blue{t: iteration}}
	\While{$ t \leq T$ \normalfont{\bf and} $ k<Size(\GB)$}{
		generate candidate sets \neighbor $\subseteq 2^{S}$ \label{alg:line:candidate} \Comment*[r]{\blue{Sect.~\ref{sec:method:search:candidate}}}
		\For{$\mathbf{each}$ candidate set $C\in\ST$ \label{alg:line:merge1}}{
			greedily merges supernodes within $C\subseteq S$ and \ \ \ \ \ \  adds new superedges selectively  \label{alg:line:merge2} \Comment*[r]{\blue{Sect.~\ref{sec:method:search:merge}}}
		}
		$t$ $\leftarrow$ $t + 1$ ;\\
	}
	\If{$Size(\GB) > ~\size$ \label{alg:line:sparsify1}}
	{greedily drops superedges from $P$ so that $Size(\GB)\leq~\size$ \label{alg:line:sparsify2} \Comment*[r]{\blue{Sect.~\ref{sec:method:search:drop}}}}
	{\bf return} \fsummary
	\caption{Overview of \method \label{algo:main3}}
\end{algorithm}

\vspace{-1mm}
\subsection{Search Method in \method}
\label{sec:method:search}
\vspace{-1mm}

Now that we have defined the cost function (i.e., Eq.~\eqref{eq:savingcost_simple}) for measuring the quality of candidate summary graphs, we present how \method performs a rapid and effective randomized greedy search over candidate summary graphs.
We first provide an overview of \method, and then we describe each step in detail.


\subsubsection{\bf Overview (Alg.~\ref{algo:main3})}
\label{sec:method:search:overview}
%
Given an input graph \fgraph, the desired size \size in bits of the summary graph, and the number $T$ of iterations, \method produces a summary graph \fsummary. 
\method first initializes $\GB$ to $G$. That is, $S=\{\{u\}:u\in V \}$ and $P=\{\{\Vu,\Vv\}\in S\times S:\{u,v\}\in E\}$ (lines \ref{alg:line:init1}-\ref{alg:line:init2}).
Then, it repeatedly merges pairs of supernodes and sparsifies the summary graph by alternatively running the following two phases until the size of the summary graph reaches \size or the number of iterations reaches $T$:


\begin{compactitem}[$\bullet$]
	\item \textbf{Candidate generation (line~\ref{alg:line:candidate}):} 
	To rapidly and effectively search promising pairs of supernodes whose merger significantly reduces the cost function, \method first divides \snode into candidate sets \neighbor each of which consists of supernodes within 2 hops.
	To take more pairs of supernodes into consideration, \method changes
	\neighbor probabilistically at each iteration $t$.
	
	\item \textbf{Merging and sparsification (lines~\ref{alg:line:merge1}-\ref{alg:line:merge2}):} 
	Within each candidate set, obtained in the previous phase, \method repeatedly merges two supernodes whose merger reduces the cost function most. 
	Simultaneously, \method sparsifies the summary graph by selectively creating superedges adjacent to newly created supernodes. Each superedge is created only when it reduces the cost function.
\end{compactitem}
After that, if the size of summary graph is still larger than the given target size \size, the following phase is executed:
\begin{compactitem}[$\bullet$]
	\item \textbf{Further sparsification (lines~\ref{alg:line:sparsify1}-\ref{alg:line:sparsify2}):} \method further sparsifies the summary graph until its size reaches the given target size \size.
	Specifically, \method repeatedly removes a superedge so that the cost function is minimized.
\end{compactitem}
Lastly, \method returns the summary graph as an output.
In the following subsections, we present each phase in detail. 


\subsubsection{\bf Candidate generation phase} 
\label{sec:method:search:candidate}

The objective of this step is to find candidate sets of supernodes.
\method uses the candidate sets in the next merging and sparsification phase, and specifically, it searches pairs of supernodes to be merged  within each candidate set.
For rapid and effective search, the candidate sets should be small, and at the same time, they should contain many promising supernode pairs whose merger leads to significant reduction in the cost function, i.e., Eq.~\eqref{eq:savingcost_simple}.

To find such candidate sets, \method groups supernodes within two hops of each other.
If we define the distance between two supernodes as the minimum distance between subnodes in one supernode and those in the other,
merging supernodes within two hops tends to reduce the cost function more than merging those three or more hops away from each other, as formalized in Lemmas~\ref{lemma:2hop} and \ref{lemma:3hop}, where 
\begin{align}
	Reduction(A,B)  := & ~\CSTARSUB{A}{S}+\CSTARSUB{B}{S} 
	- \CSTARSUB{AB}{S} \nonumber \\  & -\CSTARSUB{A\cup B}{S\cup\{A\cup B\}\setminus \{A,B\}} \label{eq:reduce}
\end{align}
is the reduction of the cost function, i.e., Eq.~\eqref{eq:savingcost_simple}, when two supernodes $A\neq B\in S$ are merged.

\begin{lemma}[Merger within 2 Hops]\label{lemma:2hop}	
	If two supernodes $A\in S$ and $B\in S$ are within $2$ hops, then
	\begin{equation}
		Reduction(A,B) \leq \min(\CSTARSUB{A}{S},\CSTARSUB{B}{S}), \label{eq:2hop}
	\end{equation}
	and this inequality is tight.
\end{lemma}

\begin{lemma}[Merger outside 2 Hops]\label{lemma:3hop}	
	If two supernodes $A\in S$ and $B\in S$ are $3$ or more hops away from each other, then
	\begin{equation}
	Reduction(A,B) \leq 2\log_{2}|V|+\log_{2}|E|. \label{eq:3hop}
	\end{equation}
\end{lemma}

\noindent See Appendix~\ref{appendix:proof} for proofs of the lemmas.
Empirically, for carefully chosen  $A\neq B\in S$ within two hops, $\min(\CSTARSUB{A}{S},\CSTARSUB{B}{S})$ and $Reduction(A,B)$ are much larger than $2\log_{2}|V|+\log_{2}|E|$.

To rapidly group supernodes within two hops of each other, \method divides the supernodes into those with the same shingles \cite{chierichetti2009compressing}. 
Note that, for a random bijective function $h:V\rightarrow \{1,...,|V|\}$, if we define the shingle of each supernode $A\in S$ as 
$$f(A):=\min_{u\in A}\left(\min\left(h(u), \min_{(u,v)\in E}h(v)\right)\right),$$
then two supernodes $A\neq B\in S$ have the same shingle (i.e., $f(A) = f(B)$) only if $A$ and $B$ are within two hops.\footnote{If $f(A) = f(B)=h(v)$, there exist a subnode in $A$ and a subnode in $B$ within $1$-hop of $v$.}
Specifically, until each candidate set consists of at most a constant (spec., $500$) number of nodes, \method divides the supernodes using shingles recursively at most constant (spec., $10$) times and then randomly.
Note that computing the shingle of all supernodes takes $O(|V|+|E|)$ time if we (1) create a random hash function $h$, which takes $O(|V|)$ time \cite{knuth1969seminymerical}, (2) compute and store $\min(h(u), \min_{(u,v)\in E}h(v))$ for every subnode $u\in V$, which takes $O(|V|+|E|)$ time, and (3) compute $f(A)$ for every supernode $A\in S$, which takes $O(|V|)$ time.

\subsubsection{\bf Merging and sparsification phase}
\label{sec:method:search:merge}

In this phase, \method searches a concise and accurate summary graph by repeatedly (1) merging two supernodes within the same candidate set into a single supernode and (2) greedily sparsifying its adjacent superedges.
To this end, each candidate set $C$ obtained in the previous phase is processed as described in  
Alg.~\ref{alg:merge}. 
\method first finds two supernodes $A\neq B \in C$, among $\log_{2}|C|$ randomly chosen supernode pairs of $C$, whose merger maximizes
\begin{align}
Relative&\_Reduction(A,B) := \nonumber \\
&	1- \frac{\CSTARSUB{A \cup B}{S\cup\{A\cup B\}\setminus \{A,B\}}}{\CSTARSUB{A}{S}+\CSTARSUB{B}{S}-\CSTARSUB{AB}{S}},
\label{eq:relative_saving}
\end{align}
which  is the reduction of the cost function (i.e., Eq.~\eqref{eq:savingcost_simple}) due to the merger of $A$ and $B$ over the current cost of describing the superedges adjacent to $A$ and $B$ (line~\ref{alg:line:threshold}).
Then, if Eq.~\eqref{eq:relative_saving} exceeds a threshold (line~\ref{alg:line:threshold}), $A$ and $B$ are merged into a single supernode $A\cup B$ (line~\ref{alg:line:merge}).
Inspired by simulated annealing \cite{kirkpatrick1983optimization} and SWeG \cite{shin2019sweg}, we let the threshold decrease over iterations as follows:
\begin{align}
\anneal:=\left\{
\begin{array}{ll}
(1 + t)^{-1}  & \text{if} ~ t < T\\ 
0 & \text{if}~ t = T,
\end{array}
\right.
\label{eq: threshold}
\end{align}
where $t$ denotes the current iteration number.
Once $A$ and $B$ are merged into $A\cup B$, all superedges adjacent $A$ or $B$ are removed (line~\ref{alg:line:remove}), and then the superedges adjacent to $A\cup B$ are selectively created (or equivalently sparsified) so that the cost function given $S$ (i.e., $\CSUB{A \cup B}{\GB, G}$ defined in Eq.~\eqref{eq:cost:node}) is minimized.\footnote{Our implementation minimizes a tighter upper bound obtained by replacing $2\log_{2}|V|+\log_{2}|E|$ in $\CSUB{A \cup B}{\GB, G}$ with $2\log_{2}|S|+\log_{2}\omega_{max}$. 
Moreover, it never creates superedges that increase the reconstruction error $RE_p$.
}
Merging two supernodes in a candidate set $C$ is repeated until the relative reduction (i.e., Eq.~\eqref{eq:relative_saving}) does not exceed the threshold $\anneal$, $\max(\log_{2} |C|,1)$ times in a row (lines~\ref{alg:line:skip1}, \ref{alg:line:skip2}, and \ref{alg:line:skip3}).
Then, each of the other candidate sets is processed in the same manner.

By restricting its attention to a small number of supernode pairs in each candidate set, \method significantly reduces the search space and achieves linear scalability (see Sect.~\ref{sec:method:analysis}). However, in our experiments, this reduction does not harm the quality of the output summary graph much due to (1) careful formation of candidate sets, (2) the adaptive threshold $\anneal$, and (3) robust termination with $\max(\log_{2} |C|,1)$ chances.

\subsubsection{\bf Further sparsification phase}
\label{sec:method:search:drop}
This phase is executed only when the size of the summary graph after repeating the previous phases $T$ times still exceeds the given target size $k$ (lines \ref{alg:line:sparsify1}-\ref{alg:line:sparsify2} of Alg.~\ref{algo:main3}).
In this phase, \method sparsifies the summary graph until its size $Size(\GB)$ reaches $k$ as follows:
\begin{enumerate}[leftmargin=*]
	\item Compute the increase in the reconstruction error $RE_p$ after dropping each superedge from $P$.\footnote{If we drop $\AB$, the increase in $RE_1$ is $(2{|\EAB|}/{|\PIAB|}-1)\cdot|\EAB|$, and that in $RE_2^2$ is $|\EAB|^2/|\PIAB|$.}
	Note that $RE_p$ is directly used instead of the cost function.
	This is because the decrease in $Size(\GB)$ after dropping each superedge is a constant (spec., $2\log_{2}|S|+\log_2{\omega_{max}}$) only except for those with weight $\omega_{max}$.  
	\item Find the $\xi:=\lceil\frac{Size(\GB)-k}{2\log_{2}|S|+\log_2{\omega_{max}}}\rceil$-th smallest increase in $RE_p$, and let it be $\Delta_{\xi}$.
	\item For each superedge in $P$, drop it if the increase in $RE_p$ is smaller than or equal to $\Delta_{\xi}$.
\end{enumerate}
Note that each step takes $O(|P|)=O(|E|)$ time, and to this end, the median-selection algorithm \cite{blum1973time} is used in the second step.

\begin{algorithm}[t]
	\SetAlgoLined
	\LinesNumbered
	\KwData{(a) input graph \fgraph \\
	\quad \quad \quad  (b) current summary graph \fsummary  \\
	\quad \quad \quad (c) current iteration number $t$ \\
	\quad \quad \quad (d) a candidate supernode set $C$ \\
	}
	\KwResult{updated summary graph \fsummary}
	\caption{Merging \& Sparsification in a Candidate Set \label{alg:merge}}
		num\_skips $\leftarrow$ 0 ; \label{alg:line:skip1}  \\
		\While{\normalfont{num\_skips} < $\max\left(\log_{2} |C|,1\right)$ \label{alg:line:skip2}}{
			find a pair $\{A,B\}$ that maximizes Eq.~\eqref{eq:relative_saving} among $\log_{2}|C|$ random pairs of supernodes in $C$ ; \label{alg:line:sample} \\
			\If{$Relative\_Reduction(A,B)>\anneal$ \label{alg:line:threshold}}{
				merge $A$, $B$ into $A\cup B$ both in $S$ and $C$ \label{alg:line:merge} \Comment*[r]{\blue{merge}}
				remove the superedges adjacent to $A$ or $B$ from $P$ ;  \label{alg:line:remove}\\
				add the superedges adjacent to $A\cup B$ to $P$ selectively so that $\CSUB{A \cup B}{\GB, G}$ is minimized \label{alg:line:add} \Comment*[r]{\blue{sparsify}}
				num\_skips $\leftarrow$ 0 ;
				\label{alg:line:skip3}
			}
			\Else{
				num\_skips $\leftarrow$ num\_skips + 1 ; \label{alg:line:skip4}
			}
		}
\end{algorithm}


\begin{figure*}[t!]
	\centering
	\vspace{-4mm}
	\includegraphics[width=0.48\linewidth]{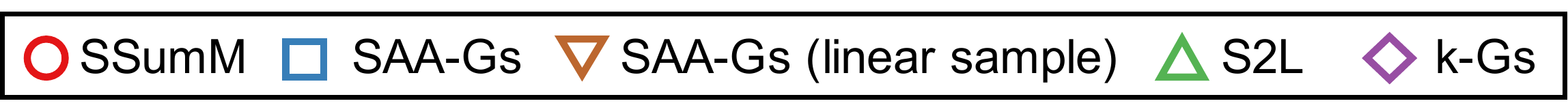} \\
	\vspace{-1.9mm}
	\subfigure[DBLP]{
		\label{fig:l1:DB}
		\includegraphics[width=0.185\textwidth]{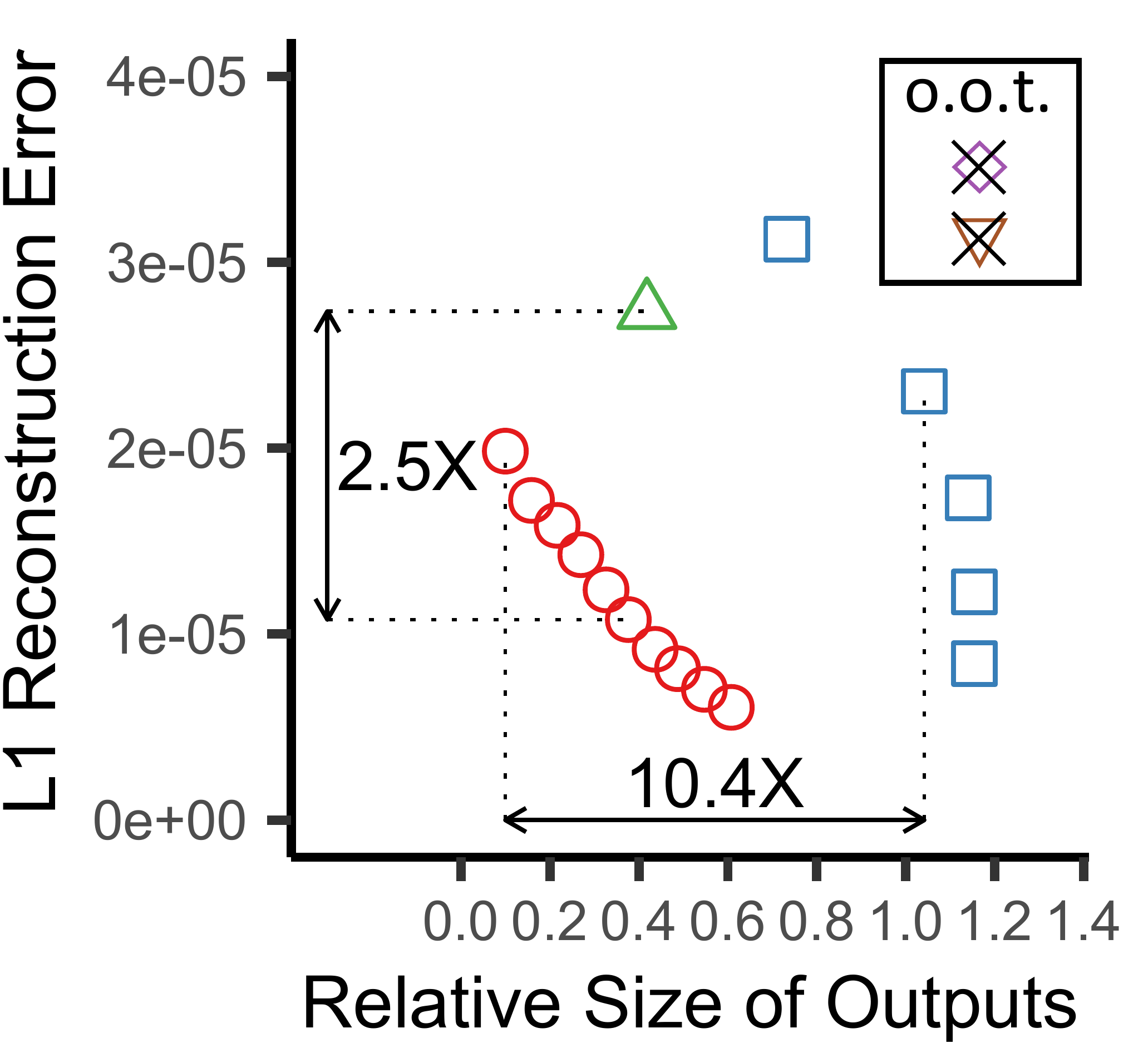}
	} 
	\subfigure[Amazon-0302]{
		\label{fig:l1:A3}
		\includegraphics[width=0.185\textwidth]{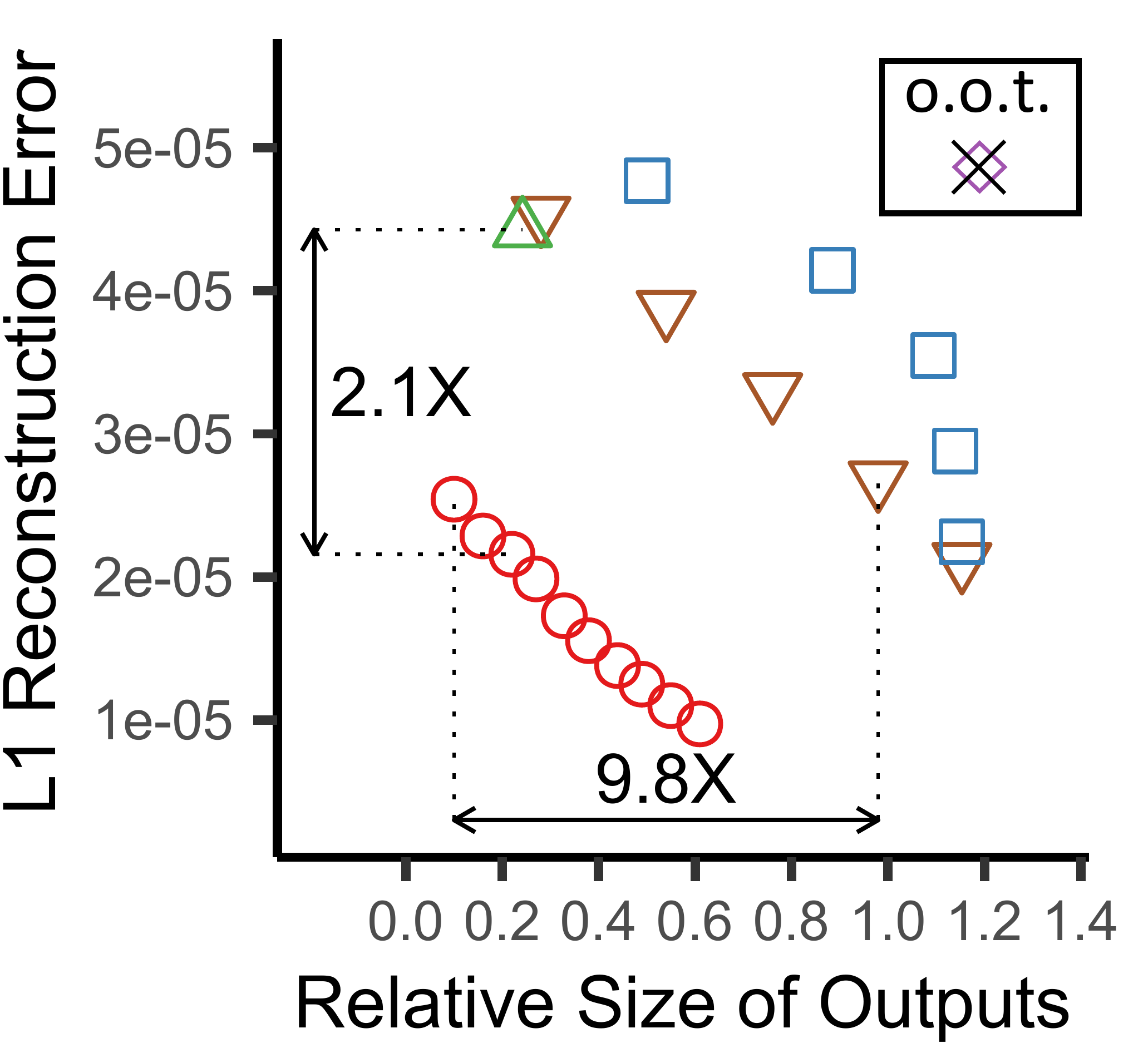}
	} 
	\subfigure[Email-Enron]{
		\label{fig:l1:EE}
		\includegraphics[width=0.185\textwidth]{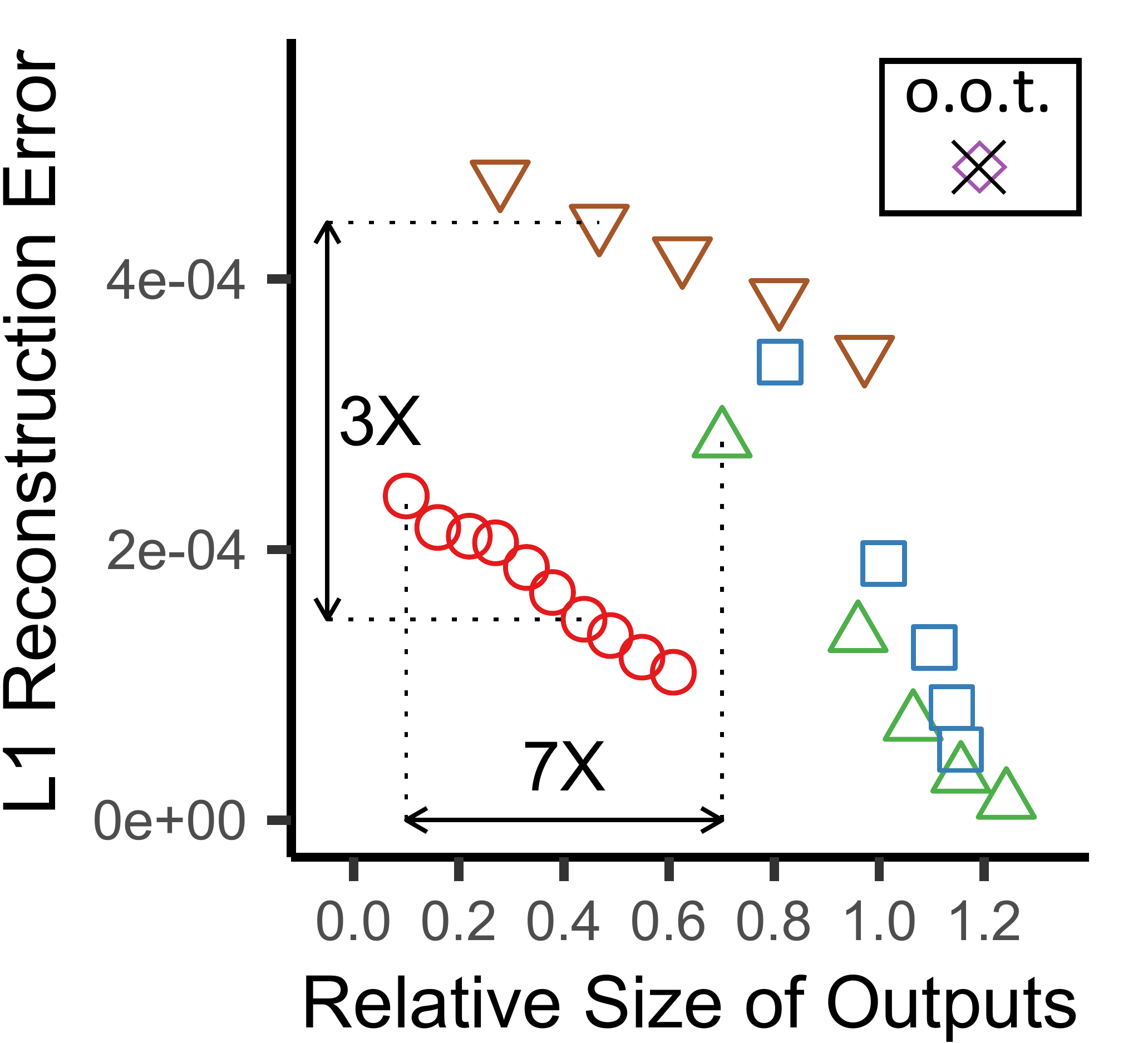}
	}
	\subfigure[Caida]{
		\label{fig:l1:CA}
		\includegraphics[width=0.185\textwidth]{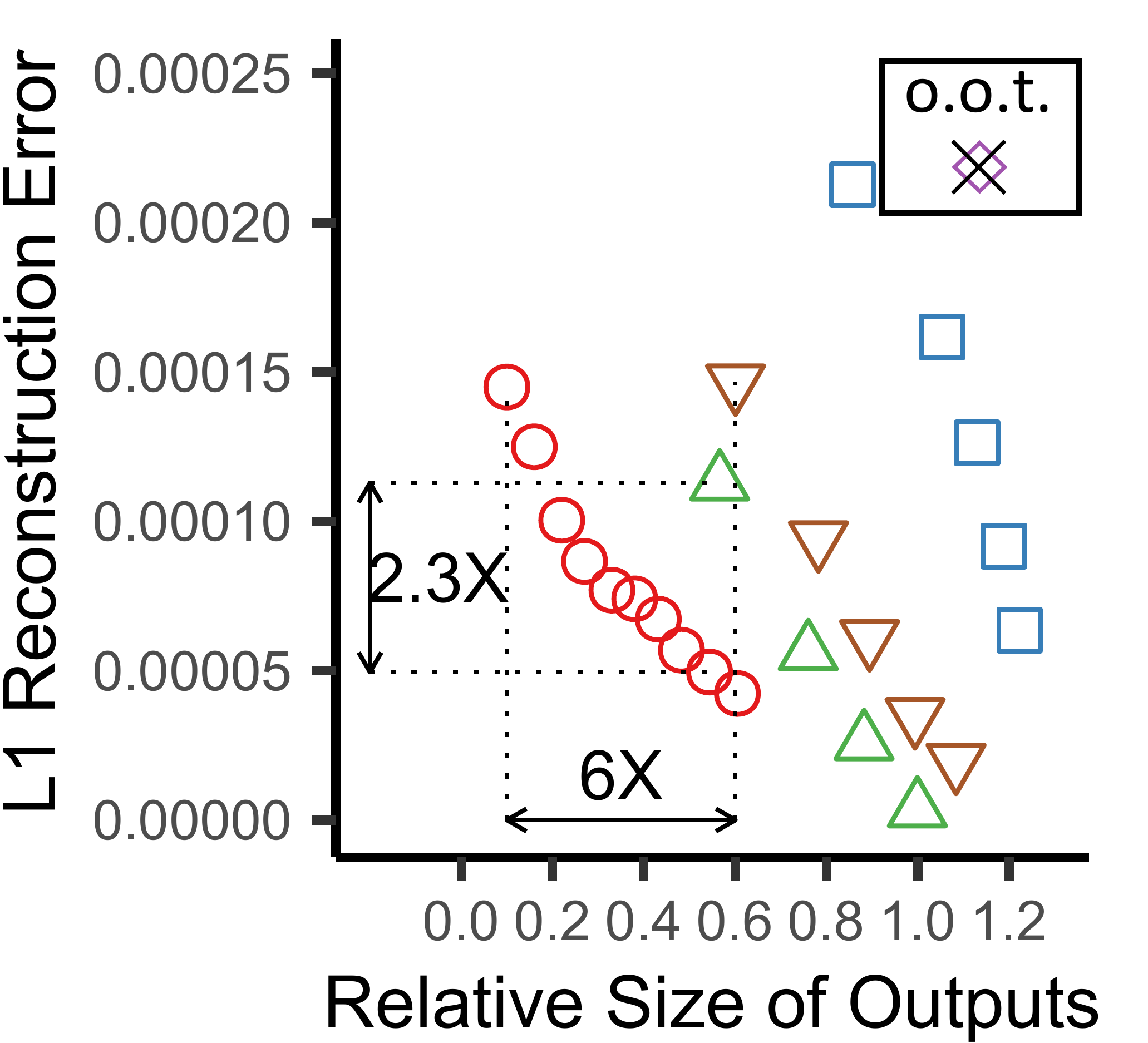}
	}
	\subfigure[Ego-Facebook]{
		\label{fig:l1:EF}
		\includegraphics[width=0.185\textwidth]{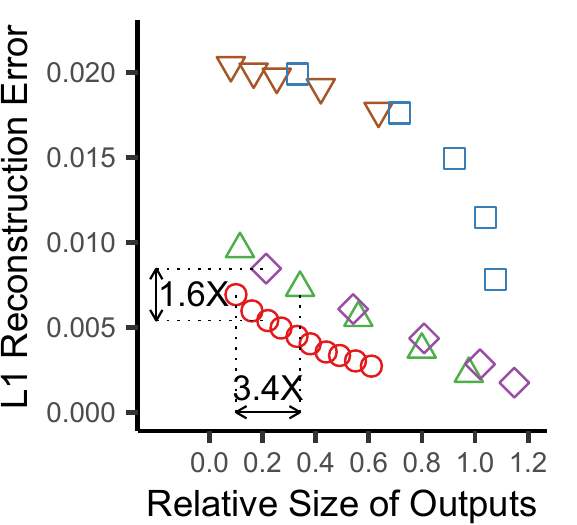}
	} \\
	\vspace{-2.5mm}
	\subfigure[Web-UK-05]{
		\label{fig:l1:W5}
		\includegraphics[width=0.185\textwidth]{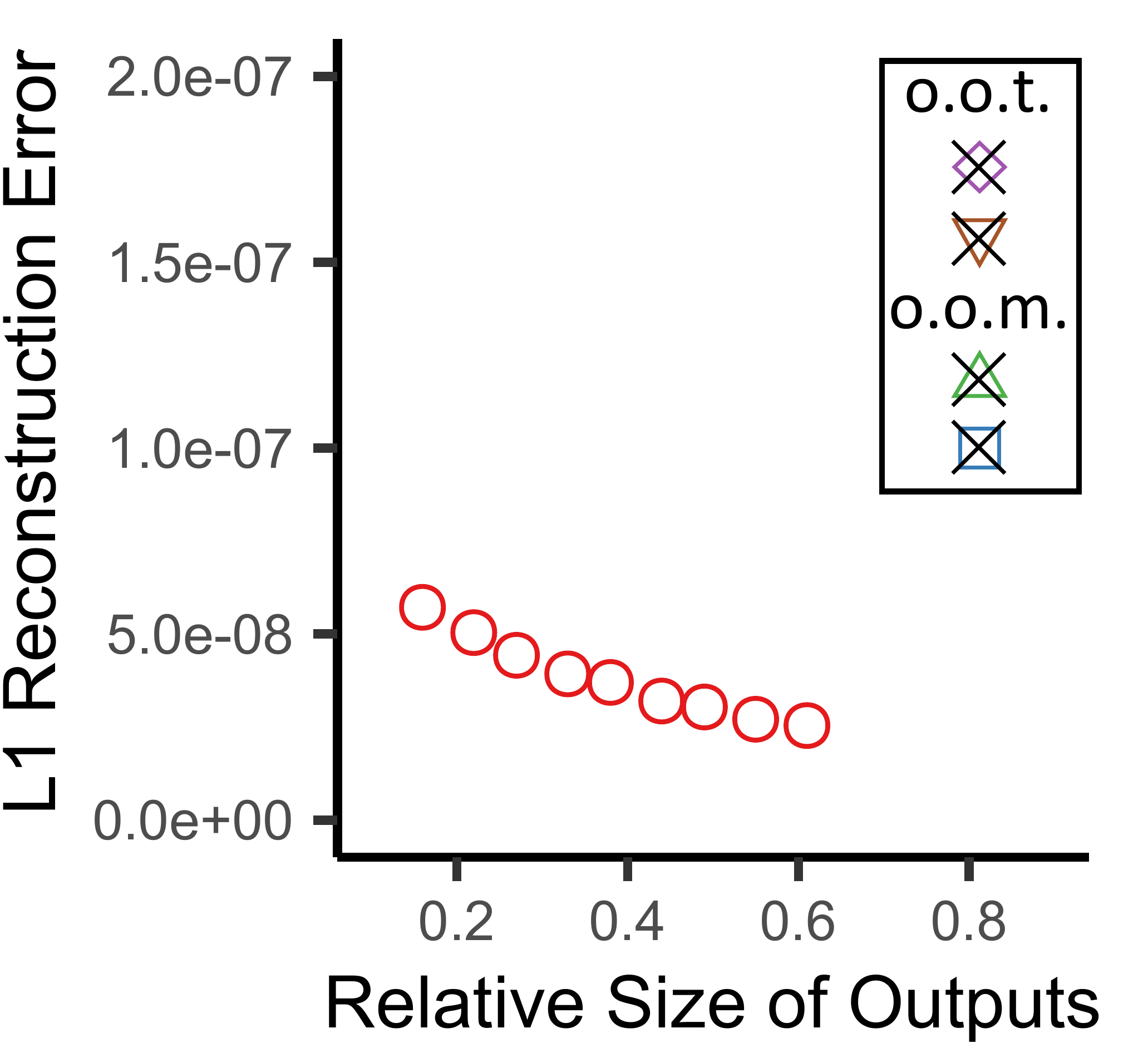}
	} 
	\subfigure[Web-UK-02]{
		\label{fig:l1:W2}
		\includegraphics[width=0.185\textwidth]{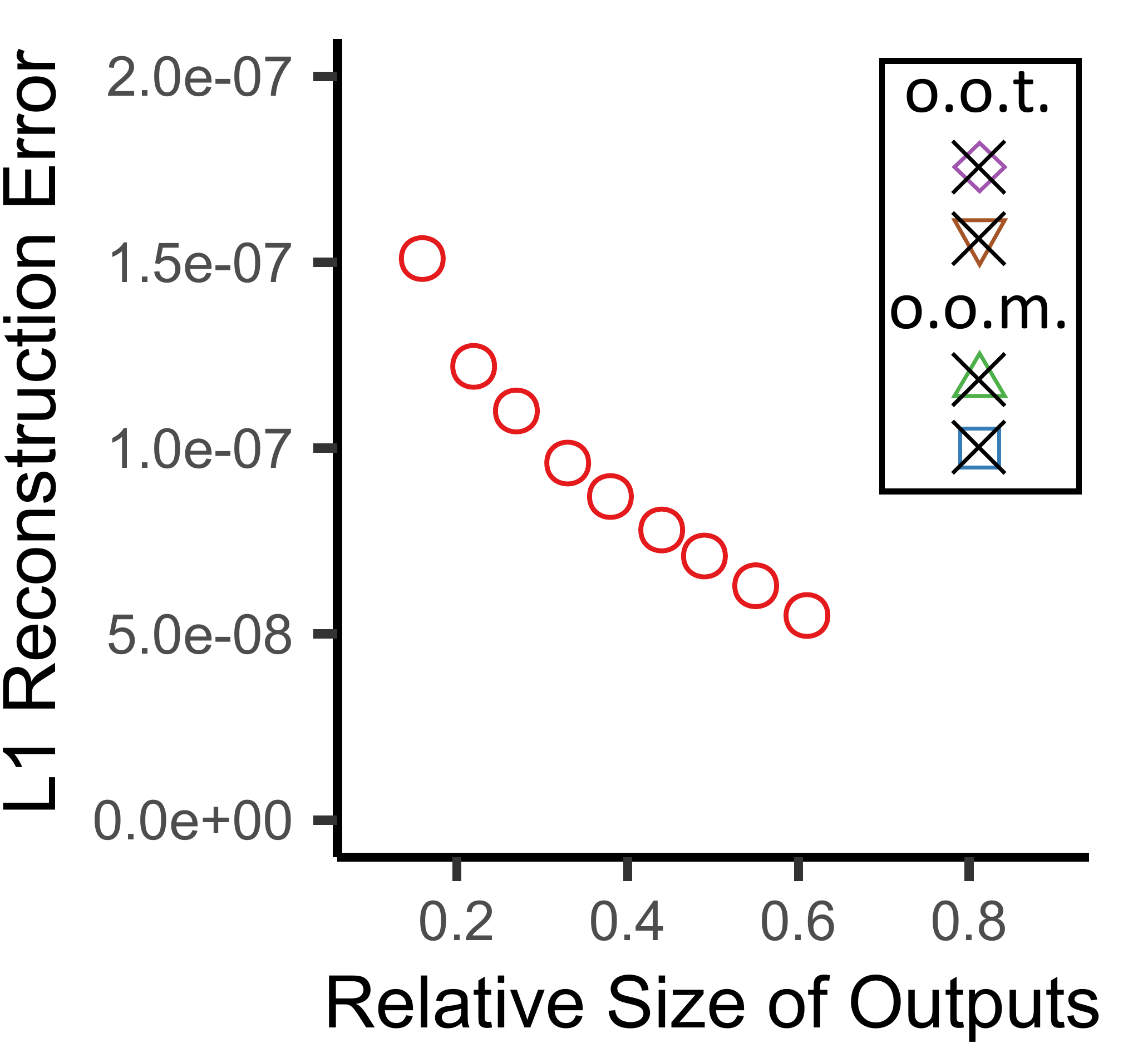}
	}
	\subfigure[LiveJournal]{
		\label{fig:l1:LJ}
		\includegraphics[width=0.185\textwidth]{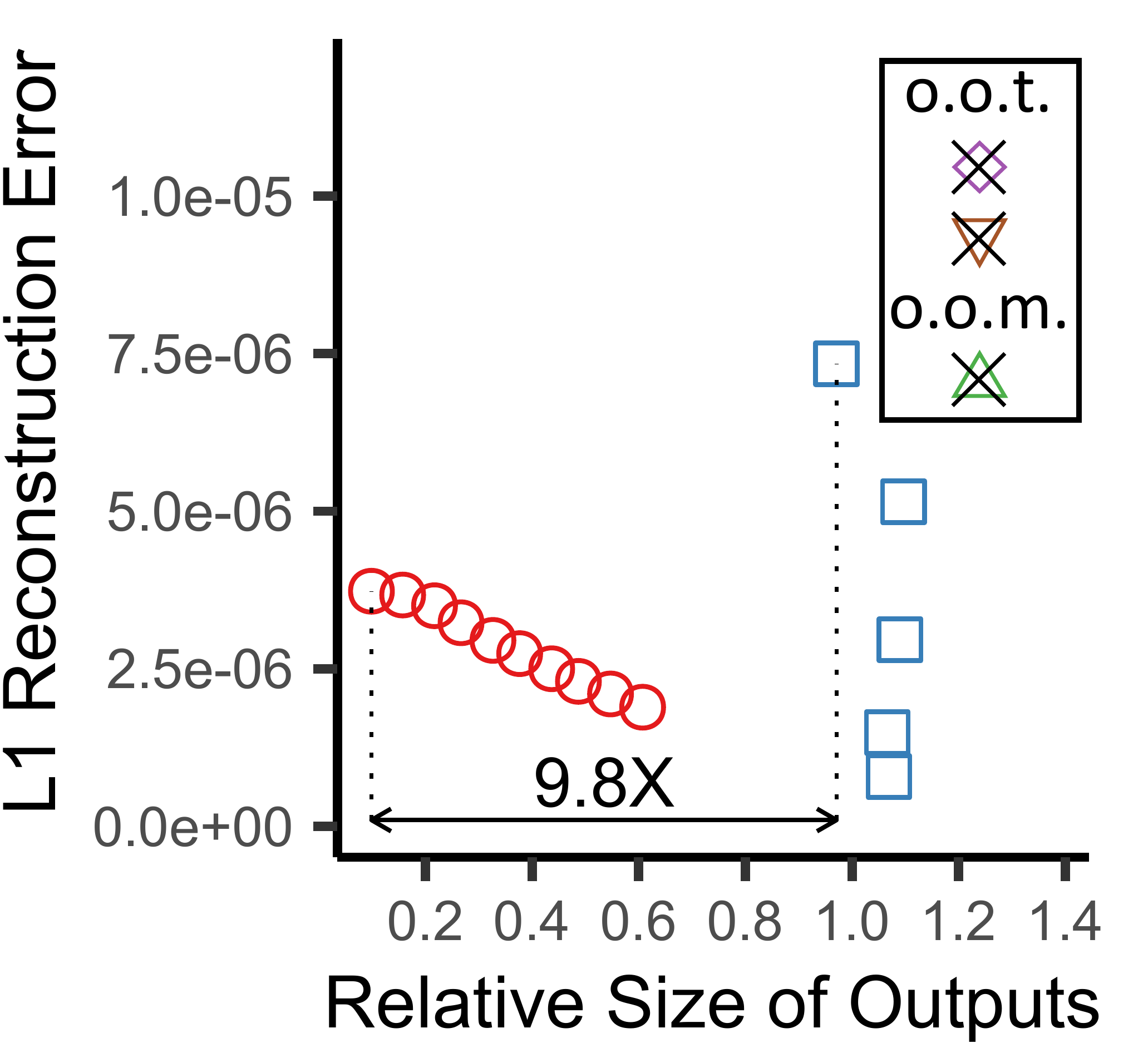}
	}
	\subfigure[Skitter]{
		\label{fig:l1:SK}
		\includegraphics[width=0.185\textwidth]{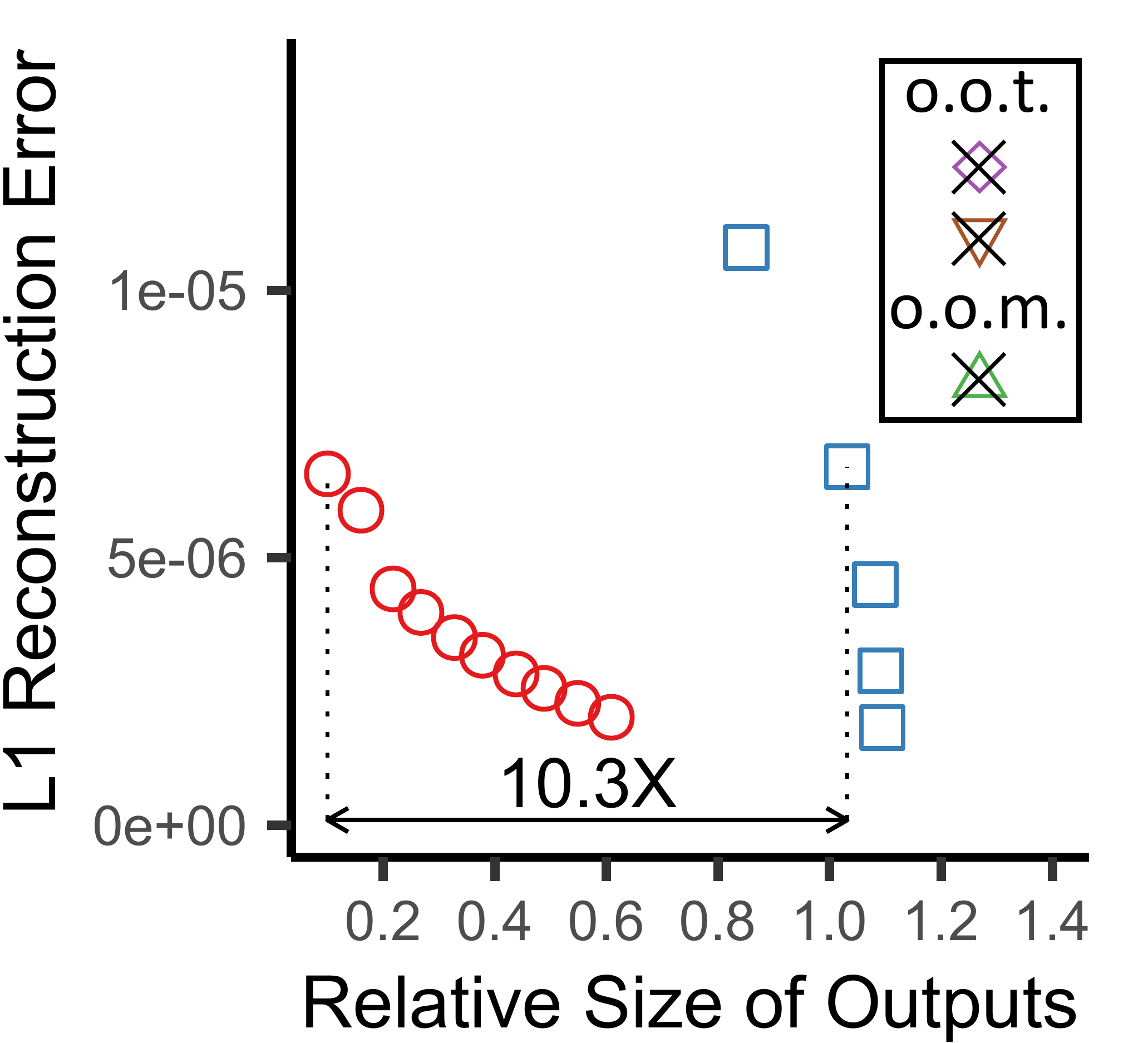}
	} 
	\subfigure[Amazon-0601]{
		\label{fig:l1:A6}
		\includegraphics[width=0.185\textwidth]{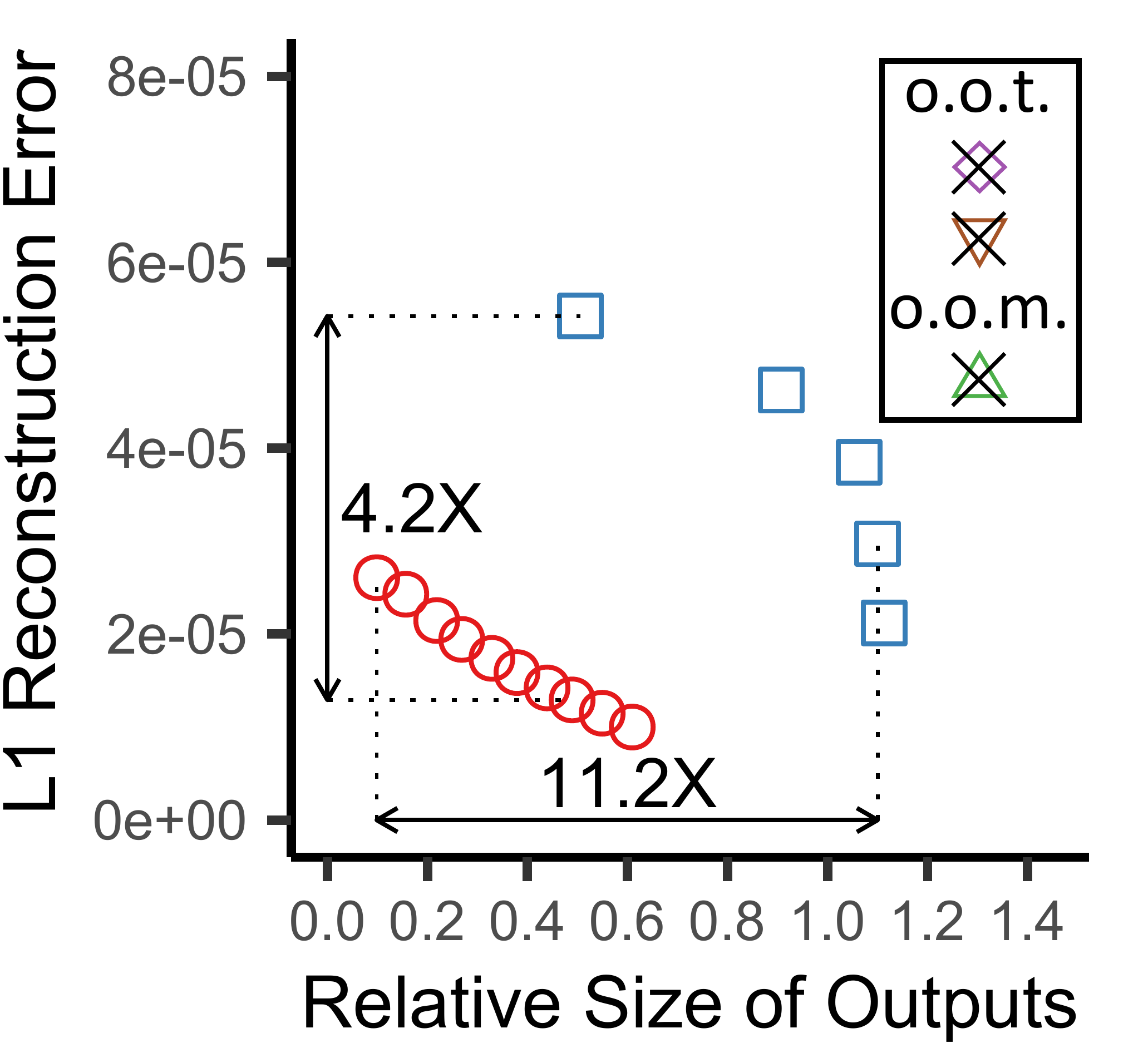}
	} 
	\\\vspace{-2mm}
	\caption{\underline{\smash{\method yields compact and accurate summary graphs.}} o.o.t.: out of time (>12hours). o.o.m.: out of memory (>64GB). \method yielded up to $\mathbf{11.2\times}$ {\bf smaller summary graphs} with similar reconstruction error. It also achieved up to $\mathbf{4.2\times}$ {\bf smaller reconstruction error} with similarly concise outputs. \label{fig:concise_accurate}}
\end{figure*}

\vspace{-1mm}
\subsection{Complexity Analysis}
\label{sec:method:analysis}
\vspace{-1mm}

We analyze the time and space complexities of \method.
To this end, we define the neighborhood of a supernode $A\in S$ as $\sneighbor{A} := \{B\in S : \exists u \in A,~ \exists v \in B \text{ s.t. } \{u,v\} \in E\}$, i.e., the set of supernodes that include a subnode adjacent to any subnode in $A$.
For simplicity, we assume $|V|=O(|E|)$, as in most real-world graphs.


\smallsection{Time complexity:} \method scales linearly with the size of the input graph, as formalized in Thm.~\ref{thm:time_complexity}, which is based on Lemma~\ref{lemma:time_complexity}.

\begin{lemma} \label{lemma:time_complexity}
	The merging and sparsification phase, i.e., lines~\ref{alg:line:merge1}-\ref{alg:line:merge2} of Alg.~\ref{algo:main3}, takes $O(|E|)$ time.
\end{lemma}
\begin{proof} 
	Consider a candidate set $C\in \ST$.
	Considering the termination condition (i.e., line~\ref{alg:line:skip2} of Alg.~\ref{alg:merge}), to merge a pair, $O(\log_{2}^{2}|C|+1)$ pairs are considered.
	Thus, finding the best pair among them takes $O((\log_{2}^{2}|C|+1)\cdot \max_{A \in C} |\sneighbor{A}|)$ time, and if Eq.~\eqref{eq:relative_saving} is greater than $\anneal$, then merging the pair and sparsifying the adjacent superedges takes additional $O(\sum_{A \in C}|\sneighbor{A}|)$ time. 
	In total, a merger takes $O((\log_{2}^{2}|C|+1)\cdot \sum_{A \in C}|\sneighbor{A}|)$ time, and since at most $|C|$ merges take place within $C$, the time complexity of processing a candidate set $C$ (i.e., Alg.~\ref{alg:merge}) is $O(|C|\cdot (\log_{2}^{2}|C|+1)\cdot \sum_{A \in C}|\sneighbor{A}|)$, which is $O(\sum_{A \in C}|\sneighbor{A}|)$ because we upper bound $|C|$ by a constant, as described in Sect.~\ref{sec:method:search:candidate}. Since $\sum_{C\in\ST}\sum_{A \in C}|\sneighbor{A}|= \sum_{A \in S}|\sneighbor{A}|\leq 2|E|$, processing all candidate sets in $\ST$ takes $O(|E|)$ time.
\end{proof}

\begin{theorem}[Linear Scalability of \method] \label{thm:time_complexity}
	The time complexity of Alg.~\ref{algo:main3} is $O(T\cdot |E|)$.
\end{theorem}
\begin{proof} 
	The initialization phase takes $O(1)$ time per subnode and subedge and thus  $O(|V|+|E|)=O(|E|)$ time in total. The candidate generation and further sparsification phases take $O(|V|+|E|)=O(|E|)$ time, as discussed in Sects.~\ref{sec:method:search:candidate} and \ref{sec:method:search:drop}. 
	The merging and sparsification phase also takes $O(|E|)$ time, as proven in Lemma~\ref{lemma:time_complexity}. 
	Since each phase is repeated at most $T$ times, the total time complexity of Alg.~\ref{algo:main3} is $O(T\cdot |E|)$. 
\end{proof}

\smallsection{Space complexity:}
\method (i.e., Alg.~\ref{algo:main3}) needs to maintain (1) the input graph \fgraph, (2) the summary graph \fsummary, (3) the neighborhood $\sneighbor{A}$ of each supernode $A\in S$, and (4) a random hash function $h(v)$ for each subnode $v \in V$. 
Since $|S|\leq |V|$, $|P|\leq |E|$, and $\sum_{A \in S}|\sneighbor{A}|\leq 2|E|$, its memory requirements are $O(|E|)$.

\vspace{-1mm}
\section{Experiments} 
\label{sec:exp}
\vspace{-1mm}

We review our experiments designed for the following questions:
\begin{enumerate}[leftmargin=*]
	\item[Q1.] \textbf{Compactness \& Accuracy}: Does \method yield more compact and accurate summary graphs than its best competitors?
	\item[Q2.] \textbf{Speed}: Is \method faster than its best competitors?
	\item[Q3.] \textbf{Scalability}: Does \method scale linearly with the size of the input graph? Can it handle graphs with about $1$ billion edges?
	\item[Q4.] \textbf{Effects of Parameters (Appendix~\ref{appendix:param})}: How does the number of iterations $T$ affect the accuracy of summary graphs?
\end{enumerate}

\vspace{-1mm}
\subsection{Experimental Settings}
\vspace{-1mm}

\smallsection{Machines}: All experiments were conducted on a desktop with a 3.7 GHz Intel i5-9600k CPU and 64GB memory. 

\smallsection{Datasets}: We used the publicly available real-world graphs listed in Table~\ref{tab:DatasetTable} after removing all self-loops and the direction of all edges.

\smallsection{Implementations}: We implemented \method and \kGs \cite{lefevre2010grass} in Java, and for \SL \cite{riondato2017graph} and \SAAGs \cite{beg2018scalable}, we used the implementation in C++ and Java, resp., released by the authors.
In \method The target summary size was set from 10\% to 60\% of the size of the input graph, at equal intervals. 
The number of iterations $T$ was fixed to $20$ unless otherwise stated (see Appendix~\ref{appendix:param} for its effects.)
For \kGs, \SL, and \SAAGs, the target number of supernodes was set from 10\% to 60\% of the number of nodes in the input graph, at equal intervals.
For \kGs, we used the \textit{SamplePairs} method with $c=1.0$, as suggested in \cite{lefevre2010grass}.
For \SAAGs and \linearSAAGs, the number of sample pairs was set to $\log n$ and $n$, resp., and the count-min sketch was used with $w=50$ and $d=2$.

\begin{table}[t!]
	\vspace{-2mm}
	\begin{center}
		\caption{Summary of the real-world datasets}
		\label{tab:DatasetTable}
		\small
		\scalebox{0.9}{
			\begin{tabular}{r|r|r|r}
				\toprule 
				\textbf{Name} & \textbf{\# Nodes} & \textbf{\# Edges}  & \textbf{Summary}\\
				\midrule
				Caida (CA) \cite{leskovec2005graphs} & 26,475 & 53,381 & Internet\\
				Ego-Facebook (EF) \cite{leskovec2012learning} & 4,039 & 88,234 & Social\\
				Email-Enron (EE) \cite{klimt2004introducing} & 36,692 & 183,831 & Email\\
				Amazon-0302 (A3) \cite{leskovec2007dynamics} & 262,111 & 899,792 & Co-purchase\\
				DBLP (DB) \cite{yang2015defining} & 317,080 & 1,049,866 & Collaboration\\
				Amazon-0601 (A6) \cite{leskovec2007dynamics} & 403,394 & 2,443,408 & Co-purchase\\
				Skitter (SK) \cite{leskovec2005graphs} & 1,696,415 & 11,095,298 & Internet\\
				LiveJournal (LJ) \cite{yang2015defining} & 3,997,962 & 34,681,189 & Social\\
				Web-UK-02 (W2) \cite{boldi2004webgraph} & 18,483,186 & 261,787,258 & Hyperlinks\\
				Web-UK-05 (W5) \cite{boldi2004webgraph}& 39,454,463 & 783,027,125 & Hyperlinks\\
				\bottomrule
			\end{tabular}
		}
	\end{center}
\end{table}

\smallsection{Evaluation Metrics}: 
We evaluated summary graphs in terms of accuracy, size, and quality. 
For accuracy, we measured \lone and \ltwo reconstruction errors, i.e., $RE_{1}$ and $RE_{2}$ (see Eq.~\eqref{eq: lpNorm}), and we normalized them by dividing them by the size of the adjacency matrix.\footnote{We ignore the diagonals, and the size of the adjacency matrix is $|V|\cdot(|V|-1)$.}
For size, we used the number of bits required to store each summary graph (i.e., Eq.~\eqref{eq: summarysize}). 
The quality of a summary graph is a metric for evaluating its accuracy and size at the same time. 
For quality, we (1) measured the reconstruction error $RE_{1}$ and size of the summary graphs obtained by all competitors, (2) normalized both so that they are between $0$ and $1$ in each dataset,\footnote{Normalizing $X_{i}$ results in ${(X_{i}-\min_{j}X_{j})}/{(\max_{j}X_{j}-\min_{j}X_{j})}$.} and (3) computed $\sqrt{\text{normalized size}^2 + \text{normalized reconstruction error}^2}$, i.e., the euclidean distance from the ideal quality.\footnote{The maximum distance is $\sqrt{2}$.}
All evaluation metrics were averaged over $5$ iterations.

\begin{figure*}[t!]
	\centering
	\vspace{-4mm}
	\includegraphics[width=0.48\linewidth]{SsumM_Legend.pdf} \\
	\vspace{-1.9mm}
	\subfigure[DBLP]{
		\label{fig:time:DB}
		\includegraphics[width=0.185\textwidth]{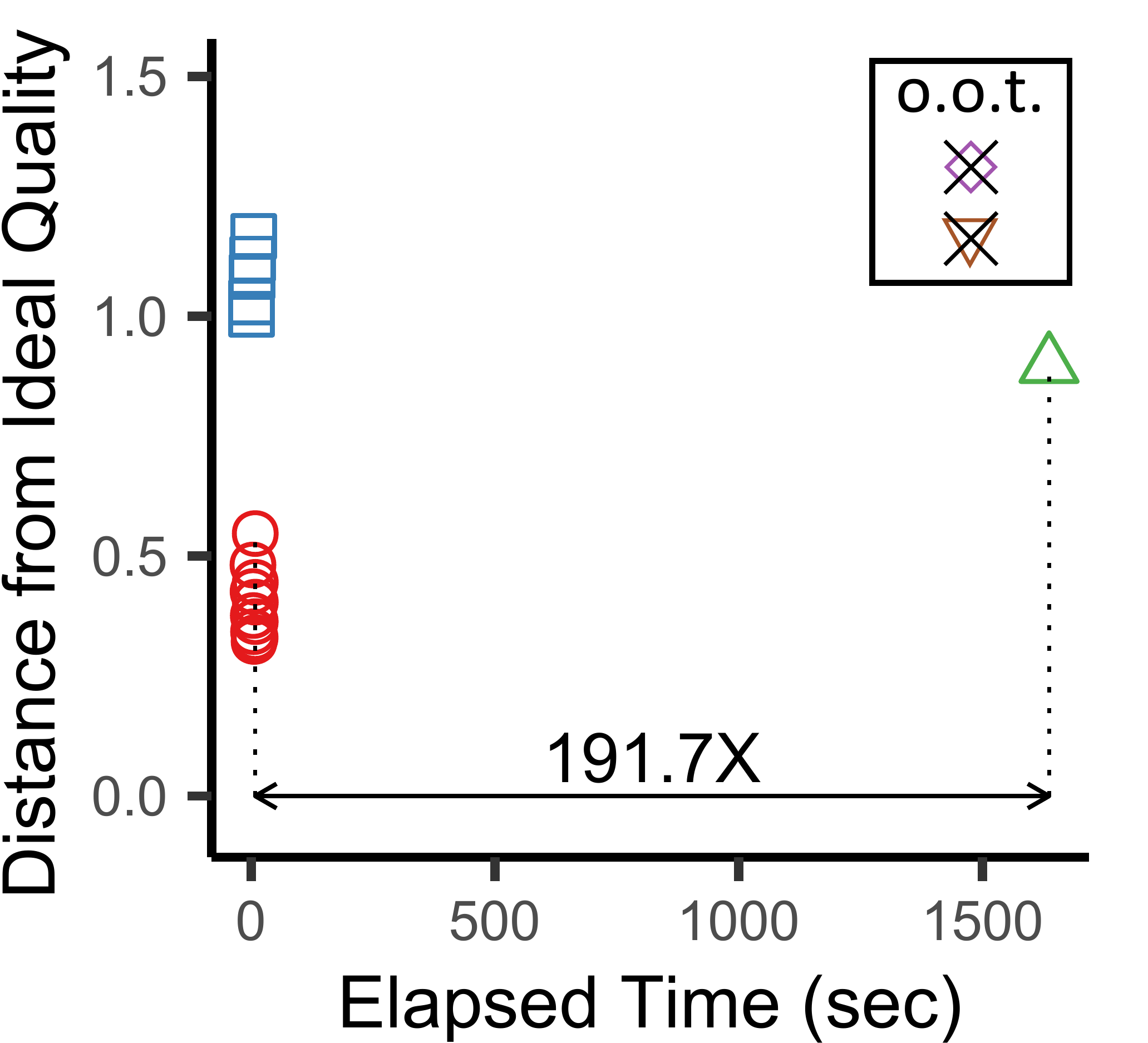}
	} 
	\subfigure[Amazon-0302]{
		\label{fig:time:A3}
		\includegraphics[width=0.185\textwidth]{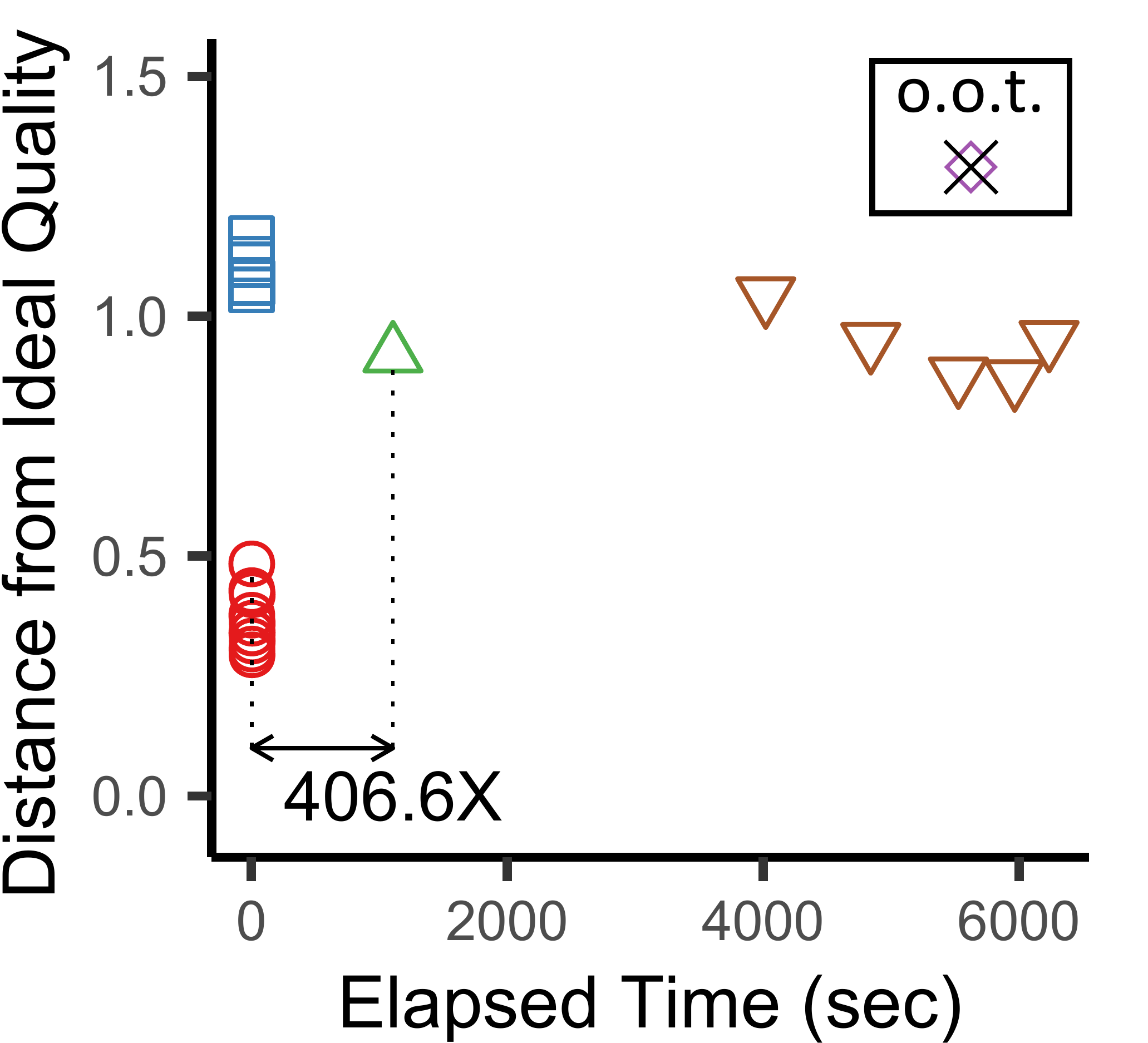}
	} 
	\subfigure[Email-Enron]{
		\label{fig:time:EE}
		\includegraphics[width=0.185\textwidth]{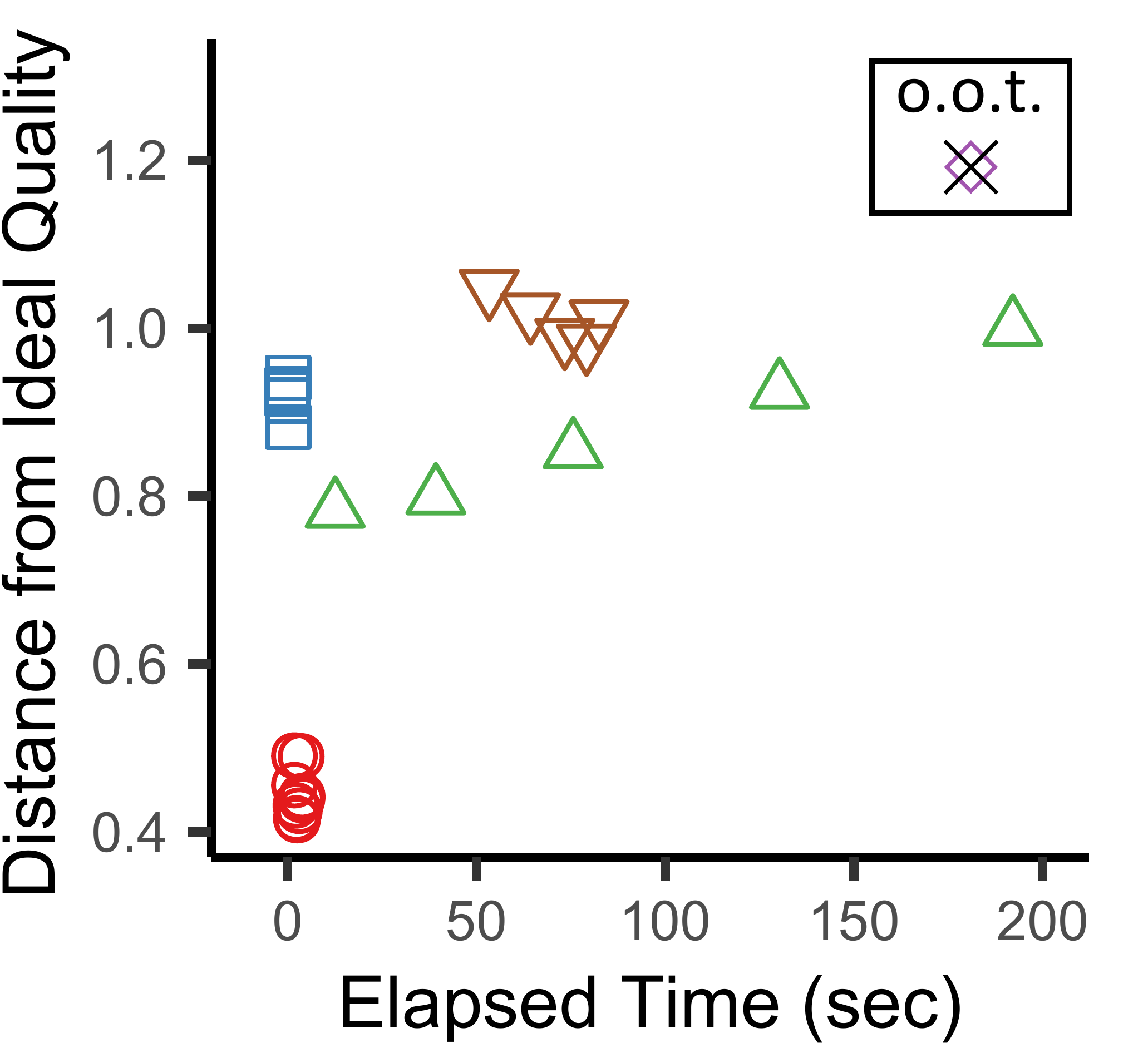}
	}
	\subfigure[Caida]{
		\label{fig:time:CA}
		\includegraphics[width=0.185\textwidth]{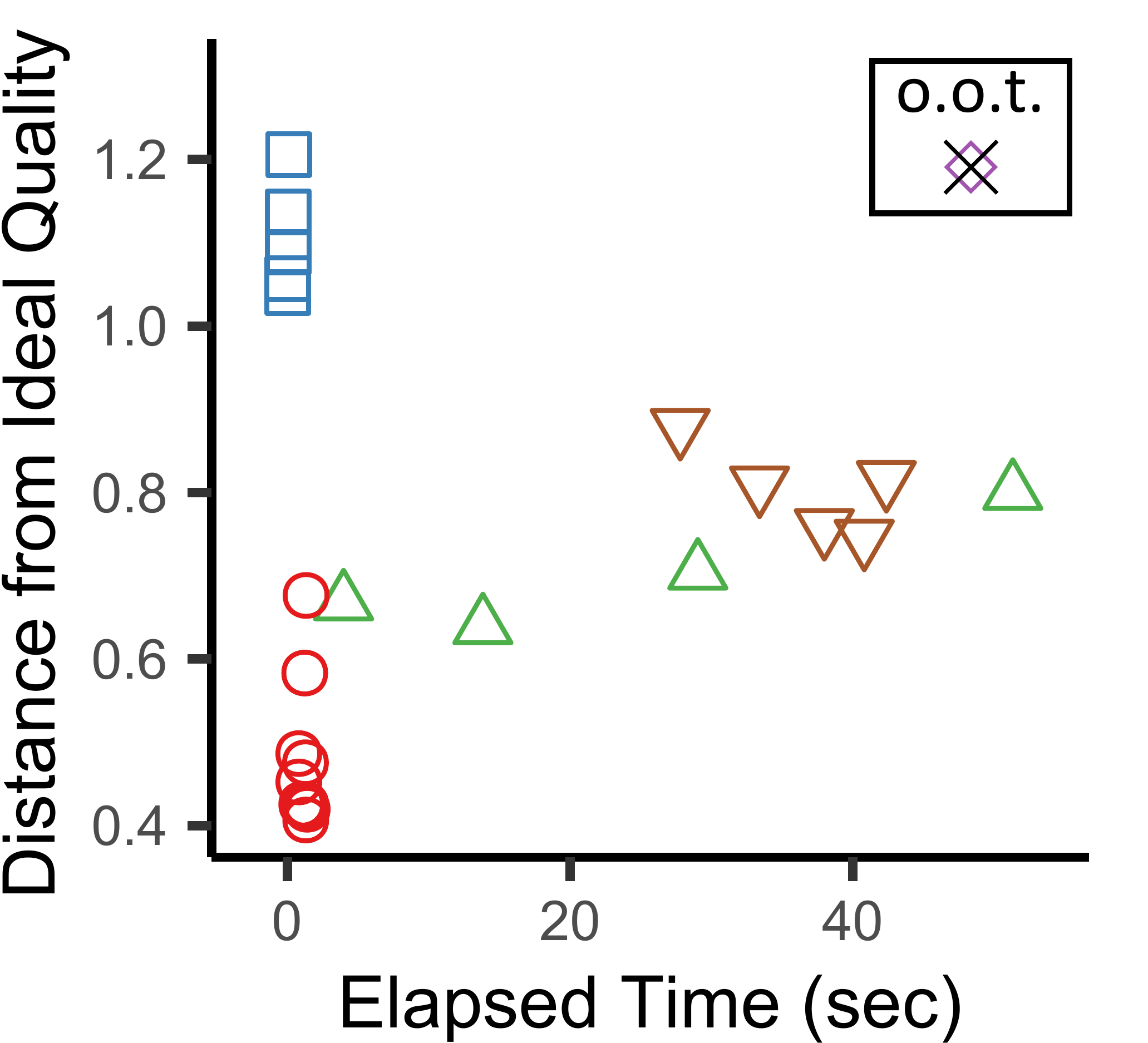}
	}
	\subfigure[Ego-Facebook]{
		\label{fig:time:EF}
		\includegraphics[width=0.185\textwidth]{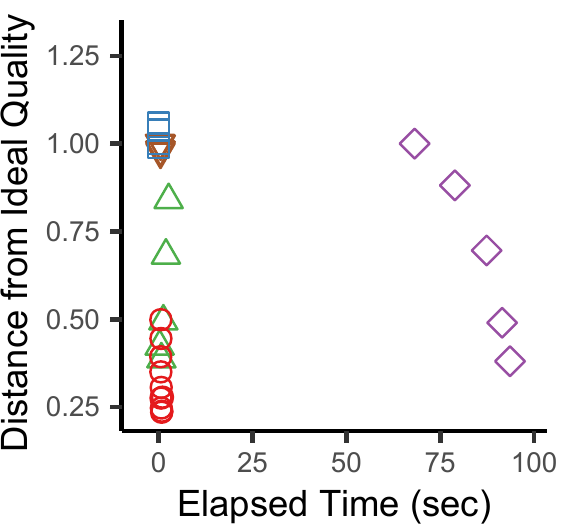}
	}  \\
	\vspace{-2.5mm}
	\subfigure[Web-UK-05]{
		\label{fig:time:W5}
		\includegraphics[width=0.185\textwidth]{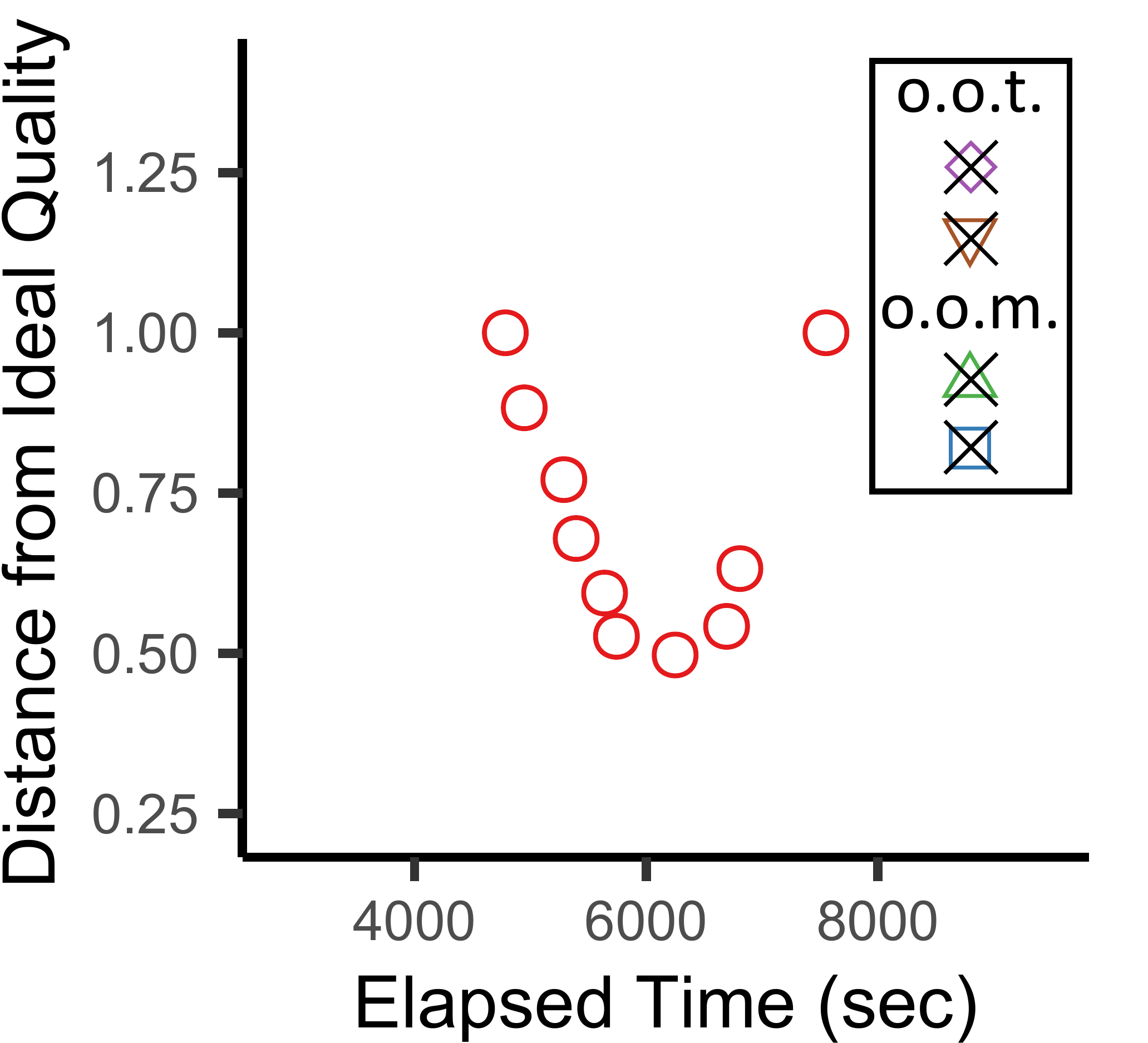}
	} 
	\subfigure[Web-UK-02]{
		\label{fig:time:W2}
		\includegraphics[width=0.185\textwidth]{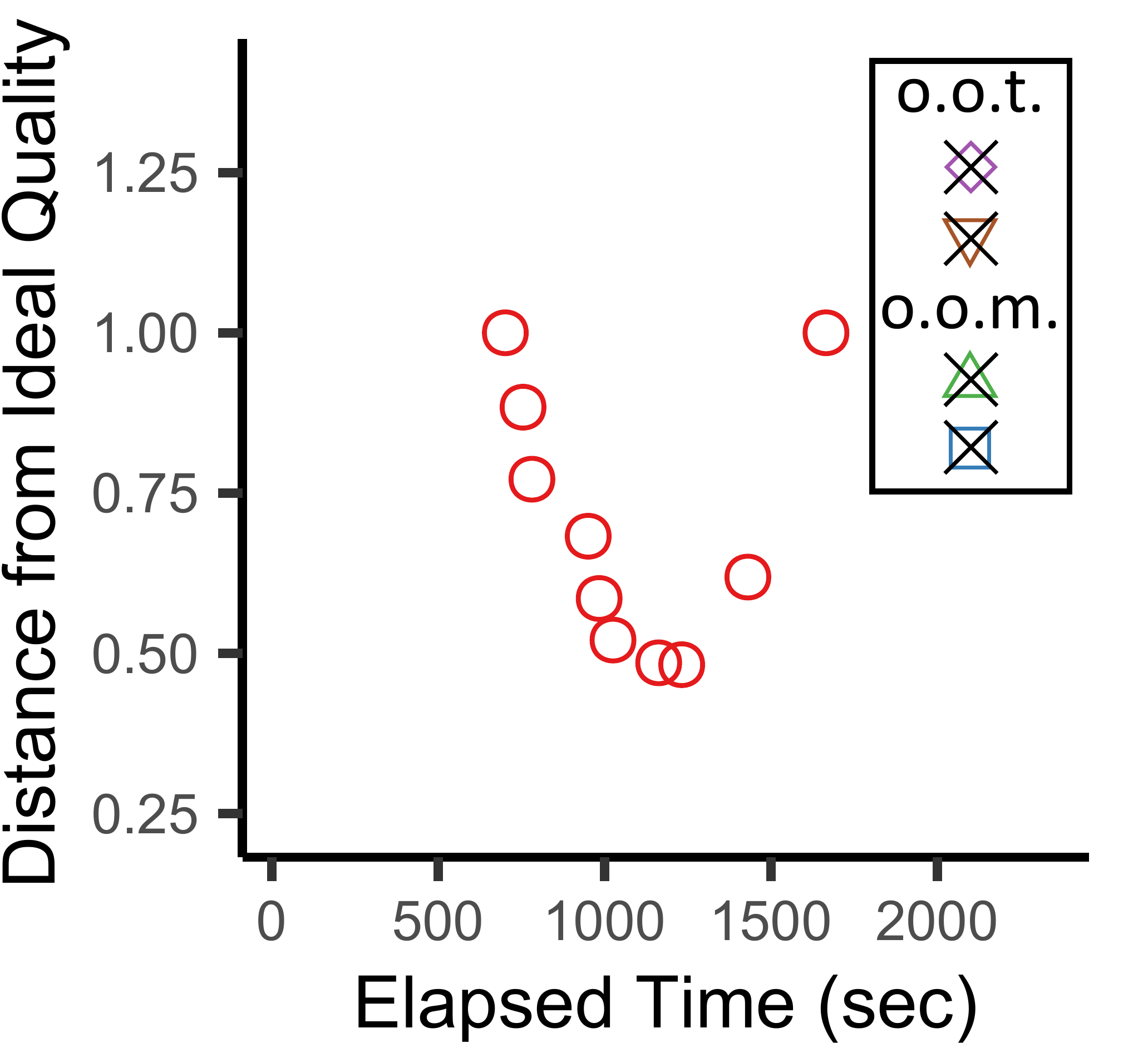}
	}
	\subfigure[LiveJournal]{
		\label{fig:time:LJ}
		\includegraphics[width=0.185\textwidth]{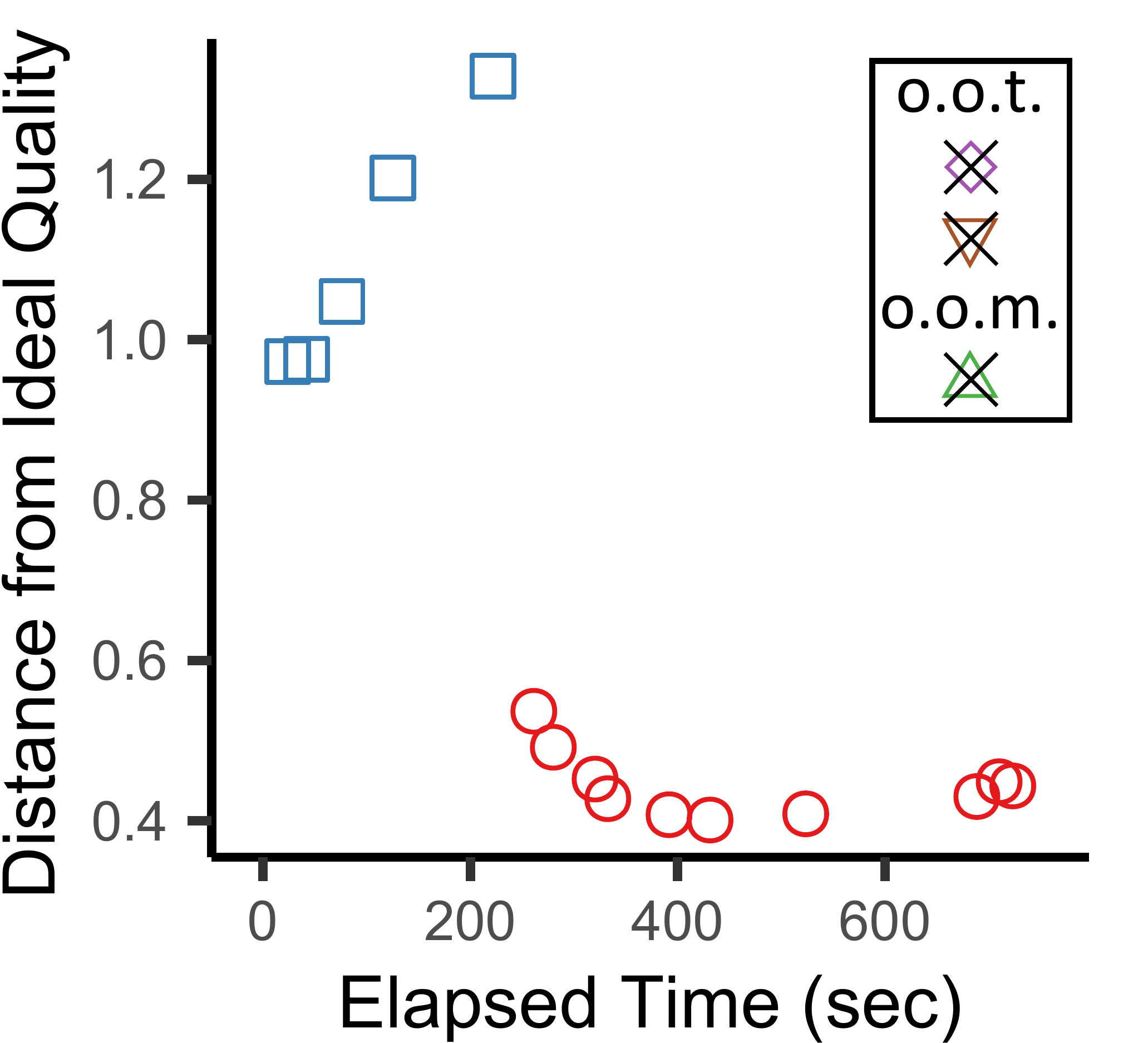}
	}
	\subfigure[Skitter]{
		\label{fig:time:SK}
		\includegraphics[width=0.185\textwidth]{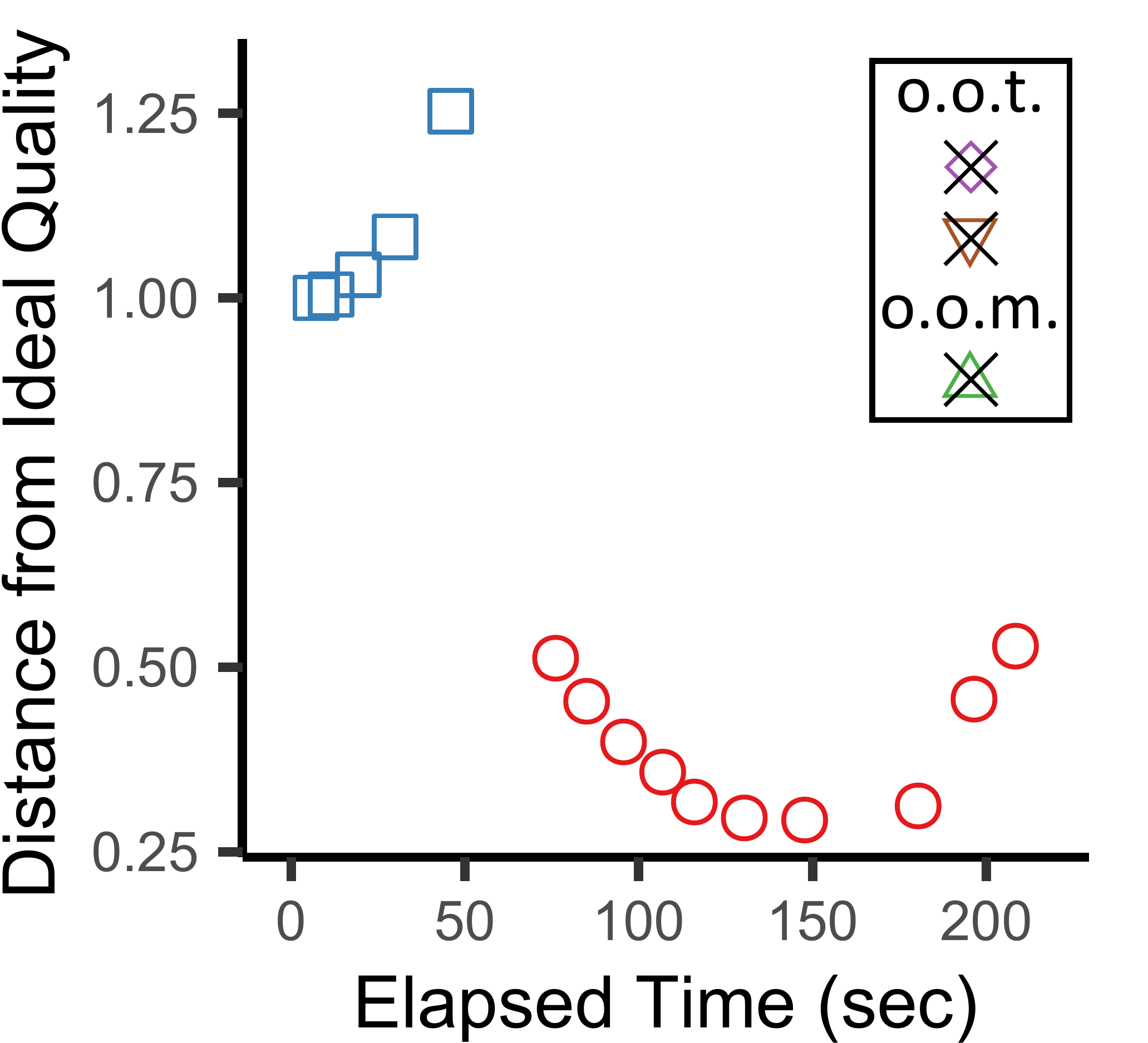}
	} 
	\subfigure[Amazon-0601]{
		\label{fig:time:A6}
		\includegraphics[width=0.185\textwidth]{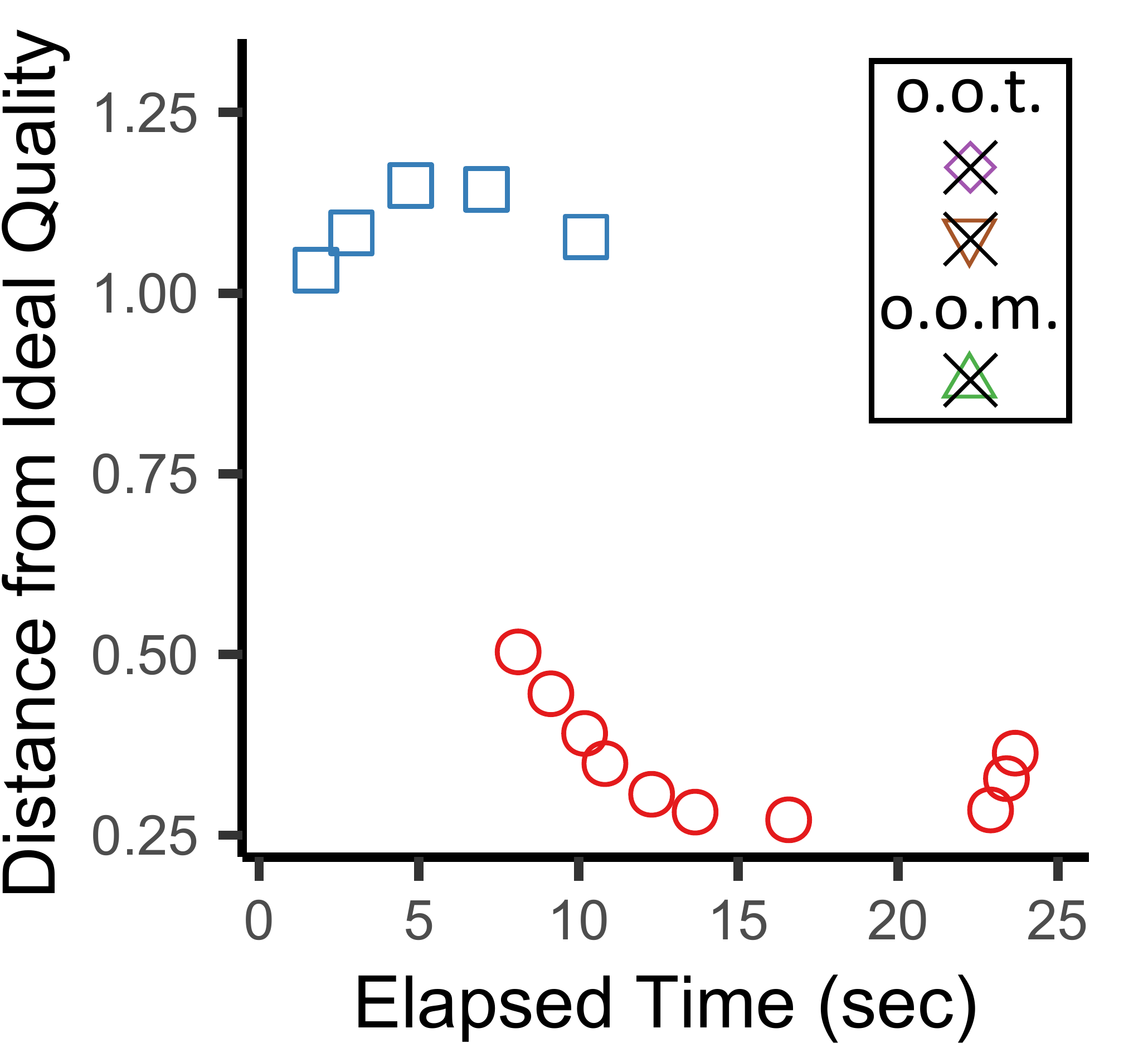}
	} \\
	\vspace{-2mm}
	\caption{\label{fig:time}\underline{\smash{\method is fast with high-quality summary graphs.}} o.o.t.: out of time (>12hours). o.o.m.: out of memory (>64GB). \method was up to $\mathbf{406.6\times}$ faster than the competitors with outputs of better quality.  Only \method scaled to the largest datasets. 
	}
\end{figure*}

\vspace{-1mm}
\subsection{Q1. Compactness and Accuracy}
\label{sec:exp:compact}
\vspace{-1mm}

We compared the size and \lone \re ($RE_{1}$) of the summary graphs obtained by \method and its competitors.
As seen in Fig.~\ref{fig:concise_accurate}, \textbf{\method yielded the most compact and accurate summaries} in all the considered datasets.
Specifically, \method gave a $\mathbf{11.2\times}$ \textbf{smaller summary graph} with similar or smaller $RE_{1}$ than its competitors in the Amazon-0601 dataset. 
It also gave a summary graph with $\mathbf{4.2\times}$ \textbf{smaller} $RE_{1}$ but similar or smaller sizes than its competitors in the Amazon-0601 dataset.
We obtained consistent results when $RE_{2}$ was used instead of $RE_{1}$ (see Appendix~\ref{appendix:L2}).

Note in Fig.~\ref{fig:concise_accurate} that competitors often gave summary graphs whose (relative) size is greater than $1$. That is, they failed to reduce the size in bits of the input graph since they focused solely on reducing the number of nodes. On the contrary, \method always gives a summary graph whose size does not exceed a given size in bits. 

In Fig.~\ref{fig:vis}, we visually compared the summary graphs obtained by different methods in the Ego-Facebook dataset. While they have similar \lone \re (spec., $(5.9 \pm{0.3}) \times 10^{-3}$), one provided by \method is more concise with fewer edges.


%
%

\vspace{-1mm}
\subsection{Q2. Speed (Fig.~\ref{fig:time})}
\vspace{-1mm}

We compared \method and its competitors in terms of speed and the quality of summary graphs. 
As seen in Fig.~\ref{fig:time},  
\textbf{\method gave the best trade-off between speed and the quality of the summary} on all datasets. 
Specifically, \method was $\mathbf{406.6\times}$ {\bf faster} than \SL while giving summary graphs with better quality in the Amazon-0302 dataset.
While \SAAGs was faster than \method, \method gave outputs of much higher quality than \SAAGs.
\linearSAAGs and \kGs were slower with lower-quality outputs than \method, and they did not scale to large datasets, taking more than $12$ hours.

\begin{figure}[t!]
	\centering
	\vspace{-3mm}
	\subfigure[Skitter]{
		\label{fig:scal:A6}
		\includegraphics[width=0.43\linewidth]{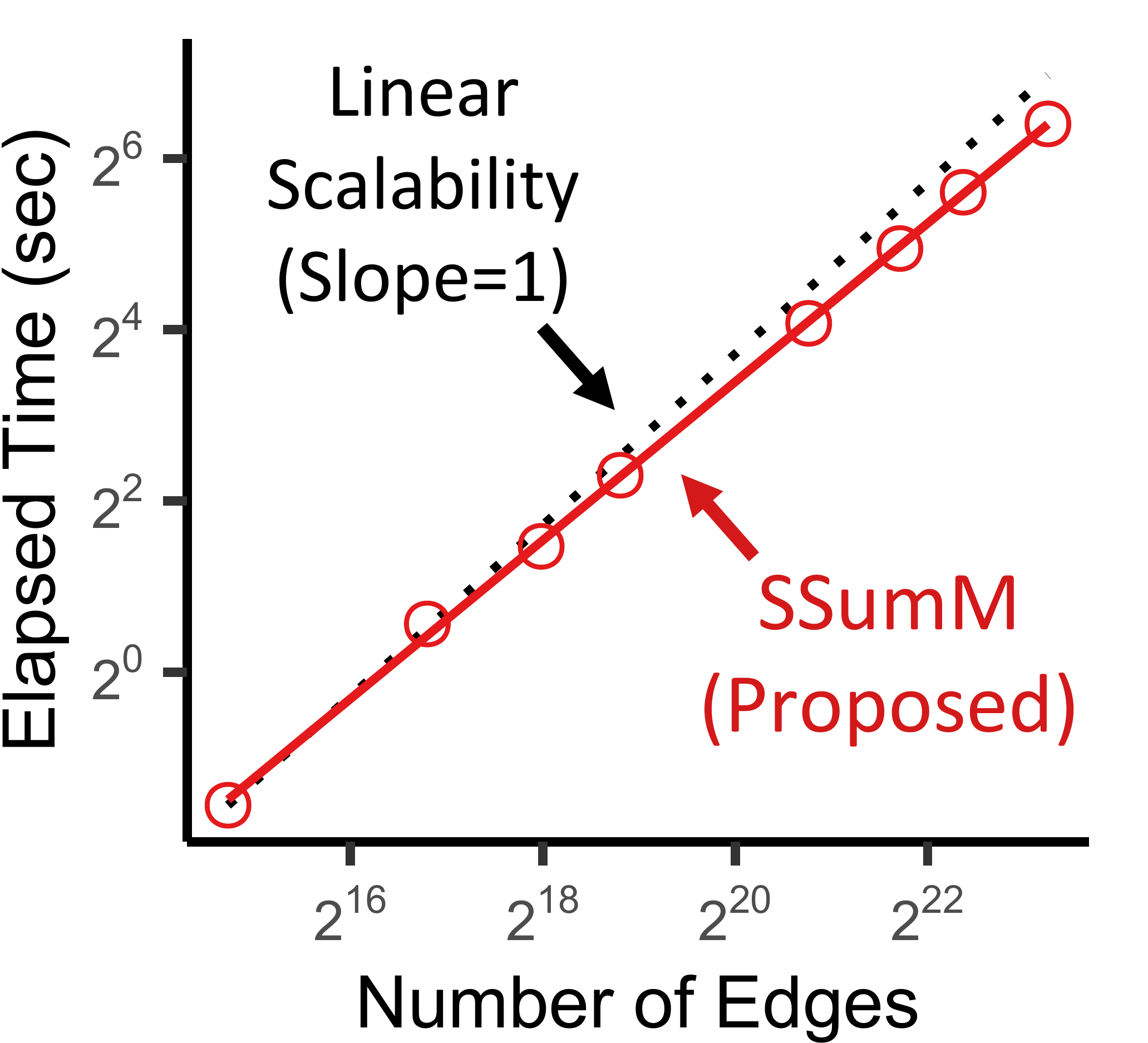}
	}
	\subfigure[Livejournal]{
		\label{fig:scal:LJ}
		\includegraphics[width=0.43\linewidth]{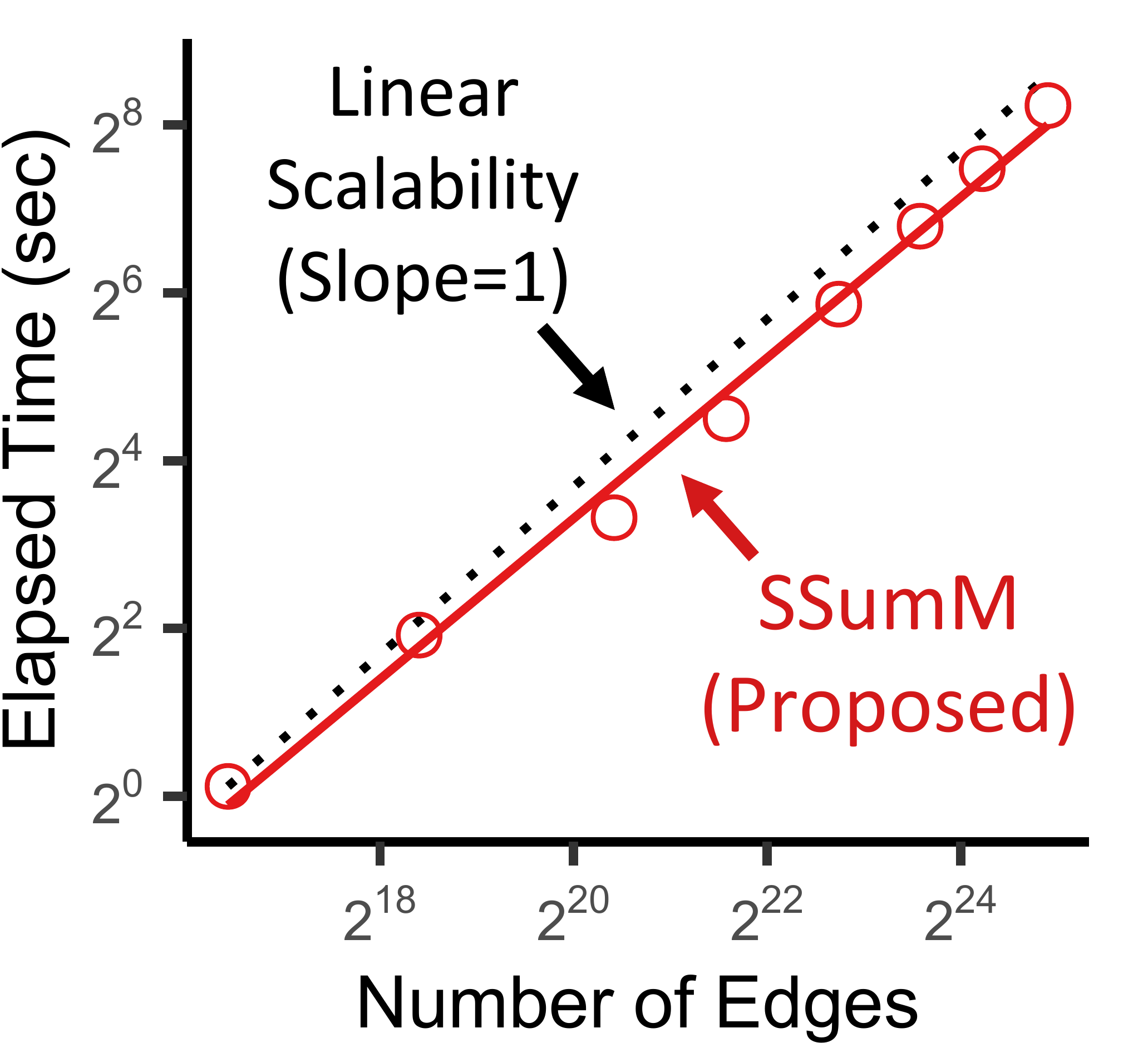}
	} \\
	\vspace{-2mm}
	\caption{\underline{\smash{\method is scalable.}} \method scaled \textbf{linearly} with the number of edges in the input graph.}
	\label{fig:scal}
\end{figure}

\vspace{-1mm}
\subsection{Q3. Scalability (Fig.~\ref{fig:scal})}
\vspace{-1mm}

We evaluated the scalability of \method by measuring how its runtime changes depending on the size of the input graph. To this end, we used a number of graphs that are obtained from the Skitter and Livejournal datasets by randomly sampling different numbers of nodes. 
As seen in Fig. \ref{fig:scal}, \textbf{\method scaled linearly with the size of the input graph}, as formulated in Thm.~\ref{thm:time_complexity}.
In addition, \method successfully processed $26\times$ larger datasets (with about $\mathbf{0.8}$ {\bf billion edges}) than its best competitors, as seen Fig.~\ref{fig:crown:scal}.

\vspace{-1mm}
\section{Related Work}
\label{sec:rel}
\vspace{-1mm}

Graph summarization have been studied extensively for various objectives, including efficient queries \cite{lefevre2010grass,riondato2017graph,tian2018efficient}, compression \cite{toivonen2011compression,navlakha2008graph,khan2014set}, and visualization \cite{dunne2013motif,koutra2014vog,shen2006visual,shah2015timecrunch,lin2008summarization}. See \cite{liu2018graph} for a survey. 
Below, we focus on previous studies directly related to Problem~\ref{problem}. 

Given the target number of supernodes, \kGs \cite{lefevre2010grass} aims to minimize \lone \re by repeatedly merging a pair of supernodes that decrease the \lone \re most among candidate pairs. 
While several sampling methods are proposed to reduce the number of candidate pairs from  $O(|V|^2)$ to $O(|V|)$, \kGs still takes $O(|V|^3)$ time.
\Gs \cite{lefevre2010grass} aims to minimize its loss function, which takes both reconstruction error and the number of supernodes into consideration.
\Gs greedily merges supernodes, as in \kGs, until the loss function increases.


\SL \cite{riondato2017graph} uses geometric clustering for summarizing a graph with a given number of supernodes.
Specifically, \SL considers each row (or column) in the adjacency matrix as a point in the $|V|$-dimensional space, and it employs $k$-means and $k$-median clustering to obtain clusters, each of which is considered as a supernode.
It is shown that \SL provides a theoretical guarantee in terms of the \lpp \re of its output summary graph.
To speed up clustering, which incurs expensive computation of the pairwise distances between many high-dimensional points, \SL also adopts dimensionality reduction \cite{indyk2006stable} and adaptive sampling \cite{aggarwal2009adaptive} techniques.
The scalability of \SL is still limited due to high memory requirements for clustering and high time complexity. Its time complexity, $O(|E| + k|V|)$, becomes $O(|V|^2)$ if $k=O(\lvert V \rvert)$.

\SAAGs \cite{beg2018scalable} is a more scalable algorithm for the same problem. 
Like \kGs, \SAAGs repeatedly merges the best supernode pair among some candidate pairs.
When finding the candidate pairs, \SAAGs uses a weighted sampling method designed to increase the probability that promising pairs are sampled.
To speed up the candidate search, \SAAGs maintains a tree storing the weights defined on each supernode, and it approximates \re using the count-min sketch \cite{cormode2005improved}.
Although it has lower time complexity (spec., $O(|V| \log^{2}{|V|})$), the scalability of \SAAGs is limited due to its high memory requirements for maintaining the tree.

Different from the aforementioned algorithms, which focus solely on reducing the number of supernodes by merging nodes, our proposed algorithm \method aims to minimize the size in bits of summary graphs by merging nodes and also sparsifying superedges.

A number of algorithms were developed for variants of the graph summarization problem \cite{toivonen2011compression,navlakha2008graph,shin2019sweg,ko2020incremental,khan2015set,koutra2014vog}. As outputs, \cite{navlakha2008graph,shin2019sweg,khan2015set,ko2020incremental} yield an unweighted summary graph and edge corrections (i.e., edges to be added to or removed from the restored graph).


\section{Conclusion}
\label{sec:con}
In this work, we consider a new practical variant of the graph summarization problem where the target size is given in bits rather than the number of nodes so that outputs easily fit target storage.
Then, we propose \method, a fast and scalable algorithm for concise and accurate graph summarization.
While balancing conciseness and accuracy, \method greedily combines two strategies: merging nodes and sparsifying edges. Moreover,
\method achieves linear scalability by significantly but carefully reducing the search space without sacrificing the quality of outputs much.
Throughout our extensive experiments on $10$ real-world graphs, we show that \method has the following advantages over its best competitors:
\begin{itemize}[leftmargin=*]
	\item \textbf{Concise and Accurate}: yields up to $\mathbf{11.2\times}$ {\bf more concise summary graphs} with similar reconstruction error (Fig.~\ref{fig:concise_accurate}).
	\item \textbf{Fast}: gives outputs of better quality up to $\mathbf{406.6\times}$ faster (Fig.~\ref{fig:time}).  
	\item \textbf{Scalable}: summarizes graphs with about {\bf 0.8 billion edges} (Fig.~\ref{fig:crown}), scaling linearly with the size of the input graph (Thm.~\ref{thm:time_complexity}, Fig.~\ref{fig:scal}).
\end{itemize}
\noindent \textbf{Reproducibility}: The source code and datasets used in the paper can be found at \url{http://dmlab.kaist.ac.kr/ssumm/}.

\begin{figure*}[t!]
	\centering
	\vspace{-4mm}
	\includegraphics[width=0.48\linewidth]{SsumM_Legend.pdf} \\
	\vspace{-1.9mm}
	\subfigure[DBLP]{
		\label{fig:l2:DB}
		\includegraphics[width=0.185\textwidth]{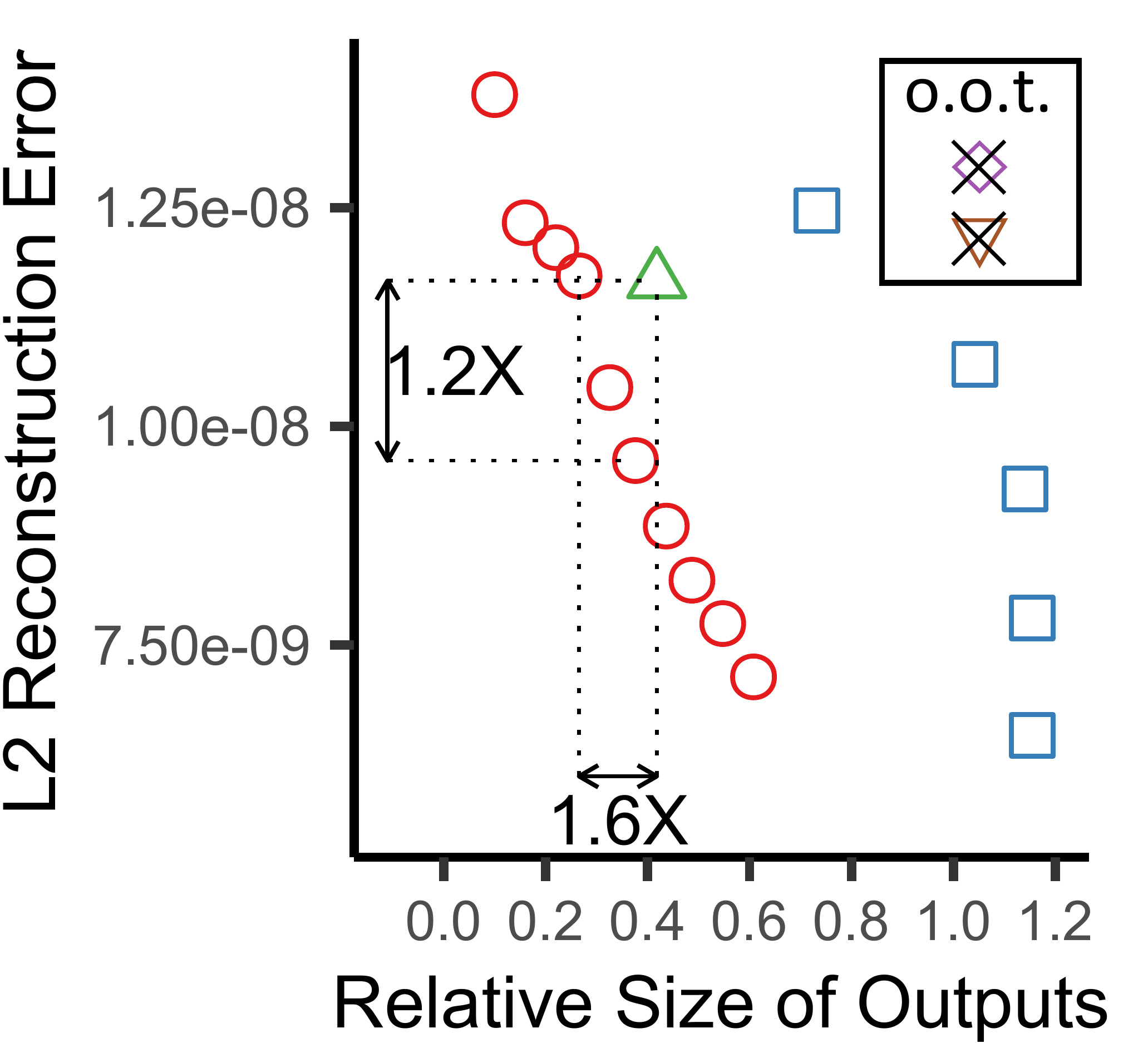}
	} 
	\subfigure[Amazon-0302]{
		\label{fig:l2:A3}
		\includegraphics[width=0.185\textwidth]{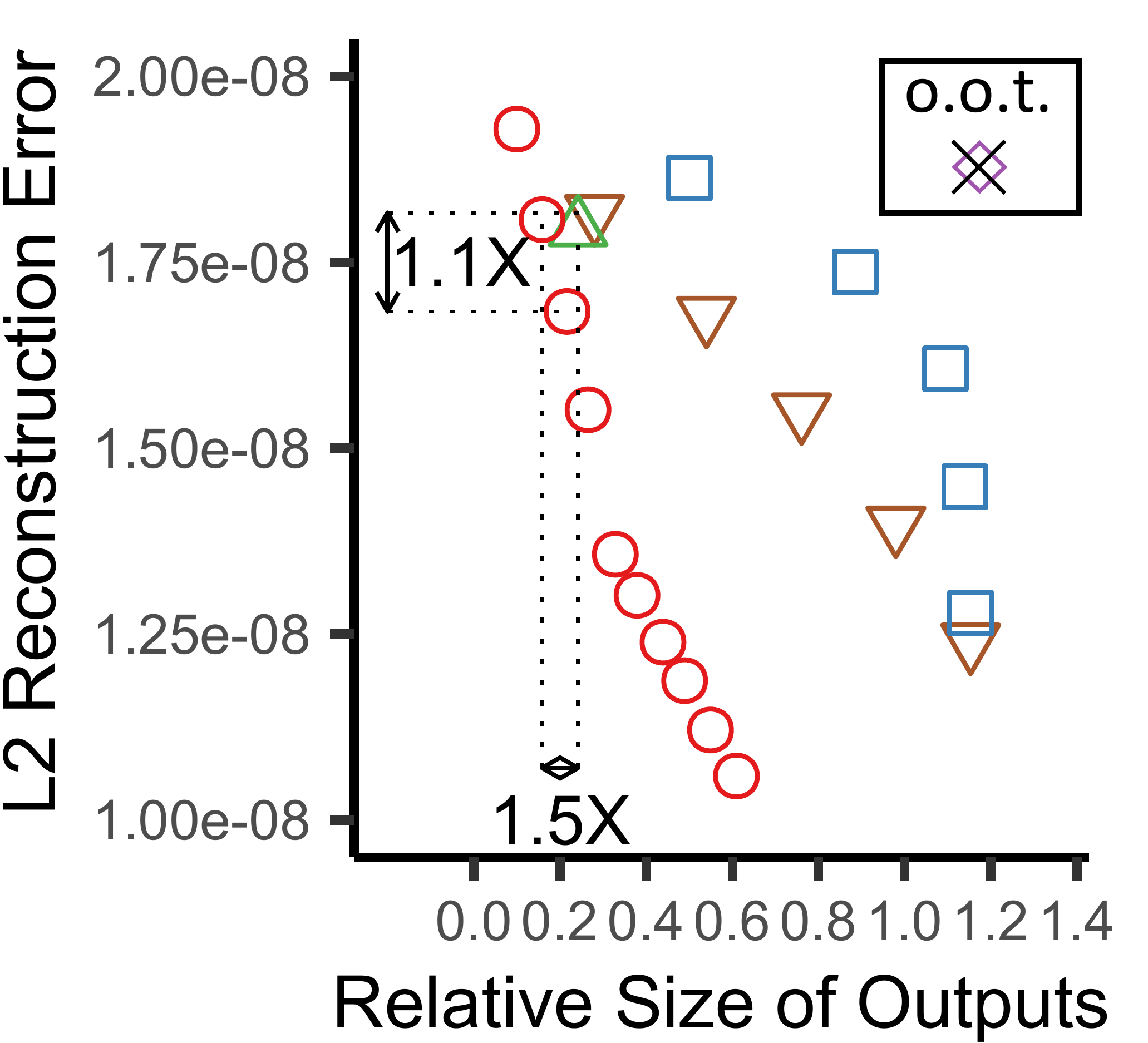}
	} 
	\subfigure[Email-Enron]{
		\label{fig:l2:EE}
		\includegraphics[width=0.185\textwidth]{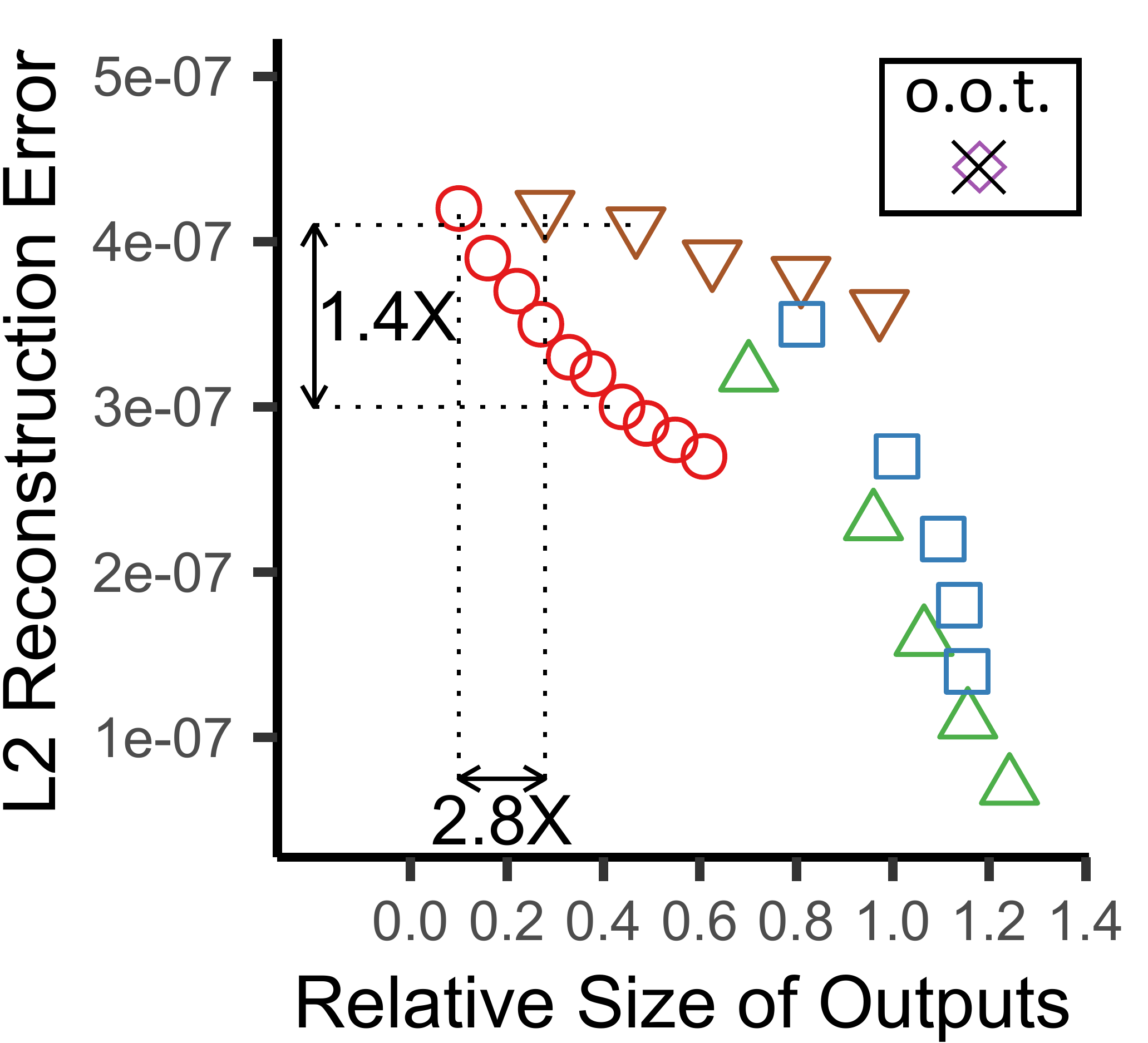}
	}
	\subfigure[Caida]{
		\label{fig:l2:CA}
		\includegraphics[width=0.185\textwidth]{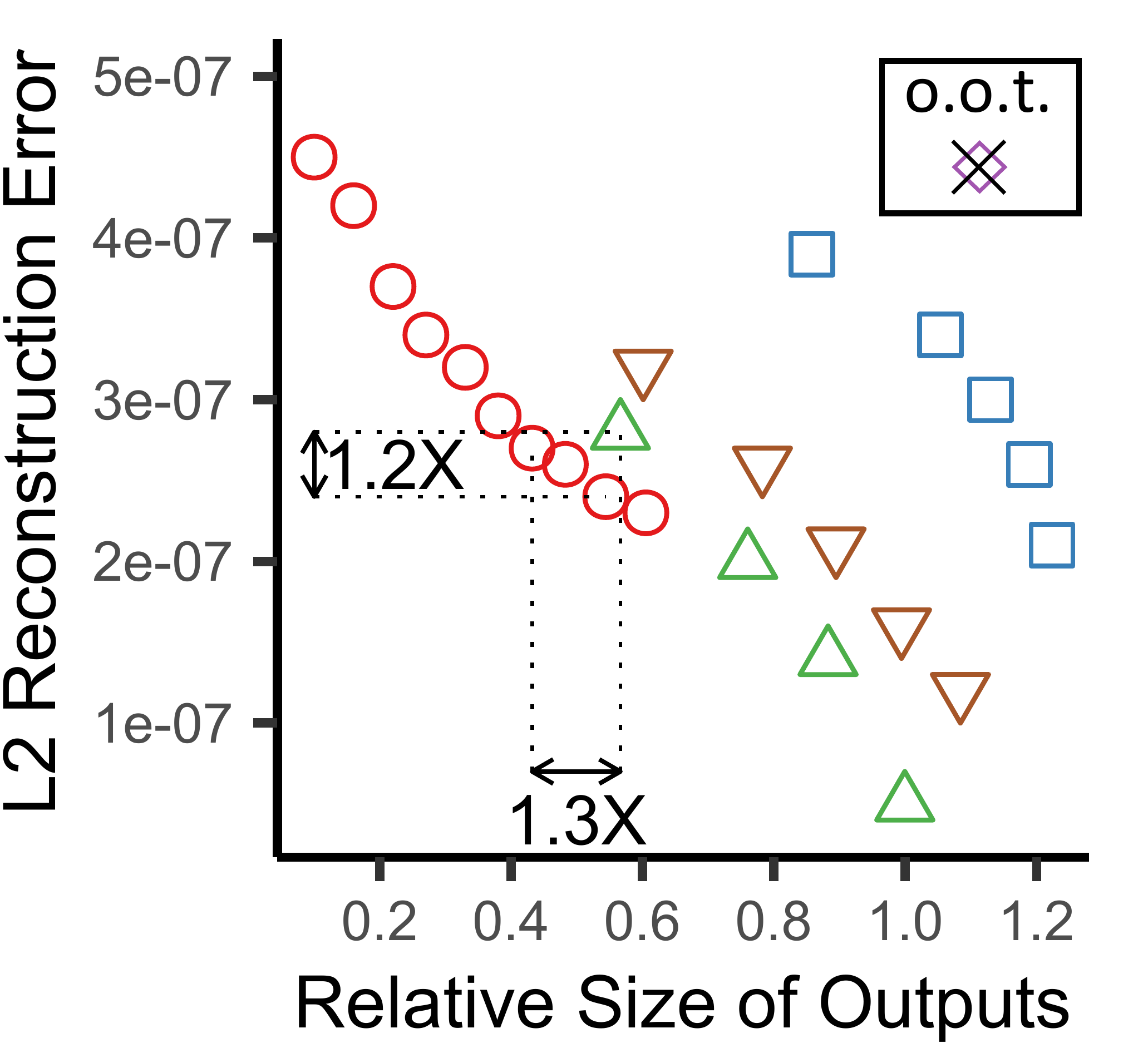}
	}
	\subfigure[Ego-Facebook]{
		\label{fig:l2:EF}
		\includegraphics[width=0.185\textwidth]{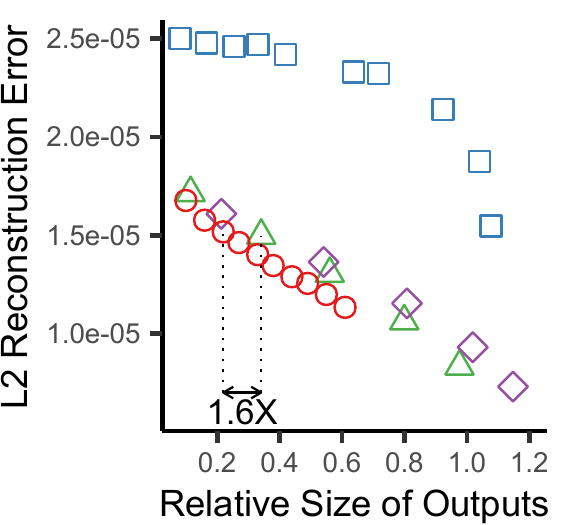}
	} \\
	\vspace{-2.5mm}
	\subfigure[Web-UK-05]{
		\label{fig:l2:W5}
		\includegraphics[width=0.185\textwidth]{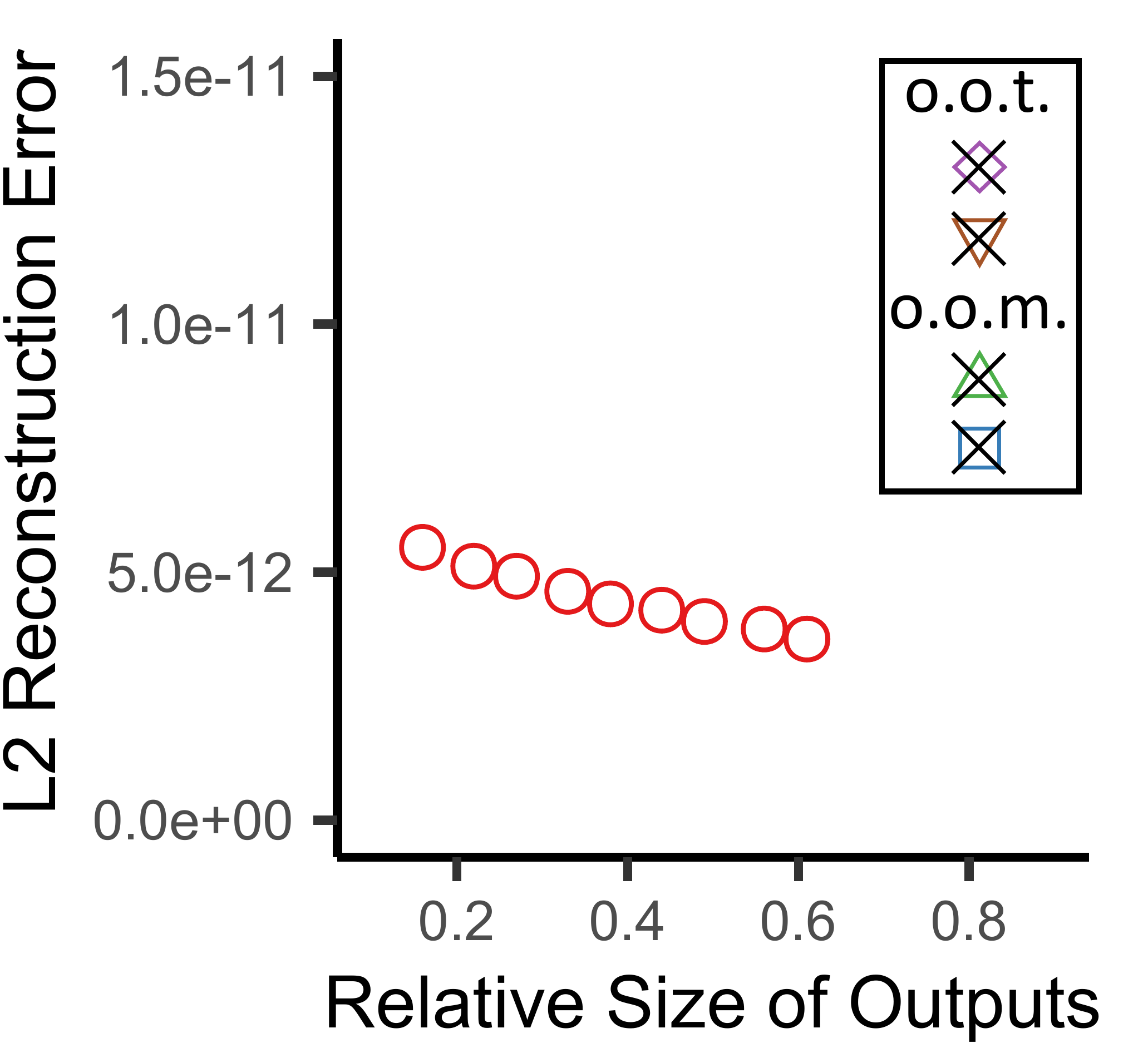}
	} 
	\subfigure[Web-UK-02]{
		\label{fig:l2:W2}
		\includegraphics[width=0.185\textwidth]{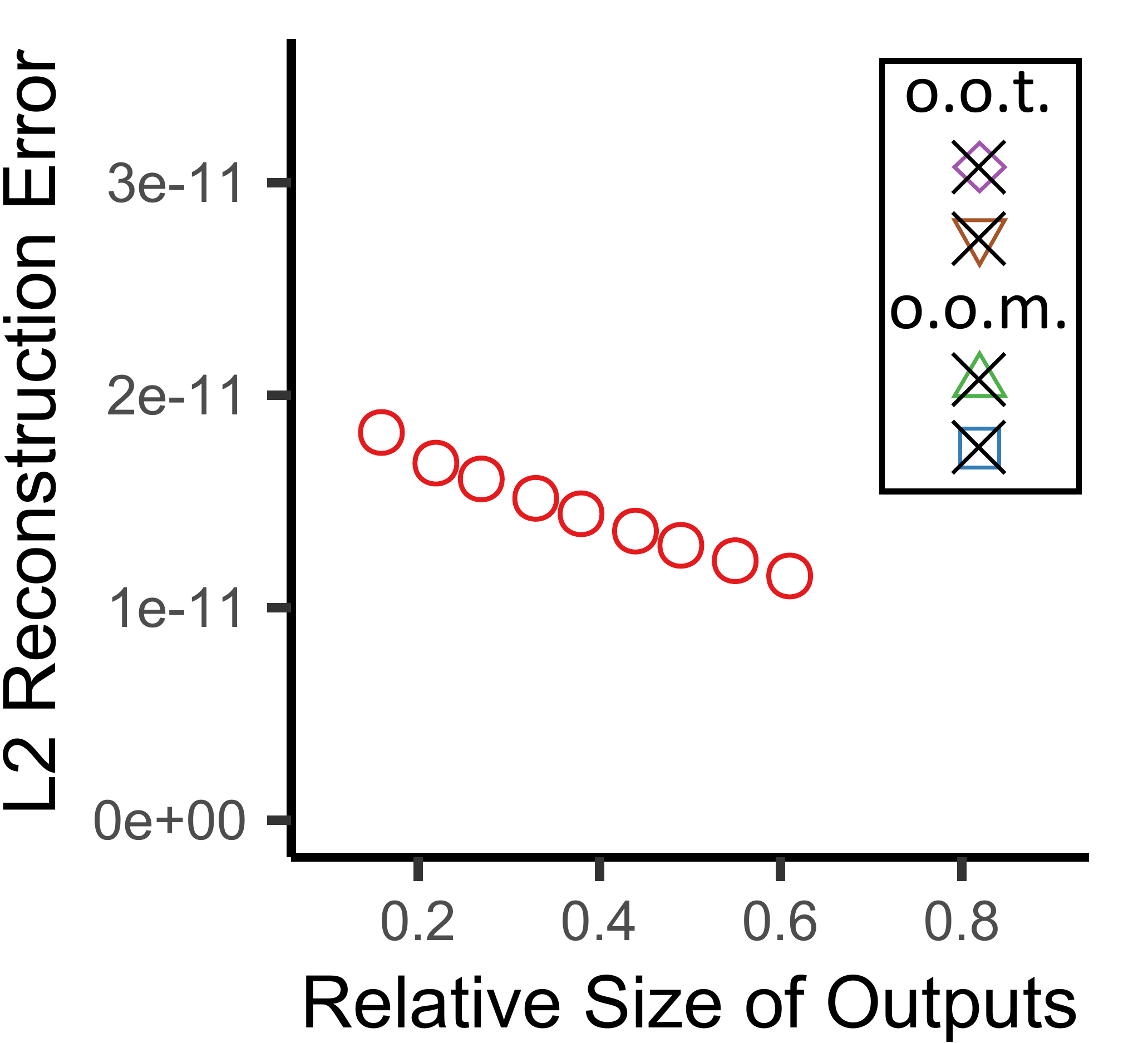}
	}
	\subfigure[LiveJournal]{
		\label{fig:l2:LJ}
		\includegraphics[width=0.185\textwidth]{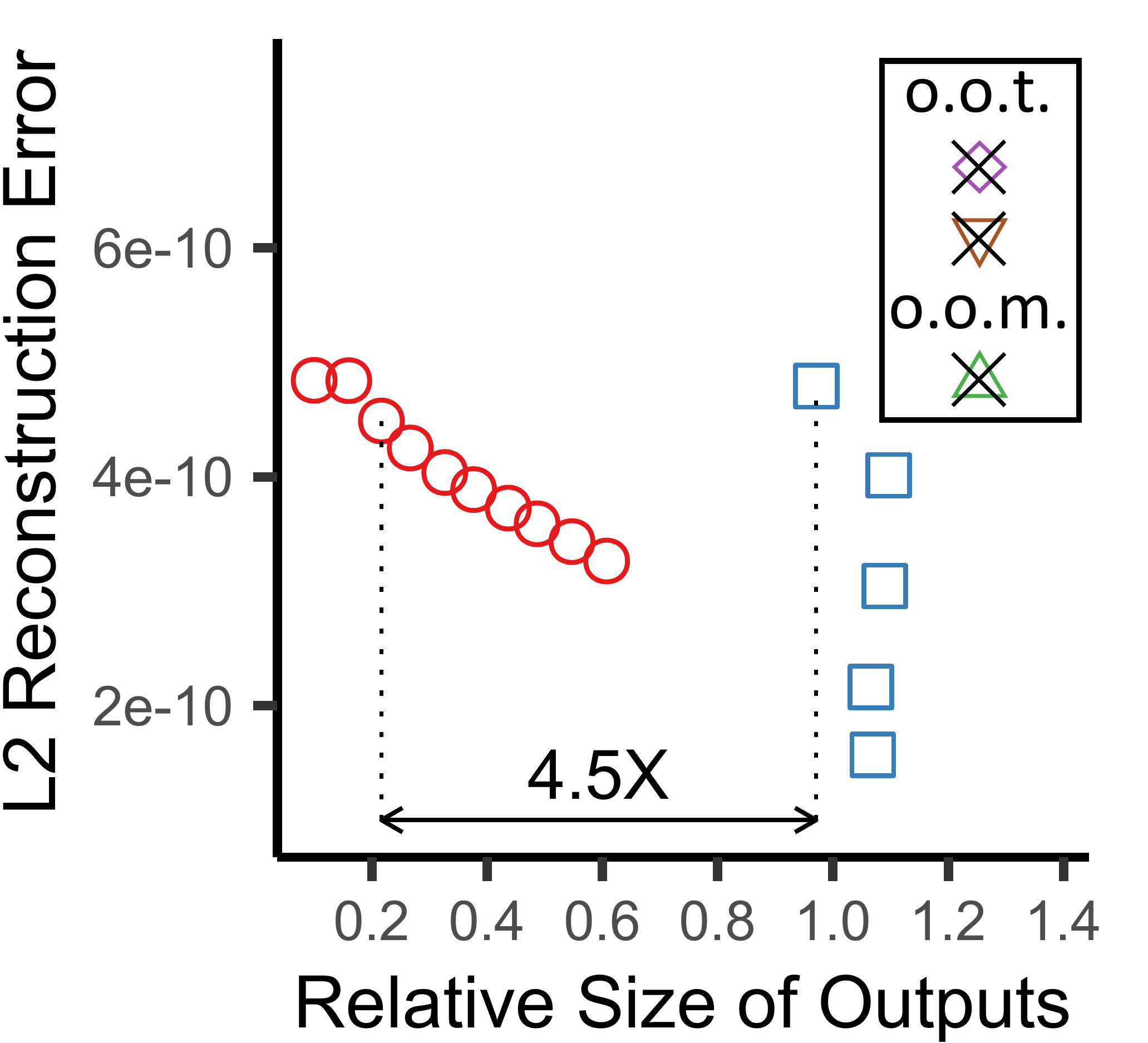}
	}
	\subfigure[Skitter]{
		\label{fig:l2:SK}
		\includegraphics[width=0.185\textwidth]{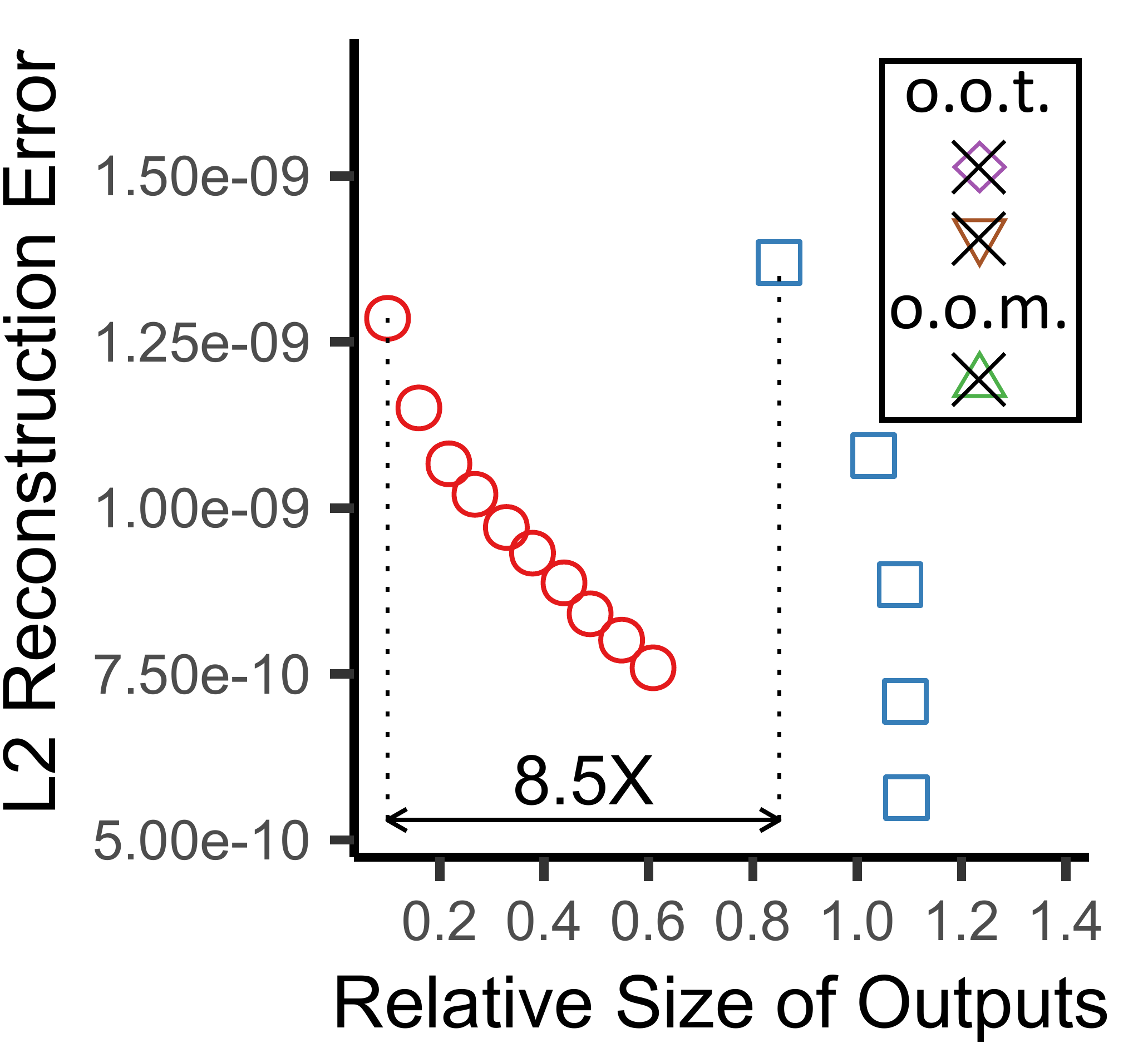}
	} 
	\subfigure[Amazon-0601]{
		\label{fig:l2:A6}
		\includegraphics[width=0.185\textwidth]{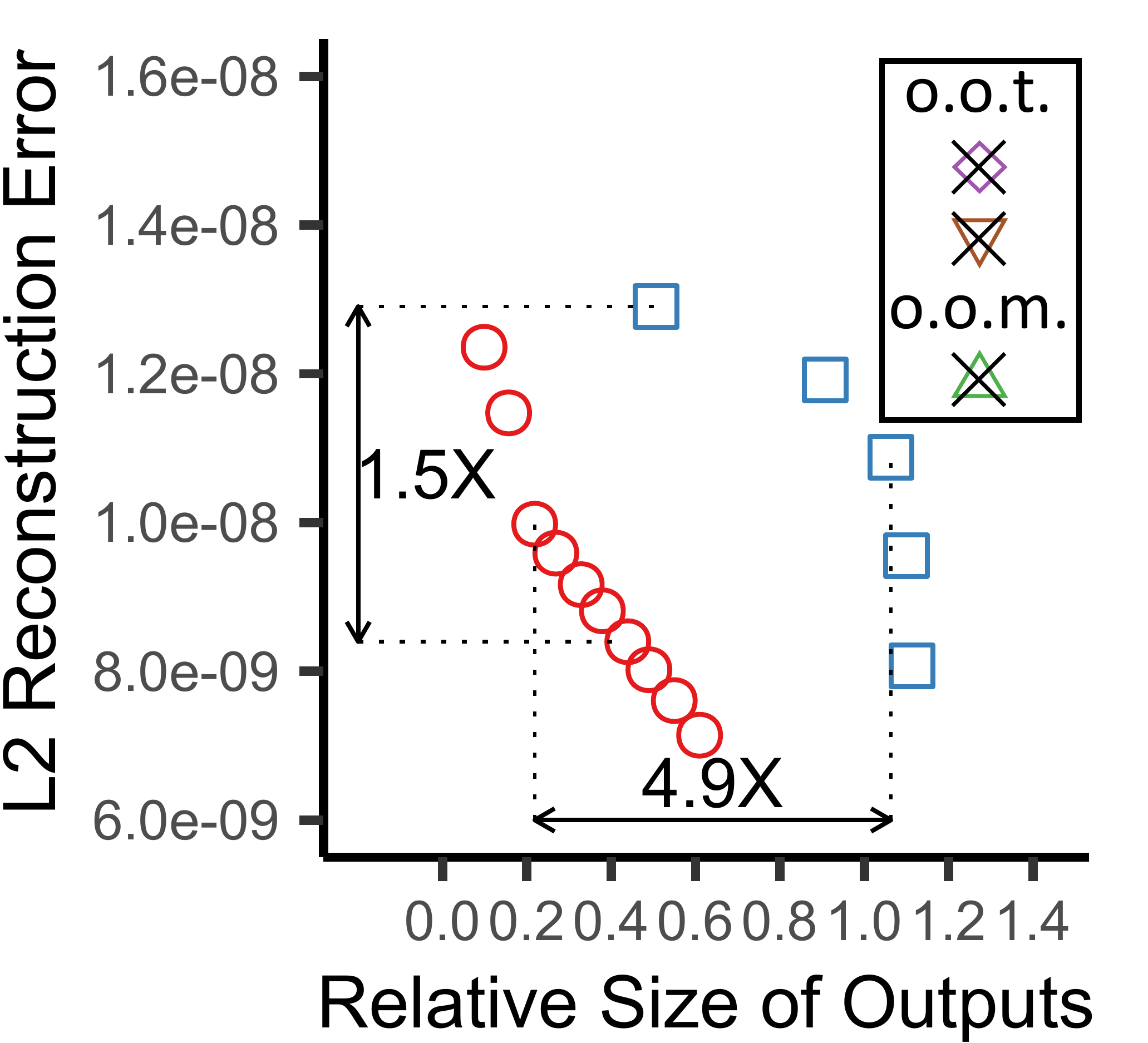}
	} 
	\\
	\vspace{-2mm}
	\caption{\underline{\smash{\method yields compact and accurate summaries.}} o.o.t.: out of time (>12hours). o.o.m.: out of memory (>64GB). Specifically, \method yielded up to $\mathbf{8.5\times}$ {\bf smaller summary graphs} with similar reconstruction error (spec., $RE_{2}$). It also achieved up to $\mathbf{1.5\times}$  {\bf smaller reconstruction error} with similarly concise outputs.\label{fig:concise_accurate:l2}}
\end{figure*}

\vspace{-1mm}
\subsection*{Acknowledgements}
\vspace{-1mm}
{\small This work was supported by National Research Foundation of Korea (NRF) grant funded by the
	Korea government (MSIT) (No. NRF-2019R1F1A1059755) and Institute of Information \& Communications
	Technology Planning \& Evaluation (IITP) grant funded by the Korea government (MSIT) (No. 2019-0-00075, Artificial Intelligence Graduate School Program (KAIST)).
}

\balance
\bibliographystyle{ACM-Reference-Format}
\bibliography{ms.bib}

\begin{figure}[t!]
	\centering
	\vspace{-1mm}
	\includegraphics[width=0.85\linewidth]{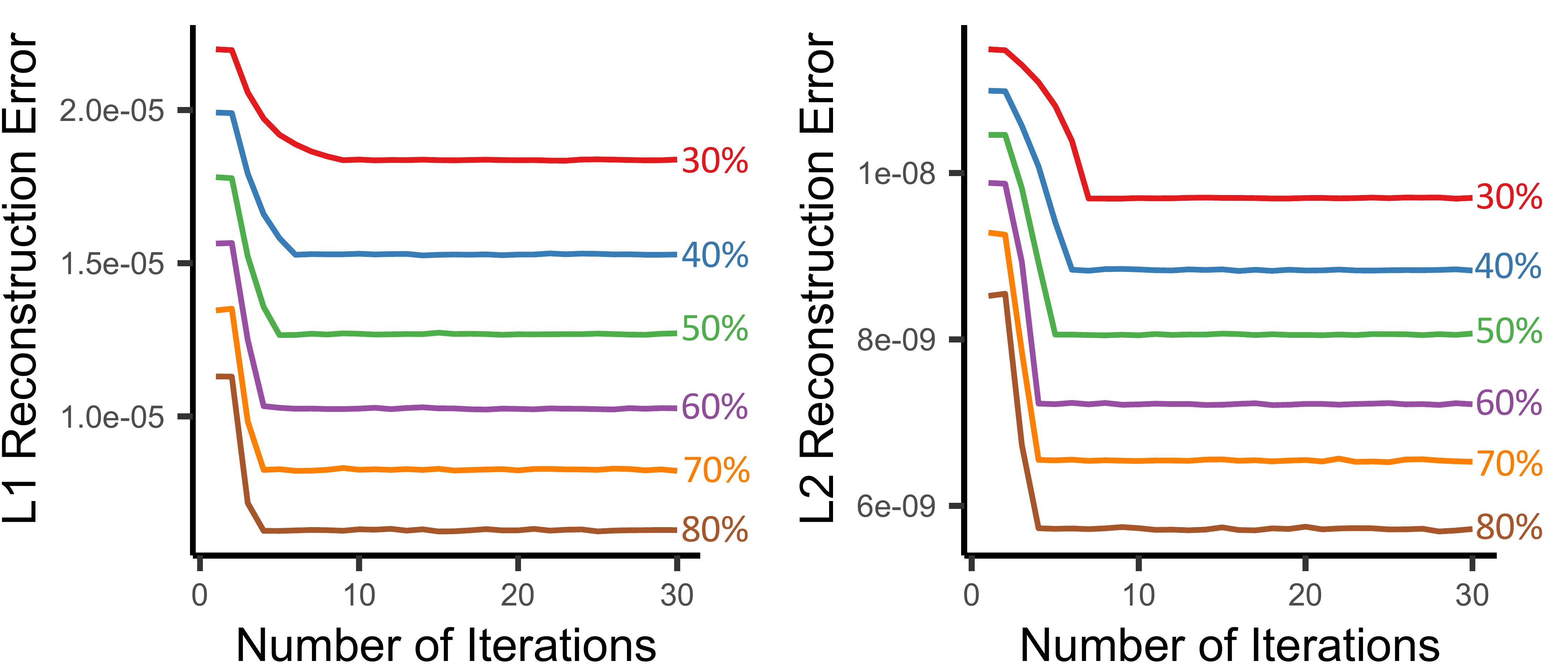} \\
	\caption{\underline{\smash{The effects of the iteration number $T$ in \method.}} Regardless of the target size, the reconstruction error of the output summary graph converged within $20$ iterations. \label{fig:paramt}}
\end{figure}

\appendix
\section{Appendix: Extra Experiments}

\subsection{Compactness and Accuracy (Fig.~\ref{fig:concise_accurate:l2})}
\label{appendix:L2}
\vspace{-1mm}

We compared the size and \ltwo \re ($RE_{2}$) of the summary graphs obtained by \method and its competitors in Fig.~\ref{fig:concise_accurate:l2}.
As in Sect.~\ref{sec:exp:compact}, where $RE_{1}$ was used, \method consistently produced more concise and accurate summary graphs than its competitors.

\vspace{-1mm}
\subsection{Effects of Parameters (Fig.~\ref{fig:paramt})}
\label{appendix:param}
\vspace{-1mm}

We measured how the number of iteration $T$ in \method affects the \re of its summary graph in the Amazon-0601 dataset by changing the target size of summary graph evenly from 30\% to 80\%. As seen in Fig.~\ref{fig:paramt}, the reconstruction error decreased over iterations and eventually converged. As the target size decreased, more iterations were needed for convergence. In all settings, however, $20$ iterations were enough for convergence.

\vspace{-1mm}
\section{Appendix: Proofs}
\label{appendix:proof}

In this section, we provide proofs of Lemmas~\ref{lemma:2hop} and \ref{lemma:3hop} in Sect.~\ref{sec:method:search:candidate}.
The proofs are based on Lemmas~\ref{lemma:merge_cost_ineq} and \ref{lemma:dense_cost_ineq}.

\begin{lemma} \label{lemma:merge_cost_ineq}
	If two supernodes $A\neq B \in S$ are merged into a single supernode $\APRIME := A\cup B$, then 
	\begin{equation}
	\CSTARSUB{AC}{S} \leq \CSTARSUB{\APRIME C}{\SPRIME}, \;\;\; \forall C \in S \setminus \{A,B\}, \label{eq:bound:edge}
	\end{equation}
	where $\SPRIME := S\cup\{\APRIME\}\setminus \{A,B\}$.
	\begin{proof}\
		Let $\CBAR:=2\log_{2}|V|+\log_{2}|E|$.
		From Eqs.~\eqref{eq:min_cost}, \eqref{eq:cost:edge}, \eqref{eq:cost:edge_simple},
		\begin{equation*}
		\CSTARSUB{\APRIME C}{\SPRIME} = \left\{
		\begin{array}{ll}
		\CBAR + Cost_{(1)}(E_{\APRIME C}|\GBSTAR{\SPRIME}) & \text{if} ~\{\APRIME ,C\} \in \PSTAR{\SPRIME} \\ 
		Cost_{(2)}(E_{\APRIME C}|\GBSTAR{\SPRIME}) & \text{otherwise}.\\
		\end{array}
		\right.
		\end{equation*}	
		We show that Eq.~\eqref{eq:bound:edge} holds by dividing into $4$ cases as follows:
		\begin{enumerate}[leftmargin=*]
		\item {\bf Case 1.} $\NPedge$ and $\ANPedge$: 
		\begin{align}
		& \CSTARSUB{AC}{S}  = Cost_{(2)}(\EAC|\GBSTAR{S}) = 2|\EAC|\log_{2}|V|  \nonumber \\ 
		& ~~
		\leq 2|\EAPRIMEC|\log_{2}|V| = Cost_{(2)}(E_{\APRIME C}|\GBSTAR{\SPRIME})=\CSTARSUB{\APRIME C}{\SPRIME}. \label{eq:bound:edge:1}
		\end{align}
		\item {\bf Case 2.} $\Pedge$ and $\APedge$: \\
		Let $\PAC := \frac{|\EAC|}{|\PIAC|}$ and $\PAPRIMEC:= \frac{|\EAPRIMEC|}{|\PIAPRIMEC|}$. Then,
		\begin{align}
		&\CSTARSUB{AC}{S}  = \CBAR + Cost_{(1)}(\EAC|\GBSTAR{S}) \nonumber\\ 
		&~~ =\CBAR -|\PIAC|(\PAC\log_{2}\PAC + (1-\PAC)\log_{2}(1-\PAC)) \nonumber\\
		&~~ \leq \CBAR -|\PIAC|(\PAPRIMEC\log_{2} \PAPRIMEC + (1-\PAPRIMEC)\log_{2}(1-\PAPRIMEC)) \nonumber \\
		&~~ \leq \CBAR -|\PIAPRIMEC|(\PAPRIMEC\log_{2} \PAPRIMEC + (1-\PAPRIMEC)\log_{2}(1-\PAPRIMEC))  \nonumber \\
		&~~  = \CBAR + Cost_{(1)}(\EAPRIMEC|\GBSTAR{S}) =\CSTARSUB{\APRIME C}{\SPRIME}, \label{eq:bound:edge:2}
		\end{align}
		where the first inequality holds by Shannon's source coding theorem \cite{shannon1998mathematical}.
		\item {\bf Case 3.} $\NPedge$ and $\APedge$:
		\begin{align*}
		& \CSTARSUB{AC}{S}  = Cost_{(2)}(\EAC|\GBSTAR{S})  
		\leq \CBAR + Cost_{(1)}(\EAC|\GBSTAR{S}) \\ 
		& ~~ \leq \CBAR + Cost_{(1)}(\EAPRIMEC|\GBSTAR{S}) =\CSTARSUB{\APRIME C}{\SPRIME},
		\end{align*}
		where the first inequality holds from the optimality of $\PSTAR{S}$, and the second one can be shown as exactly in Eq.~\eqref{eq:bound:edge:2}.
		\item {\bf Case 4.}  $\Pedge$ and $\ANPedge$:
		\begin{align*}
		& \CSTARSUB{AC}{S} = \CBAR + Cost_{(1)}(\EAC|\GBSTAR{S}) \leq Cost_{(2)}(\EAC|\GBSTAR{S}) \\
		& ~~  \leq Cost_{(2)}(\EAPRIMEC|\GBSTAR{S}) =\CSTARSUB{\APRIME C}{\SPRIME},
		\end{align*}
		where the first inequality holds from the optimality of $\PSTAR{S}$, and the second one can be shown as exactly in Eq.~\eqref{eq:bound:edge:1}. \qedhere
		\end{enumerate} 
	\end{proof}
\end{lemma}
 
\begin{lemma} \label{lemma:dense_cost_ineq} 
	If two supernodes $A \neq B \in S$ are merged into a single supernode $\APRIME := A\cup B$, then the following inequalities hold: 
	\begin{enumerate}
		\item 
		$~~ Cost_{(1)}(\EAA|\GBSTAR{S})   + Cost_{(1)}(\EBB|\GBSTAR{S})$
		\begin{equation}
		\qquad \leq Cost_{(1)}(E_{\APRIME \APRIME}|\GBSTAR{\SPRIME}), \label{eq:bound:loop:1}
		\end{equation}
		\item $~~ Cost_{(2)}(\EAA|\GBSTAR{S})  + Cost_{(2)}(\EBB|\GBSTAR{S})$
		\begin{equation}
		\qquad \leq Cost_{(2)}(E_{\APRIME \APRIME}|\GBSTAR{\SPRIME}),  \label{eq:bound:loop:2}
		\end{equation}
		\item $ $
		\vspace{-5mm}
		\begin{equation}
		\CSTARSUB{AA}{S} + \CSTARSUB{BB}{S} \leq \CBAR + \CSTARSUB{\APRIME \APRIME}{\SPRIME}, \qquad \quad \  \label{eq:bound:loop}
		\end{equation}
		\item $ $
		\vspace{-4.8mm}
		\begin{equation}
		\CSTARSUB{AA}{S} \leq \CSTARSUB{\APRIME \APRIME}{\SPRIME},\qquad \qquad \qquad \qquad \quad \quad  \label{eq:bound:loop2}
		\end{equation}
	\end{enumerate}
	where $\SPRIME := S\cup\{\APRIME\}\setminus \{A,B\}$.
\end{lemma}
\begin{proof}
		Let $\CBAR:=2\log_{2}|V|+\log_{2}|E|$. From Eqs.~\eqref{eq:min_cost}, \eqref{eq:cost:edge}, \eqref{eq:cost:edge_simple},
		\begin{equation}
		\CSTARSUB{\APRIME \APRIME}{\SPRIME} = \left\{
		\begin{array}{ll}
		\CBAR + Cost_{(1)}(E_{\APRIME \APRIME}|\GBSTAR{\SPRIME}) & \text{if} ~\{\APRIME ,\APRIME\} \in \PSTAR{\SPRIME} \\ 
		Cost_{(2)}(E_{\APRIME \APRIME}|\GBSTAR{\SPRIME}) & \text{otherwise}.
		\end{array}
		\right. \label{eq:bound:loop:3}
		\end{equation}	
		
		First, we show Eq.~\eqref{eq:bound:loop:1} holds. Let $\PAPRIME := \frac{|\EAPRIME|}{|\PIAPRIME|}$. Then, Eq.~\eqref{eq:entropy} and Shannon's source coding theorem \cite{shannon1998mathematical} imply
		\begin{align*}
		& Cost_{(1)}(\EAA|\GBSTAR{S}) + Cost_{(1)}(\EBB|\GBSTAR{S}) \\
		& \leq -(|\PIAA|+|\PIBB|)\cdot(\PAPRIME\log_{2} \PAPRIME + (1-\PAPRIME)\log_{2}(1-\PAPRIME)) \\
		& \leq -|\PIAPRIME|(\PAPRIME\log_{2} \PAPRIME + (1-\PAPRIME)\log_{2}(1-\PAPRIME)) \\
		& = Cost_{(1)}(E_{\APRIME \APRIME}|\GBSTAR{\SPRIME}), 
		\end{align*}
		 
		Second,	we show Eq.~\eqref{eq:bound:loop:2} holds.
		Eq.~\eqref{eq:entropy} and $|\EAA|+|\EBB|\leq |\EAPRIME|$ imply
		\begin{align*}
		& Cost_{(2)}(\EAA|\GBSTAR{S}) + Cost_{(2)}(\EBB|\GBSTAR{S}) \\
		& = 2*(|\EAA|+ |\EBB|)\log_{2}|V| \leq 2*|\EAPRIME|\log_{2}|V| \qquad \qquad \qquad\\
		& = Cost_{(2)}(E_{\APRIME \APRIME}|\GBSTAR{\SPRIME}).
		\end{align*}
		
		Third, we show Eq.~\eqref{eq:bound:loop} holds.
		The optimality of $\PSTAR{S}$ and Eqs.~\eqref{eq:bound:loop:1} and \eqref{eq:bound:loop:2} imply
		\begin{align}
		\CSTARSUB{AA}{S} & + \CSTARSUB{BB}{S}  \nonumber \\
		& \leq 2\CBAR + Cost_{(1)}(\EAA|\GBSTAR{S}) + Cost_{(1)}(\EBB|\GBSTAR{S}) \qquad \ \nonumber \\
		& \leq 2\CBAR + Cost_{(1)}(E_{\APRIME \APRIME}|\GBSTAR{\SPRIME}) \label{eq:bound:loop:4}
		\end{align}
		\begin{align}
		\CSTARSUB{AA}{S} & + \CSTARSUB{BB}{S} \nonumber \\
		&  \leq Cost_{(2)}(\EAA|\GBSTAR{S}) + Cost_{(2)}(\EBB|\GBSTAR{S}) \qquad \qquad \quad \nonumber \\
		&  \leq Cost_{(2)}(E_{\APRIME \APRIME}|\GBSTAR{\SPRIME}). \label{eq:bound:loop:5}
		\end{align}
		The optimality of $\PSTAR{\SPRIME}$ and Eqs.~\eqref{eq:bound:loop:4} and \eqref{eq:bound:loop:5} imply
		\begin{align*}
		\CSTARSUB{AA}{S} & + \CSTARSUB{BB}{S} \\
		& \leq \min(2\CBAR + Cost_{(1)}(E_{\APRIME \APRIME}|\GBSTAR{\SPRIME}),
		Cost_{(2)}(E_{\APRIME \APRIME}|\GBSTAR{\SPRIME})) \\
		& \leq \CBAR + \CSTARSUB{\APRIME \APRIME}{\SPRIME}.
		\end{align*}
		
		Lastly, we show Eq.~\eqref{eq:bound:loop2} holds. The optimality of $\PSTAR{S}$ and Eqs.~\eqref{eq:bound:loop:1} and \eqref{eq:bound:loop:2} imply
		\begin{align}
		\CSTARSUB{AA}{S} & \leq \CBAR + Cost_{(1)}(\EAA|\GBSTAR{S}) \nonumber \\
		& \leq \CBAR + Cost_{(1)}(E_{\APRIME \APRIME}|\GBSTAR{\SPRIME}), \label{eq:bound:loop:6}
		\end{align}
		\begin{align}
		\CSTARSUB{AA}{S} & \leq Cost_{(2)}(\EAA|\GBSTAR{S})  \quad \quad \ \ \ \nonumber \\
		& \leq Cost_{(2)}(E_{\APRIME \APRIME}|\GBSTAR{\SPRIME}). \label{eq:bound:loop:7}
		\end{align}
		The optimality of $\PSTAR{\SPRIME}$ and Eqs.~\eqref{eq:bound:loop:3}, \eqref{eq:bound:loop:6}, and \eqref{eq:bound:loop:7} imply 
		\begin{align*}
		&\CSTARSUB{AA}{S}  \\
		&~~~ \leq \min(\CBAR + Cost_{(1)}(E_{\APRIME \APRIME}|\GBSTAR{\SPRIME}),
		Cost_{(2)}(E_{\APRIME \APRIME}|\GBSTAR{\SPRIME})) \\
		&~~~ \leq \CSTARSUB{\APRIME \APRIME}{\SPRIME}. \qedhere
		\end{align*}
\end{proof}

\subsection{Proof of Lemma~\ref{lemma:2hop}}
\label{appendix:proof:2hop}

\begin{figure}[t]
	\centering
		\includegraphics[width=0.5\linewidth]{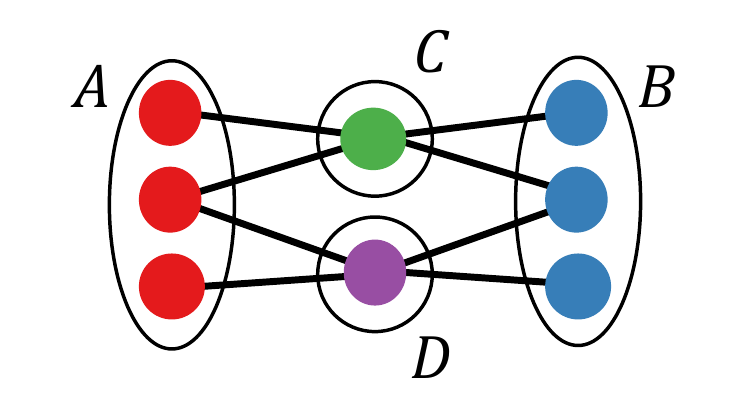}
		\label{fig:proof:2-hop}
	\caption{\label{fig:proof} An example pair of supernodes $\boldsymbol{\{A, B\}}$ which are $\boldsymbol{2}$ hops away from each other.}
\end{figure}

\begin{proof}
	Suppose two  $A\neq B\in S$ that are within $2$ hops are merged into a single supernode $\APRIME:=A\cup B$, and
	without loss of generality, $\CSTARSUB{A}{S} \geq \CSTARSUB{B}{S}$.
	We let 
	$\SPRIME := S\cup\{\APRIME\}\setminus \{A,B\}$.
	
	We first show that Eq.~\eqref{eq:2hop} holds.
	From Eqs.~\eqref{eq:bound:edge} and \eqref{eq:bound:loop2}, 
	\begin{align}
	& \CSTARSUB{A}{S} - \CSTARSUB{AB}{S} = \CSTARSUB{AA}{S} + \sum\nolimits_{C \in S\setminus \{A,B\}}\CSTARSUB{AC}{S} \nonumber \\
	& \leq \CSTARSUB{\APRIME\APRIME}{\SPRIME} + \sum\nolimits_{C \in S\setminus \{A,B\}}\CSTARSUB{\APRIME C}{\SPRIME}  = \CSTARSUB{\APRIME}{\SPRIME}. \label{eq:2hop:1}
	\end{align}
	 Eq.~\eqref{eq:reduce}, Eq.~\eqref{eq:2hop:1}, and $\CSTARSUB{A}{S} \geq \CSTARSUB{B}{S}$ imply
	Eq.~\eqref{eq:2hop}.
	
%
%
	
	Now, we show that Eq.~\eqref{eq:2hop} is tight. That is, we show that there exists $A\neq B\in S$ within $2$ hops where
	\begin{equation}
	\reductionAB = \min(\CSTARSUB{A}{S},\CSTARSUB{B}{S}). \label{eq:2hop:tight}
	\end{equation}
	Fig.~\ref{fig:proof}, where $A$ and $B$ are 2 hops away from each other, provides such an example.
	In the example, 
	\begin{align*}
	\CSTARSUB{A}{S} = \CSTARSUB{B}{S} = 2\cdot(2\log_{2}|V|+\log_{2}|E|) = \CSTARSUB{\APRIME \APRIME}{\SPRIME},
	\end{align*} and
	$\CSTARSUB{AB}{S} = 0$.
	Hence, 
	\begin{align*}
	\reductionAB & = 2\cdot(2\log_{2}|V|+\log_{2}|E|) \\
	& = \min(\CSTARSUB{A}{S},\CSTARSUB{B}{S}). \qedhere
	\end{align*} 
\end{proof}

\subsection{Proof of Lemma~\ref{lemma:3hop}}
\label{appendix:proof:3hop}

\begin{proof}
	Suppose two supernodes $A\neq B\in S$ that are $3$ or more hops away from each other are merged into a single supernode $\APRIME:=A\cup B$. Then, the following equalities hold:
	\begin{equation}
	\CSTARSUB{AB}{S}= 0,
	\label{eq:2hop_zero}
	\end{equation}
	\begin{equation}
	\CSTARSUB{AC}{S}= 0\;\; or \;\; \CSTARSUB{BC}{S} = 0, \;\;\; \forall C \in S \setminus \{A,B\}. \label{eq:zero}
	\end{equation}
	Eqs.~\eqref{eq:bound:edge}, \eqref{eq:2hop_zero}, and \eqref{eq:zero} imply
	\begin{equation}
	\CSTARSUB{AC}{S} + \CSTARSUB{BC}{S} \leq  \CSTARSUB{\APRIME C}{\SPRIME}, \;\;\; \forall C \in S \setminus \{A,B\}, \label{eq:bound:edge:sum}
	\end{equation}
	where 
	$\SPRIME := S\cup\{\APRIME\}\setminus \{A,B\}$.
	Then, Eqs.~\eqref{eq:bound:loop}, \eqref{eq:2hop_zero} and \eqref{eq:bound:edge:sum} imply
	\begin{align}
	 \CSTARSUB{A}{S} & + \CSTARSUB{B}{S} - \CSTARSUB{AB}{S} \nonumber \\
	& =  \CSTARSUB{AA}{S} + \CSTARSUB{BB}{S} + \CSTARSUB{AB}{S} \nonumber\\
	& \quad + \sum\nolimits_{C \in S\setminus \{A,B\}}(\CSTARSUB{AC}{S}+\CSTARSUB{BC}{S}) \nonumber \\
	& \leq \CBAR + \CSTARSUB{\APRIME\APRIME}{\SPRIME} + \sum\nolimits_{C \in S\setminus \{A,B\}}\CSTARSUB{\APRIME C}{\SPRIME}  \nonumber \\
	& = \CBAR + \CSTARSUB{\APRIME}{\SPRIME}, \label{eq:3hop:1}
	\end{align}
	where $\CBAR:=2\log_{2}|V|+\log_{2}|E|$. Eqs.~\eqref{eq:3hop:1} and \eqref{eq:reduce} imply Eq.~\eqref{eq:3hop}.
	\end{proof}

\end{document}